\documentclass{aa}  
\usepackage{graphicx}
\usepackage{txfonts}
\usepackage{hyperref}
\usepackage{natbib}
\usepackage{xcolor}
\usepackage{verbatim}
\usepackage{soul}
\usepackage{textgreek}
\usepackage{CJKutf8}

\newcommand{\ms}{\,\mathrm{M}_\odot}
\newcommand{\msol}{M_\odot}

\newcommand{\lsol}{L_\odot}
\newcommand{\kms}{\,\mathrm{km/s}}
\newcommand{\days}{\,\mathrm{d}}
\newcommand{\unit}[1]{\,\mathrm{#1}}
\newcommand{\combine}{\textsc{ComBinE}}

\newcommand{\vrot}{\varv_\mathrm{rot}} 
\newcommand{\vcr}{\varv_\mathrm{cr}}
\let\Omega\varOmega

\def\simle{\mathrel{\hbox{\rlap{\hbox{\lower4pt\hbox{$\sim$}}}\hbox{$<$}}}}
\def\simgr{\mathrel{\hbox{\rlap{\hbox{\lower4pt\hbox{$\sim$}}}\hbox{$>$}}}}

\begin{document} 

%\title{Massive Be black hole binaries in the Small Magellanic Cloud}
%\title{X-ray silent compact object binaries in the Small Magellanic Cloud. I. Prediction from rapid binary evolution}
%\title{Spins and compact companions of massive main-sequence stars in the SMC I: predictions from the rapid binary evolution code \combine}
\title{Populations of evolved massive binary stars in the Small Magellanic Cloud II: Predictions from rapid binary evolution} 
%code \combine}

\author{
C.~Schürmann\inst{1}\fnmsep\thanks{Mail to \texttt{chr-schuermann@uni-bonn.de}}, 
X.-T.~Xu \begin{CJK}{UTF8}{gkai}(徐啸天)\end{CJK}\inst{1}\fnmsep\thanks{Mail to \texttt{xxu.astro@outlook.com}}, 
N.~Langer\inst{1}\fnmsep\inst{2}, 
D.~Lennon\inst{3}\fnmsep\inst{4}, 
M.~U.~Kruckow\inst{5}\fnmsep\inst{6}, 
J.~Antoniadis\inst{7}\fnmsep\inst{2}, 
F.~Haberl\inst{8}, 
A.~Herrero\inst{3}\fnmsep\inst{4}, 
M.~Kramer\inst{2}, 
A.~Schootemeijer\inst{1},
T.~Shenar\inst{9}, 
T.~M.~Tauris\inst{10}, 
C.~Wang\inst{11}
}

\institute{
Argelander-Institut für Astronomie, Universität Bonn, Auf dem Hügel 71, 53121 Bonn, Germany \and 
Max-Planck-Institut für Radioastronomie, Auf dem Hügel 69, 53121 Bonn, Germany \and
Instituto de Astrofísica de Canarias, 38200 La Laguna, Tenerife, Spain \and
Departamento de Astrofísica, Universidad de La Laguna, 38205 La Laguna, Tenerife, Spain \and
Département d’Astronomie, Université de Genève, Chemin Pegasi 51, CH-1290 Versoix, Switzerland \and
Gravitational Wave Science Center (GWSC), Université de Genève, 24 quai E. Ansermet, CH-1211 Geneva, Switzerland \and
Institute of Astrophysics, Foundation for Research \& Technology Hellas (FORTH), GR-70013 Heraklion, Greece \and
Max-Planck-Institut für Extraterrestrische Physik, Gießenbachstraße 1, 85748 Garching, Germany \and
Tel Aviv University, The School of Physics and Astronomy, Tel Aviv 6997801, Israel \and
Department of Materials and Production, Aalborg University, Fibigerstr{\ae}de 16, 9220 Aalborg, Denmark \and
Max-Planck-Institut für Astrophysik, Karl-Schwarzschild-Straße 1, 85748 Garching, Germany
%\and Yunnan Obervatories, Chinese Academy of Sciences, Kunming 650216, China
}

\authorrunning{C.~Schürmann et al.}
\titlerunning{Synthetic populations of evolved massive binary stars in the SMC -- Part II}

\date{Submitted 30/3/2025, accepted ???}

\abstract
%context
{Massive star evolution plays a crucial role in astrophysics but bares large uncertainties. This problem becomes more severe by the majority of massive stars being born in close binary systems, whose evolution is affected by the interaction of their components.}
%aims
{We want to constrain major uncertainties in massive binary star evolution, in particular the efficiency and the stability of the first mass transfer phase.}
%methods
{We use the rapid population synthesis code \combine\ to generate synthetic populations of post-interaction binaries, assuming constant mass-transfer efficiency. We employ a new merger criterion that adjusts self-consistently to any prescribed mass-transfer efficiency. We tailor our synthetic populations to be comparable to the expected binary populations in the Small Magellanic Cloud (SMC).}
%results
{We find that the observed populations of evolved massive binaries can not be reproduced with a single mass-transfer efficiency. Instead, a rather high efficiency ($\simgr 50$\%) is needed to reproduce the number of Be~stars and Be/X-ray binaries in the SMC, while a low efficiency ($\sim 10$\%) leads to a better agreement with the observed number of Wolf-Rayet stars. We construct a corresponding mass-dependent mass-transfer efficiency recipe to produce our fiducial synthetic SMC post-interaction binary population. It reproduces the observed number and properties of the Be/X-ray and WR-binaries rather well, and is not in stark disagreement with the observed OBe~star population. It further predicts two large, yet unobserved populations of OB+BH binaries, that is $\sim 100$ OB+BH systems with rather small orbital periods ($\simle 20\days$) and $\sim 40$ longer period OBe+BH systems.}
%conclusions
{Searches for these system may strongly advance our understanding of massive binary evolution.}


\maketitle

\section{Introduction}

Massive stars are powerful cosmic engines. They produce spectacular events like supernovae and gamma-ray bursts, drive the chemical evolution of galaxies \citep{1995ApJS...98..617T} and shape their interstellar medium \citep{2014MNRAS.445..581H}. Massive stars are preferentially born in close binary systems \citep{2012Sci...337..444S} which produce X-ray binaries and gravitational wave mergers \citep{2006csxs.book..623T,2023pbse.book.....T}, but our understanding of their evolution is far from complete \citep{2012ARA&A..50..107L,2022arXiv220904165L}.

\begin{figure}
 \centering
 \includegraphics[width=\columnwidth]{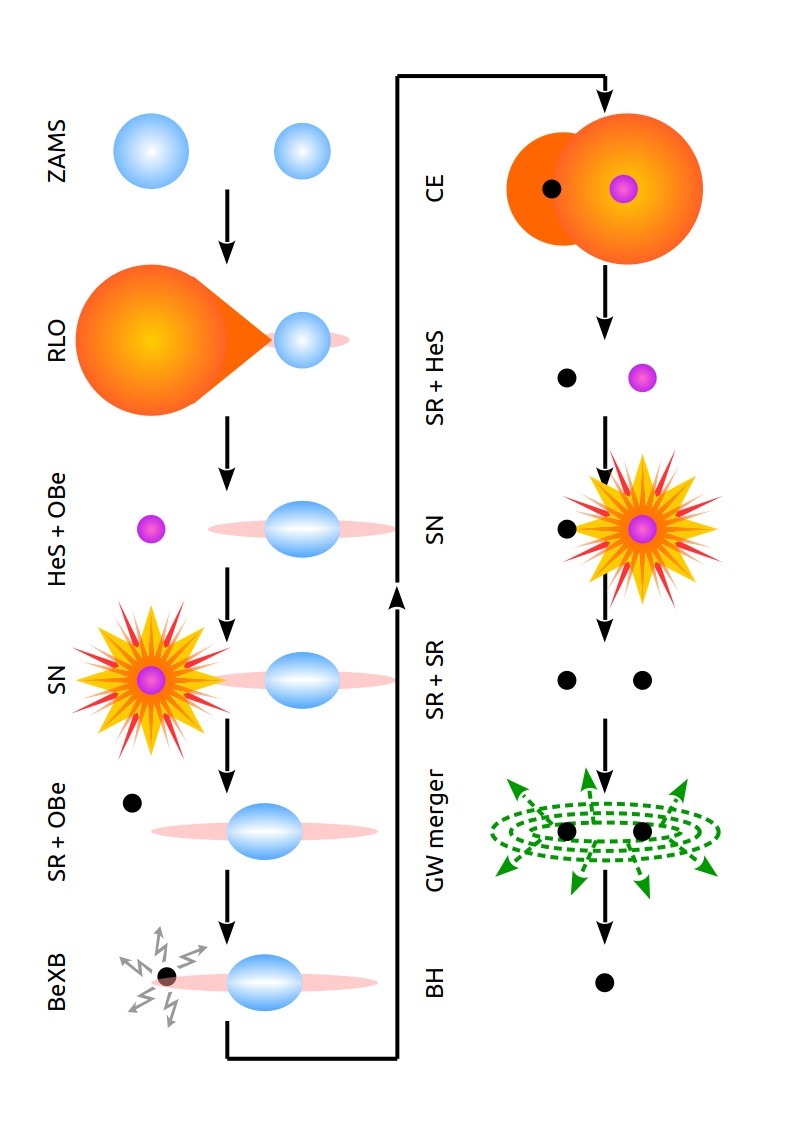}
 \caption{A schematic evolutionary path from ZAMS to gravitational wave merger, adapted from \citet{2018MNRAS.481.1908K}. We focus on the BH+OBe phase. ZAMS = zero age main-sequence, RLO = Roche-lobe overflow, HeS = helium star, OBe = O/B~type emission line star, SN = supernova, SR = stellar remnant BeXB = Be/X-ray binary, CE = common envelope, GW = gravitational wave, BH = black hole.}
 \label{path}
\end{figure}

The different types of massive binaries can be ordered \citep[e.g.][]{2006csxs.book..623T} in a simplified evolution scheme, from zero-age main-sequence binaries up to compact object mergers (Fig.~\ref{path}). However, the vast majority of massive binary star will exit the evolutionary path, either by breaking up as a consequence of core collapse \citep{1998A&A...330.1047T}, or by merging to a single object due to the interaction between the two stellar components \citep{1992ApJ...391..246P,2012ARA&A..50..107L,2014ApJ...782....7D}. A critical step in this evolutionary scheme is the Roche-lobe overflow (RLO), where matter is transferred from one star (the "donor") to the other (the "accretor") \citep{1967ZA.....65..251K}. In the first RLO, usually, the heavier star is the donor as it burns its fuel faster and expands first. This can either happen during its main-sequence phase (Case~A) or during shell burning \citep[Case~B,][]{1967ZA.....65..251K}. The physics of the RLO is not well understood, especially under which conditions decide between to donor stripping or merger \citep{2013A&ARv..21...59I}, which fraction of material lost by the donor is deposited on the accretor \citep{2007A&A...467.1181D}, and how much angular momentum the ejected material removes from the system \citep{2024ARA&A..62...21M}.

Stable mass transfer is expected to lead to an almost complete removal of the hydrogen-rich envelope of the donor. While this stripped star may retain a thin hydrogen layer \citep{2020A&A...637A...6L} or may only partially strip \citep{2024A&A...685A..58E}, it is commonly referred to as a helium star (HeS). The accretor receives mass, which may rejuvenate the star \citep{1995A&A...297..483B}, and gains spin angular momentum. \cite{1981A&A...102...17P} showed that only a small amount of accreted material is enough to bring the accretor to rotate close to critical rotation.

%, where the centrifugal forces equate the star's gravity \citep{2013A&ARv..21...69R}. The critical rotation rate $\omega_\mathrm{cr}$ and the critical rotation velocity $\vcr$ are given by 
%\begin{equation}
% \omega_\mathrm{cr} = \sqrt{\frac{GM}{R_\mathrm{eq,cr}^3}}, \quad \vcr = \omega_\mathrm{crit} R_\mathrm{eq,cr}
%\end{equation}
%with $M$ the star's mass and $R_\mathrm{eq,crit}$ its critical equatorial radius \citep{2013ApJ...764..166D}. Assuming Roche geometry \citep{2012sse..book.....K} one finds
%\begin{equation}
%    R_\mathrm{eq,cr} = \frac32 R_\mathrm{p},
%\end{equation}
%where $R_\mathrm{p}$ is the (unchanged) polar radius.

Rotating near critical rotation, such a star expels matter from its surface, forming a decretion disk around its equator. This disk causes broad emission lines to appear in the spectrum. They and the rotationally broadened absorption lines of the star serve as an explanation for the Be~star phenomenon \citep{2003PASP..115.1153P,2013A&ARv..21...69R}. It is however possible, that either the mass loss (and thus angular momentum loss) or tidal forces brake stellar rotation, or that strong stellar winds remove the disk so that the OBe~star becomes a usual OB~type star again. On the other side, it can happen that the HeS expands so much that it initiates a second mass transfer \citep[Case~BB/BC][]{2015MNRAS.451.2123T}%, which we will refer to a Case~(A)BC\footnote{"A" is used if a Case~A mass transfer occurred. We use "BC" instead of the commonly used "BB" as the donor star is helium burning, which is commonly marked by a "C"}.
The HeS terminates its life shortly after and becomes a stellar remnant (SR), i.e. a white dwarf (WD), neutron star (NS), or a black hole (BH). The following phase (OB+SR) is the focus of this work, as it is the second best observable one, since it features a long-living main-sequence star and an inert SR.

This formation channel of OBe~stars by binary interaction is supported by the fact that no OBe~star with main-sequence companions are observed \citep{2020A&A...641A..42B}, and on the other hand OBe~stars with hot subdwarf \citep[sdOB, the observationally counterpart to light HeSs,][]{2021AJ....161..248W}, WD \citep{2023RAA....23b5021Z} and NS \citep{2015MNRAS.452..969C} companions are well known. However, only a few Be+sdOB and Be+WD systems have been discovered since the light and faint companion is hard to detect \citep{2020MNRAS.497L..50C,2021MNRAS.508..781K,2024MNRAS.534.1937G,2025ApJ...980L..36M}. However they can be vary X-ray bright (Eddington luminosity) when they show nova-like outbursts. Several of these have been discovered recently \citep[][and references therein]{2025ApJ...980L..36M}. Be+NS systems appear often as Be/X-ray binaries (BeXB) as the NS accretes material from the disk and releases X-ray emission. Another proposed formation processes of OBe~stars is single star evolution which was recently studied by \citet{2008A&A...478..467E} and \citet{2020A&A...633A.165H}. 

The success of this picture let the question about the not yet observed configurations arise. Especially, as many massive stars with NS companions are known, one would assume that OBe~star systems with BH companions exist, too. Today only one system of that kind is proposed \citep[MWC~656,][]{2014Natur.505..378C} but its nature is not undisputed \citep{2022arXiv220812315R,2023A&A...677L...9J}. On the other hand, three super-giant/X-ray binaries with BHs (Cyg~X-1, LMC~X-1, and M33~X-7) are well observed \citep{2007ATel..977....1O,2009ApJ...697..573O,2011ApJ...742...84O,2021Sci...371.1046M,2022A&A...667A..77R} and even one source with an X-ray quiet BH is known \citep[VFTS~243,][]{2022NatAs...6.1085S}. Therefore, an objective of this study is to answer whether it is possible to form OBe+BH systems and to predict their properties. After the OBe+SR phase the system can evolve through a common envelope ejection to a binary SR, which may undergo a gravitational wave merger (Fig.~\ref{path}). Thus, the OB+SR phase is an important intermediate step to understand binary stellar evolution as a whole. % pulsar+BH, WR+NS/BH.

%The binary evolution scenario examined in this work involves several phases of direct interaction between the two stars and is illustrated in Figure~\ref{path}. Previous work on this channel was done by e.g. \citet{2003MNRAS.342.1169V, 2006csxs.book..623T, 2008ApJS..174..223B, 2016Natur.534..512B, 2017ApJ...846..170T, 2018MNRAS.481.1908K}. This evolutionary path contains many side-channels \citep{2018MNRAS.481.1908K} and is able to reach many pre-merger configurations from a wide range of initial systems and is therefore more potent than the chemically homogeneous channel. It contains two key features: firstly a Roche-lobe overflow (RLO), which sends matter and angular momentum to a MS star turning it into a Be-star. After that a CC is formed, which can but does not need to be an X-ray source and secondly a common envelope (CE) phase shrinking the systems and making a merger event possible.

In this study we focus on stars in the Small Magellanic Cloud (SMC). As a low metallicity satellite galaxy of the Milky Way, it is a distinctive environment. Its metallicity of about one fifth of the solar value \citep{1999ApJ...518..405V, 2000A&A...353..655K, 2008A&A...479..541H} corresponds to to the average metallicity of a redshift of about 3 \citep{2007ASSP....3..435K}. Low metallicities are interesting as the aforementioned events (supernovae, gamma-ray bursts and gravitational waves) originate predominately in low metallicity environment \citep{2019PhRvX...9c1040A, 2023PhRvX..13a1048A}. Being located at a distance of about $60\unit{kpc}$, observing individual stars is still possible. Another advantage of SMC are the weaker stellar winds caused by the low metallicity \citep{1982ApJ...259..282A, 1987A&A...173..293K, 2007A&A...473..603M}. Lower winds imply a lower loss of angular momentum and thus we expect OBe~stars to be more numerous and at higher masses and hence a higher success rate in finding OBe+BH systems.

This study is closely related to \citet[][hereafter Paper~I]{xu}, who investigate the same types of objects using detailed stellar models while we use rapid binary evolution. Our work is structured as follows. Sect.~\ref{method} describes our methods to simulate systems consisting of a hydrogen-burning star with a SR. In Sect.~\ref{sec-pq} we describe how the outcome of binary evolution code depends on the initial parameters of the binary. In Sect.~\ref{results_var} we use different physical assumptions to generate artificial populations and compare them to observational characteristics to determine a fiducial set of assumptions. The population based on them is analysed in detail in Sect.~\ref{results_fid}. In Sect.~\ref{discuss} we compare our results to observations and to previous work. A summary of our conclusions is given in Sect.~\ref{concl}.

\section{Method}\label{method}
We investigate the OB+SR population of the SMC with the Monte Carlo based rapid binary population synthesis code \combine, which was first introduced by \citet{2018MNRAS.481.1908K}. In this section we describe how this code operates and how we adjusted it for this study.

\subsection{The rapid binary population synthesis code \combine}

Evolving both binary components simultaneously, \combine\ is based on tabulated detailed single star models, in contrast to other binary population synthesis codes, which treat stellar evolution using fitting formulae. This allows for a fast exchange and modular use of the underlying stellar models. Compared to detailed codes like MESA \citep[][and references therein]{2019ApJS..243...10P}, \combine\ is much faster and enables the user to generate large model populations of different underlying binary physics in a small amount of time. Also in contrast to MESA, it follows a system's evolution through a SN event, is able to treat a common envelope phase and, in contrast to other rapid binary population synthesis codes, uses envelope binding energies calculated self consistently from stellar models \citep{2016A&A...596A..58K} rather than constant values of the envelope's binding energy or mixing results of different 1D stellar evolution codes.

The dense grid of detailed stellar models is currently available for Milky Way, Large Magellanic Cloud, SMC, and IZwicky~18 metallicity and features both hydrogen-rich and HeS models. The models were calculated with BEC \citep[][and references therein]{2010ApJ...725..940Y} using the same stellar physics as \citet{2011A&A...530A.115B} with initial masses from 0.5 to $100\msol$. For the application in \combine, the following quantities were extracted and tabulated: total stellar mass, age, photospheric radius, core mass (see Sect.~\ref{sec-ssmA} for the updated definition), luminosity, effective temperature, two versions of the $\lambda$-parameter describing the binding energy of the envelope, and (new to this version of \combine, see Sect.~\ref{sec-ssmA}) the carbon core mass and the moment of inertia factor. For hydrogen-rich models, the time of central hydrogen exhaustion is saved, too. Using initial mass and age as independent quantities, the current values of the remaining eight dependent quantities are interpolated linearly. The same HeS tracks are used for all metallicities but their masses were rescaled by the wind mass loss rate according to \citet{2015A&A...581A..21H}. While we are aware that detailed stellar models predict a small hydrogen layer to remain on the HeS and that the size of this layer has a large impact on the stellar radius \citep{2020A&A...637A...6L}, we neglect it for simplicity.

To initialise a binary star model, \combine\ draws the initial primary mass $M_1$ from an initial mass function (IMF). We use a Kroupa-like function,
\begin{equation}
    f_{M_1} = \frac{\mathrm{d}N}{\mathrm{d}M_1} \propto M_1^\alpha
\end{equation}
with a high mass slope of $\alpha = 2.3$ \citep{1955ApJ...121..161S,1995ApJ...438..188M} within the range of 3 to $100\msol$. The lower limit takes care that we consider all relevant stars to meaningfully compare the companions of NS to stars of similar mass and the upper limit is the mass of the most massive stellar model we have. The mass of the secondary is derived from a mass ratio distribution 
\begin{equation}
    f_q = \frac{\mathrm{d}N}{\mathrm{d}q} \propto q^\kappa
    \label{eq:kappa}
\end{equation}
and the initial separation from an orbital period distribution
\begin{equation}
 f_{\log P} = \frac{\mathrm{d}N}{\mathrm{d}\log P} \propto (\log P)^\pi.
 \label{eq:pi}
\end{equation}
To model the binary star population in the SMC, we rely on four sets of observationally inferred values of $\kappa$ and $\pi$. These are \citet{2012Sci...337..444S} based on Galactic O~stars ($\kappa=-0.1, \pi=-0.55$), \citet{2013A&A...550A.107S} based on LMC O~stars ($\kappa=-1.0, \pi=-0.45$), \citet{2015A&A...580A..93D} based on LMC early B~type stars ($\kappa=-2.8, \pi=0.0$), and flat distributions, which were observed by \citet{2014ApJS..213...34K} in the Cygnus OB2 association, but we will treat this mostly as an academic example. We note that these distributions might not be independent \citep{2017ApJS..230...15M}. While the lower orbital period limit is given naturally by RLO at ZAMS, one needs to be careful about the upper limit. We choose to only consider orbital periods up to $10^{3.5}\days\approx3162\days$, as this is the range considered by \citet{2012Sci...337..444S}, \citet{2013A&A...550A.107S}, and \citet{2015A&A...580A..93D}. As they note that 50\% to 70\% of massive stars have orbital periods below that value and this upper limit coincides roughly with the maximum orbital period for RLO (see Sect.~\ref{sec-pq}), the remaining fraction of stars is either single or in wide non-interacting binaries, rendering them effectively single for our purposes. We assume a binary fraction of 100\% and circular initial orbits, as tidal effects will circularise the orbit before the onset of RLO \citep{1977A&A....57..383Z}. As we implemented stellar rotation (see Sect.~\ref{sect-rot}), it is also possible to choose the initial rotation of the models. We allow for no rotation, a bimodal distribution as in \citet{2013A&A...550A.109D}, synchronised \citep[as in][]{2020A&A...638A..39L} and a flat equatorial rotation distribution between zero and a user determined value. In this study we restrict ourselves to initially synchonised stars, as tidal interaction predict this to be archived during evolution towards RLO \citep{1977A&A....57..383Z}. We used the same stellar models for all rotation rates, which is justified for stars not to close to critical rotation \citep{2011A&A...530A.115B}.

After a system is initialised, it is evolved by \combine\ until only SRs remain. In practice this means the code determines the time until the radius of one model becomes larger than the Roche radius according to the formula of \citet{1983ApJ...268..368E}, which marks the onset of a RLO, or the end of the stellar evolution. For the intermediate time, the orbital period is adjusted to account for mass loss and the rotational velocity evolves as described in Sect.~\ref{sect-rot}.

If one of the two stars is destined to overfill its Roche lobe, \combine\ evaluates whether the donor is stripped by stable mass transfer, a common envelope ejection occurs or the two stars merge. We updated the applied criteria and list them in Sect.~\ref{sec-ssmA}. In case of stable mass transfer, \combine\ employs the analytical descriptions of \citet{1997A&A...327..620S} to describe the change in orbital period. For this study, we assume that all ejected mass carries the same specific orbital angular momentum as the accretor. We assume that the mass-transfer efficiency is constant throughout the RLO and treat it as a free parameter, whose value we are going to vary in this study. After the end of the mass transfer, \combine\ searches the stellar model grid for a new model for the accretor with matching core mass and total mass. Since the envelope mass increased, the new model appears younger compared to the old model (rejuvenation). The donor is assumed to become a HeS and a matching model is determined from the HeS model grid. In case of a Case~BB/BC RLO when the donor is already a HeS, we assume an instant SN after the RLO. The duration of a Case~B mass transfer is assumed to be of the order of a thermal timescale of the donor. For the duration of Case~A mass transfer and the donor mass thereafter see Sect.~\ref{sec-ssmA}. All unstable RLOs are checked whether a common envelope ejection is possible. We assume however that all unstable RLOs lead to a merger (described in Sect.~\ref{sec-ssmA}), as common envelope ejection is expected to take a very small fraction of the population considered in this study.

When the end of stellar evolution is reached, a SR is formed. Depending on the structure of the final stellar model, this is either a WD, NS or BH. If the helium core mass exceed $6.6\msol$, we assume a BH is formed \citep{2018ApJ...860...93S}. The complete carbon core and 80\% of the helium layer above collapse into the black hole, and we reduce that mass by 20\% to account for the release of gravitational energy \citep[][and references therein]{2018MNRAS.481.1908K}. If the mass of the carbon core is above $1.435\msol$, a NS is formed by core collapse SN (CCSN), and if it is between $1.37\msol$ and $1.435\msol$ we assume an electron capture SN (ECSN) as calculated by \citet{2015MNRAS.451.2123T}. The NS masses are around $1.3\msol$ with the detailed formulae given in \citet{2018MNRAS.481.1908K}. If the carbon core is lighter than $1.37\msol$, a WD is formed which inherits the progenitor's carbon core mass as its total mass.

From the velocity distribution of young pulsars \citep{2005MNRAS.360..974H,2017A&A...608A..57V} it is known that a SN imposes a natal kick on the NSs. For ECSN our standard assumption is a uniformly distributed kick between 0 and $50\kms$ \citep{2004ApJ...612.1044P,2006ApJ...644.1063D,2006A&A...450..345K}. For CCSN we use a Maxwell-Boltzmann distribution of the kick velocity $w$
\begin{equation}
 f_w(w) = \sqrt{\frac{54}{\pi}} \frac{w^2}{w_\mathrm{rms}^3} \exp\left( -\frac32 \frac{w^2}{w_\mathrm{rms}^2} \right)
\end{equation}
with a root-mean-square velocity $w_\mathrm{rms}$ of $265\kms$ for H-rich progenitors \citep{2005MNRAS.360..974H}, $120\kms$ for HeS \citep{1996A&A...315..432T,2017ApJ...846..170T,2013ApJ...764..185C} and $60\kms$ for SN after Case~BB/BC \citep{2018MNRAS.481.1908K}. Following \citet{2017ApJ...846..170T} and \citet{2018MNRAS.481.1908K}, we assume that $20\%$ of the systems with stripped stars receive a kick of $200\kms$. We vary these assumptions by running one scenario without SN kicks and one with only 3D root-mean-square velocities of $265\kms$. We refer to these scenarios as "fiducial", "no kick" and "Hobbs". It is unknown if BHs receive a natal kick \citep{1999A&A...352L..87N,2013MNRAS.434.1355J,2015MNRAS.453.3341R,2016MNRAS.456..578M}, and so we use two extreme scenarios, either no kick or a flat kick distribution between 0 and $200\kms$ \citep{2018MNRAS.481.1908K}. Table~\ref{kick} summarises the SN kicks. We use the results of \citet{1998A&A...330.1047T} to calculate the post-SN orbit or, if the binary breaks up, the centre-of-mass velocities of the two components.

\begin{table}
 \caption{SN types, their kick velocity distribution and characteristic velocities for the fiducial scenario. For CCSN we distinguish between hydrogen-rich models, HeS, and models exploding at the end of Case~BB/BC mass transfer. *Stripped models have a 20\% probability of receiving a higher kick ($w_\mathrm{rms} = \sqrt{3} \times 200\,\mathrm{km/s}$).}
 \begin{tabular}{lll} \hline\hline
    SN type & distribution & char. velocity \\ \hline
	ECSN & flat & $w \in [0,50]\,\mathrm{km/s}$ \\ \hline
	CCSN H-rich & Maxwellian & $w_\mathrm{rms} = \sqrt{3} \times 265\,\mathrm{km/s}$ \\
	CCSN HeS* & Maxwellian & $w_\mathrm{rms} = \sqrt{3} \times 120\,\mathrm{km/s}$ \\
	CCSN BB/BC* & Maxwellian & $w_\mathrm{rms} = \sqrt{3} \times 60\,\mathrm{km/s}$ \\ \hline
	BH formation & flat & $w = 0$ \\ \hline
 \end{tabular}
 \label{kick}
\end{table}

To generate a stellar model population, a large number of binaries is calculated. Each system is assigned an age according to a predefined distribution function and the parameters of each system at its assigned age are evaluated. For this study we restrict ourselves to a flat age distribution, equivalent to a constant star formation rate. We assume a value in the SMC of $0.05\msol/\mathrm{a}$ \citep{2004AJ....127.1531H,2015MNRAS.449..639R,2017MNRAS.466.4540H,2018MNRAS.478.5017R,2021A&A...646A.106S} and scale our Monte Carlo results by it to to determine the expected number of binary systems. To do so, we sum the total simulated stellar mass (accounting for the additional stars below $3\msol$), calculate from that and the age distribution the simulated star formation rate, and compare this number to the assumed star formation rate.

\subsection{Updates to ComBinE: single stars, mergers and Case A}\label{sec-ssmA}

We implemented several updates to \combine{}. This section deals with the more minor ones.

Compared to \citet{2018MNRAS.481.1908K} we use a new definition of the core mass $M_\mathrm{c}$, which reads $M_\mathrm{c} = M - M_\mathrm{H} / X_\mathrm{env}$, where $M$ is the total mass of the model, $M_\mathrm{H}$ is the model's hydrogen mass and $X_\mathrm{env}$ is the average hydrogen mass fraction of the stellar profile with $X>0.1$. %chemically unprocessed envelope. 
In the previous version the mass coordinate at which $X=0.1$ was used, which lead to inconsistencies for hydrogen burning models. The new definition turned out to be a better predictor for the stripped stellar mass. Additionally, it helps numerical stability as it causes $M_\mathrm{c}$ to be more monotonically increasing with time. This is important when finding a stellar model based on mass and core mass in the grids.

Another change is that the carbon core mass of the hydrogen-rich models is now considered, too, to better predict the SN properties and the remaining lifetime of the HeS.

A notable fraction of binary stars merges to a single star throughout their live \citep{1992ApJ...391..246P,2014ApJ...782....7D}. While the true nature of a stellar merger is complicated, we simplify the process by assuming that 10\% of the stellar mass is lost in the process \citep{2013ApJ...764..166D} and use ordinary single star models to describe the merger product. The model of the merger process rejuvenates according to its central helium content. We assume that the merger process lasts one thermal time-scale. Following \citet{2019Natur.574..211S} we set the rotation rate of the product to 10\% of the critical rotation. \combine\ is neither able to model the changed surface abundance patterns nor the expected magnetic fields of the merger product.

In the previous version, it was assumed that the duration of Case~A mass transfer and the final mass of the donor star could be modelled in the same manner as for Case~B. We improved the treatment by adopting the results of \citet{2024A&A...690A.282S}, namely implementing fit formulae for these two quantities depending on the initial donor mass and the initial orbital period.

%\begin{figure}
% \centering
% \includegraphics[width=\columnwidth]{pic/Pq.png}
% \caption{Outcome of the first RLO occurring in the simulated systems depending on initial orbital period $P$ and initial mass ratio $q$. The primary mass is always $\sim 10\ms$. Each dot is one system. Light blue and light red mean successful RLO Case~A and B, respectively. Dark blue and dark red stand for a merger after an unstable RLO Case~A or B. Non interacting systems are shown in grey. If a star starts its post-MS expansion while being an accretor of a RLO initiated by its companion we mark them in yellow and assume they merge.}
% \label{pq}
%\end{figure}

%It is an open question under which condition during a RLO the mass transfer is stable the two binary components remain separated or the mass transfer is unstable and the two stars merge to one object \citep{2013A&ARv..21...59I,2024ARA&A..62...21M}. Examples for proposed criteria are the radius evolution of the models and the Roche lobe \citep{2010ApJ...717..724G,2015ApJ...812...40G,2020ApJ...899..132G}, the ability of the system to eject the non accreted material \citep{handle:20.500.11811/7507,xu}, limits in mass ratio \citep{2014ApJ...782....7D}, or thermal equilibrium limited accretion \citep{2002MNRAS.329..897H}. 
To determine whether the RLO leads to donor stripping, we use the results of \citet{2024A&A...691A.174S} by modelling the evolution of the size of the L\textsubscript{2}~sphere and the radius evolution of the accretor, which depends on the mass dependent ratio of the accretor's thermal time-scale and the mass transfer time-scale of the system modelled by the donor's thermal time-scale and the mass-transfer efficiency. If the accretor grows larger than the L\textsubscript{2} sphere, we assume unstable mass transfer, which leads in general to a merger.

There are further conditions under which we assume the RLO leads to a common envelope phase or merger. If accretion onto a hydrogen-rich model after central hydrogen exhaustion occurs, the accretor expands fast. Thus a contact system forms, which is believed to be unstable and merge fast. We assume a similar fate if mass is transferred onto a HeS, as a hydrogen envelope of 10 to 50\% of the stellar mass produces a bloated star \citep{2012sse..book.....K} leading to a contact configuration as above. This fate can be challenged, if the mass-transfer efficiency is small enough to keep the accretor within its Roche lobe. We assume that mass transfer is not stable if the donor has developed a convective envelope. However this approach is challenged by \citet{2010ApJ...717..724G,2015ApJ...812...40G,2020ApJ...899..132G} for binaries with mass ratios close to unity \citep[see also][]{2019A&A...628A..19Q,2021A&A...650A.107M,2024A&A...685A..58E}. Lastly it is assumed that systems with mass rations smaller than 0.1 merge immediately.

In case of a HeS donor we employ the same criteria and additionally the orbital period limit of \citet{2015MNRAS.451.2123T}. We did not use the criterion of \citet{2003MNRAS.344..629D} by which the donor is assumed to become convective if it is lighter than some threshold mass leading to a merger. However it turns out (see Sect.~\ref{sec-pq}) that if a merger is avoided during the first RLO, it is also avoided in a subsequent RLO Case BB/BC.

%No Eddington limit for stars

\subsection{Updates to ComBinE: rotation}\label{sect-rot}

\begin{figure}
 \centering
 \includegraphics[width=\columnwidth]{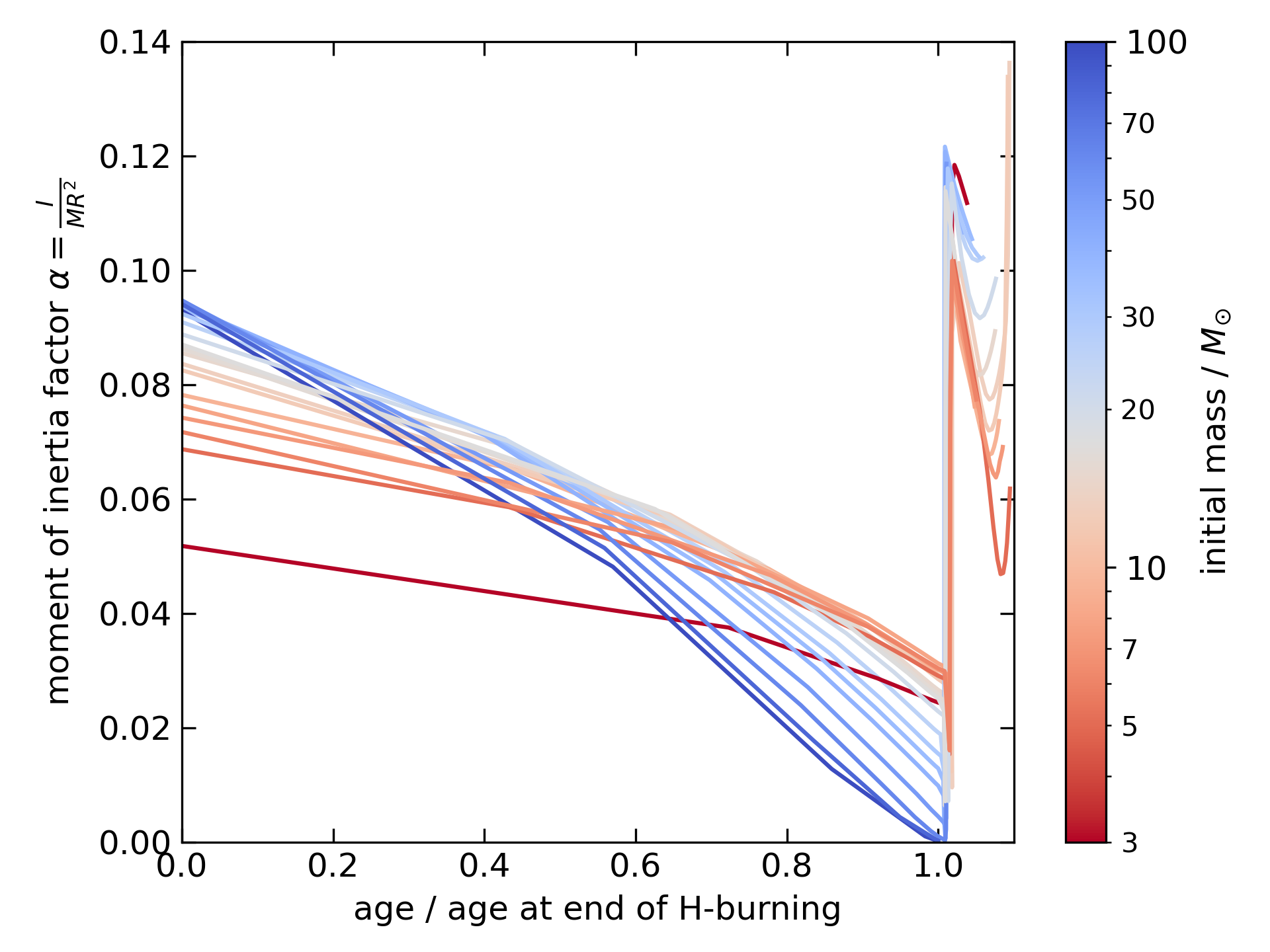}
 \caption{Evolution of the moment of inertia factor of our SMC models. The abscissa is scaled in units of hydrogen burning times.}
 \label{fig:inertia}
\end{figure}

We implemented stellar rotation into \combine. For that we added the stellar moment of inertia $I$ to the tabulated non-rotating single star models \combine\ relies on in form of the moment of inertia factor $\alpha = I/MR^2$. A homogeneous sphere would have $\alpha=0.4$ and if its density increases towards the centre, one finds $\alpha<0.4$. We show the moment of inertia factor of our stellar models in Fig.~\ref{fig:inertia}. Even though a rotating star deforms, leading to a change of moment of inertia, we assume this effect is of second order and use the single star moment of inertia in this work.

Knowing a star's rotational angular velocity $\omega$, one can calculate its rotational velocity at the equator $\vrot$ by multiplication with the star's equatorial radius $R_\mathrm{eq}$ \citep{2013ApJ...764..166D}. Since the star deforms due to rotation, the equatorial radius can be up to 1.5 times larger than the polar radius $R_\mathrm{p}$, which remains almost equal to the non-rotating radius. The ratio of the equatorial radius to the polar radius is a function of $\omega/\omega_\mathrm{cr} \in [0,1)$, where $\omega_\mathrm{cr}$ is the critical angular velocity of the model before it breaks up due to the centrifugal forces at the equator being larger than gravity. One finds
\begin{equation}
    \omega_\mathrm{cr} = \sqrt{\frac{GM(1-\varGamma)}{(1.5 R_\mathrm{p})^3}},
\end{equation}
where $\varGamma$ is the Eddington factor \citep{2012sse..book.....K}. Following \citet{2013A&ARv..21...69R} the radius ratio is %Using a similar approach as \citet{2023A&A...672A..60H} we find the radius ratio as
\begin{equation}
%    \frac{R_\mathrm{eq}}{R_\mathrm{p}} = 3\frac{\omega_\mathrm{cr}}{\omega} \cos\left( \frac13 \arccos\left( -\frac{\omega}{\omega_\mathrm{cr}} \right) - \frac{2\pi}{3} \right).
    \frac{R_\mathrm{eq}}{R_\mathrm{p}} = 3\frac{\omega_\mathrm{cr}}{\omega} \sin\left( \frac13 \arcsin\left( \frac{\omega}{\omega_\mathrm{cr}} \right) \right),
\end{equation}
where we simplified their expression. %Notice that \citet{2023A&A...672A..60H} uses the Keplerial angular velocity instead of the critical.
The critical rotational velocity is then given as $\vcr=\omega_\mathrm{cr}R_\mathrm{eq}$. The rotational deformation does not lead to an earlier RLO, since we assume that before Roche-lobe filling the model's rotation synchronises to the orbit, which leads to only mild radius ratios of a factor of about 1.1.

Using this, we can trace the rotation of a stellar model over its evolution given an initial rotation and assuming the same nuclear evolution as for a non-rotating single star model. If the spin angular momentum $S = I\omega$ of the star is conserved (neither mass loss nor tidal torques), the angular velocity of the next time step is given by
\begin{equation}
    \omega' = \omega \frac{\alpha}{\alpha'} \frac{M}{M'} \left(\frac{R}{R'}\right)^2,
\end{equation}
where we marked the quantities of the next time step with a prime and left those of the previous time step unmarked. We assume rigid rotation, which is well enough fulfilled for main-sequence models, but after central hydrogen exhaustion stellar models clearly rotate differentially \citep[e.g.][]{2010ApJ...725..940Y,2022A&A...667A.122S}. We are not able to trace differential rotation meaningfully with our means and resign from computing it from there on. Also, the rotational decoupling process between core and envelope is not trivial, as \citet{2022A&A...667A.122S} showed.

If stellar winds become important, the change of spin angular momentum is
\begin{equation}\label{eq.sdot}
    \dot S = \frac23 \dot M \omega R^2
\end{equation}
neglecting deformation and assuming isotropic winds (\citet{2011A&A...527A..52G}, see however \citet{2023A&A...672A..60H}). The wind mass loss rate $\dot M$ can be calculated from our tabulated stellar models. Following \citet{1998A&A...329..551L}, wind mass loss is amplified by rotation in form of $\dot M_\mathrm{amp} = x \dot M$ with 
\begin{equation} \label{eq.x}
    x = \left( 1-\frac{\vrot}{\vcr} \right)^{-0.43}
\end{equation}
for $\vrot/\vcr \in (0,0.8)$. Since this expression diverges for $\vrot/\vcr \to 1$, we extrapolate it linearly by assuming the slope of Eq.~\eqref{eq.x} at $x=0.8$ giving 
\begin{equation}
    x = 4.3 \frac{\vrot}{\vcr} - 1.44.
\end{equation}
With that we can write
\begin{equation}
    \frac{\dot S}{S} = \frac{2x}{3\alpha} \frac{\dot M}{M}
\end{equation}
leading to
\begin{equation}\label{eq.om}
    \omega' = \omega \frac{\alpha}{\alpha'} \left(\frac{M}{M'}\right)^{1-\frac{2x}{3\alpha}} \left(\frac{R}{R'}\right)^2.
\end{equation}

In a binary system the stars are subject to tides. To account for this we calculate the synchronisation time scales $\tau_\mathrm{sync}$ of the models as eq.~44 in \citet{2002MNRAS.329..897H} for hydrogen burning models with radiative envelopes ($M>1.2\msol$) and eq.~27 for hydrogen burning models with convective envelopes ($M<1.2\msol$), see also \citet{1981A&A....99..126H}. This gives us
\begin{equation}\label{eq.tides}
    \omega' = (\omega-\varOmega) \exp\left(-\frac{\Delta t}{\tau_\mathrm{sync}}\right) + \varOmega
\end{equation}
with $\varOmega$ as the angular velocity of the orbit, \citep[e.g.][]{2008A&A...484..831D}. Eq.~\eqref{eq.om} and~\eqref{eq.tides} applied after one another yield the models's angular velocity at the next time step. Since the orbital angular momentum of a model is much larger than its spin angular momentum, we can neglect changes of the orbit.

If this scheme leads to a model spinning over-critically, i.e. $\omega>\omega_\mathrm{cr}$, we let the model lose additional mass at its equator. As the specific spin angular momentum there is $\omega R^2$ (notice that in contrast to Eq.~\eqref{eq.sdot} the factor $\frac23$ disappears) and thus $\Delta S = \Delta M \omega R^2$. We can assume that the mass loss to bring the model down to critical rotation is small, as \citet{1981A&A...102...17P} showed that large changes in stellar rotation require only small changes in mass, and therefore the change of radius is small, too, so we write $\Delta S = \alpha M R^2 \Delta\omega$. Equating these two expressions gives the additional mass loss to bring a model from rotating over-critically back to sub-critical rotation as
\begin{equation}
    \frac{\Delta M}{M} = \alpha \cdot \left( 1-\frac{\omega}{\omega_\mathrm{cr}}\right).
\end{equation}

It is generally assumed that accretion leads to the spin-up of stars. We calculate the accreted angular momentum depending on whether or not an accretion disk forms by calculating 
\begin{equation}
    R_\mathrm{min} = 0.0425 a \sqrt[4]{q+q^2}, \quad 0.0667 \leq q \leq 15,
\end{equation}
where $a$ is the semi-major axis \citep{1975ApJ...198..383L,1976ApJ...206..509U,2015ApJS..220...15P}. If the accretor radius is larger than this value, we assume ballistic accretion with a specific angular momentum of $\sqrt{1.7 G M R_\mathrm{min}}$ and otherwise accretion from a Keplerian rotating disk at its equator and thus with a specific angular momentum of $\sqrt{GMR}$, where $M$ and $R$ refer to the accretor's mass and radius. We assume that a critically rotating accretor can keep accreting mass but limit its rotation to the critical value as the viscosity inside the disk transport mass inwards but transport angular momentum outwards \citep[see][for a discussion]{1991ApJ...370..597P,1991ApJ...370..604P,2013ApJ...764..166D}. An alternative view, where the accretion is limited by the accretors spin-up \citep{2005A&A...435.1013P,handle:20.500.11811/7507}, is investigated in Paper~I.

%\begin{equation}
% L_i = I_{ij} \omega_j
%\end{equation}

%\begin{eqnarray}
% I_{xx} &=& I_{yy} = I{zz}
% \\
% I &=& \alpha M R^2
%\end{eqnarray}

%\begin{displaymath}
% I_{ij} = 0 \quad\text{if}\quad i \neq j
%\end{displaymath}

%\subsection{ComBinE -- mass transfer}
%
%Case A: resolves Af, As, AB (and ABB)
%
%Accretion now accretion limited. Isotropic reemission only used for accretion on compact objects.

%\section{Binary evolution depending on initial mass ratio and orbital period}\label{sec-pq}
\section{Dependance of binary evolution on mass ratio, orbital period, and mass-transfer efficiency}\label{sec-pq}
\begin{figure*}
 \centering
 \includegraphics[width=0.8\textwidth]{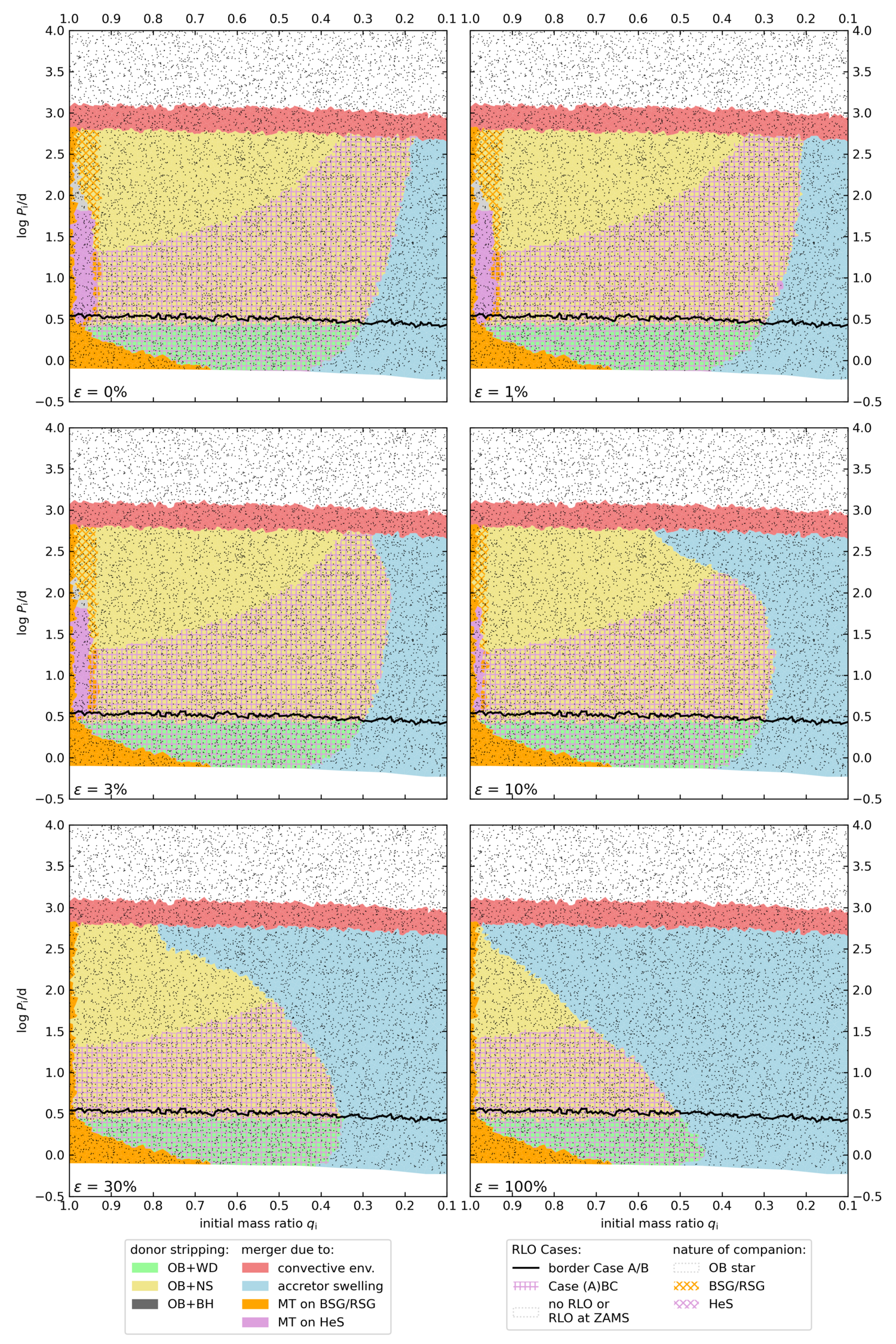}
 \caption{Evolutionary outcome of our binary models with a primary mass of $10\msol$ at the time when the first SR has formed, or when the system merges, as a function of initial orbital period $P$ and initial mass ratio $q$, for six different mass-transfer efficiencies $\varepsilon$. Each black dot represents one model system, and their total number in each plot is $10^4$. We coloured the plane according to the result of the closest model (Voronoi diagram). We mark the boundary between RLO Case~A and~B by a black line. The occurrence of a later RLO with a HeS donor (i.e. Case~BB or~BC) is indicated by pink hatches. If the companion to a SR is not a MS star but a post-MS star, we mark it by orange cross-hatches in case it is H-rich (BSG/RSG), or by pink cross-hatches if it is H-poor (HeS). White regions are those with no RLO (top) or RLO at ZAMS (bottom). %The boundary to the latter region is much smoother than the others as \combine\ does not produce systems overflowing at ZAMS so we estimated the boundary by eye. 
 In grey areas, the outcome is not none of the above (often two SN immediately after each other due to the final temporal resolution).}
 \label{fig:Pq}
\end{figure*}

Before we construct synthetic populations from our binary models, we need to understand how different initial masses, initial mass ratios, initial orbital periods and mass-transfer efficiencies affect the evolution of a binary in our code. Therefore we drew $10^4$ binary models with uniform distribution in initial mass ratio $q_\mathrm{i}$ (mass of initially less massive star divided by mass if initially more massive star) and logarithmic initial orbital period $\log P_\mathrm{i}$ for several fixed primary masses and mass-transfer efficiencies (with the non-accreted material carrying the same specific angular momentum as the accretor), and evolved them. We classify the evolutionary path either by the first SR that is produced or of a merger happens before that, and colour the $\log P_\mathrm{i}$-$q_\mathrm{i}$ plane according to the nearest model. For merging systems, we differentiate between the different reasons for that. In Fig.~\ref{fig:Pq} we assume an initial mass of $10\msol$ and Appendix~\ref{app:Pq} contains similar diagrams for other masses. For simplicity, we assume no SN kick.

In Fig.~\ref{fig:Pq} we find that for low mass-transfer efficiencies most binaries can evolve to OB+NS systems (yellow) and in case of high mass-transfer efficiencies mergers dominate because of accretor swelling and subsequent L2 overflow (blue). At high orbital periods we expect mergers due to the donor star developing a convective envelope (red) and at even wider orbits no RLO takes place at all (white). A large fraction of systems experience a Case~BB/BC mass transfer ($+$-like pink hatching) except for wide systems, as the HeS is not able to fill such a large Roche lobe. Very close systems, which undergo Case~A mass transfer end up as OB+WD systems (green) since the mass of the stripped donor decreases with decreasing initial orbital period for Case~A. If the initial mass ratio for Case~A is too close to unity (orange) the slow phase of the mass transfer lasts so long that the accretor ends hydrogen burning, expands and fills its Roche lobe. So, a contact system is formed which we assume will merge fast. We discuss systems of initial mass ratios close to unity in Sect.~\ref{app-q1}

If we consider higher donor masses, we find that Case~BB/BC RLOs soon disappear from the evolution as the HeSs do not grow to large radii for higher masses. The regions where mergers occur only shift slightly. Around masses of $20\msol$ (Fig.~\ref{fig-pq20}) the donor becomes a BH (dark grey). For even higher masses we find that the Case~A/B boundary shifts to higher orbital periods and that the merger region at $q\approx1$ becomes larger as the mass-lifetime exponent decreases.

Comparing these diagrams to the predictions of \citet{2024A&A...691A.174S}, we find about the same boundary between donor stripping and merger, as expected. In contrast to them, we find that our models merge at very low periods and high mass ratios and strip the donor (for high mass-transfer efficiencies) at high periods and high mass ratios. The differences are, as \citet{2024A&A...691A.174S} already predicted, due to our consideration of the evolution of the accretor. Compared to Paper~I, we predict mergers only at mass ratios far from unity if the mass-transfer efficiency is not too high. Only for very high mass-transfer efficiency, systems at intermediate mass ratios and high periods merge. The models of Paper~I merge for intermediate mass ratios at low periods (independently for Case~A and~B), as they use a completely different criterion to decide the stability of a mass transfer. As we show in Sect.~\ref{results_fid}, this leads to strong differences for the period distributions of BeXB and OB+BH systems.

\section{Influence of the accretion efficiency on the synthetic populations}\label{results_var}
\begin{figure*}
 \centering
 \includegraphics[width=\columnwidth]{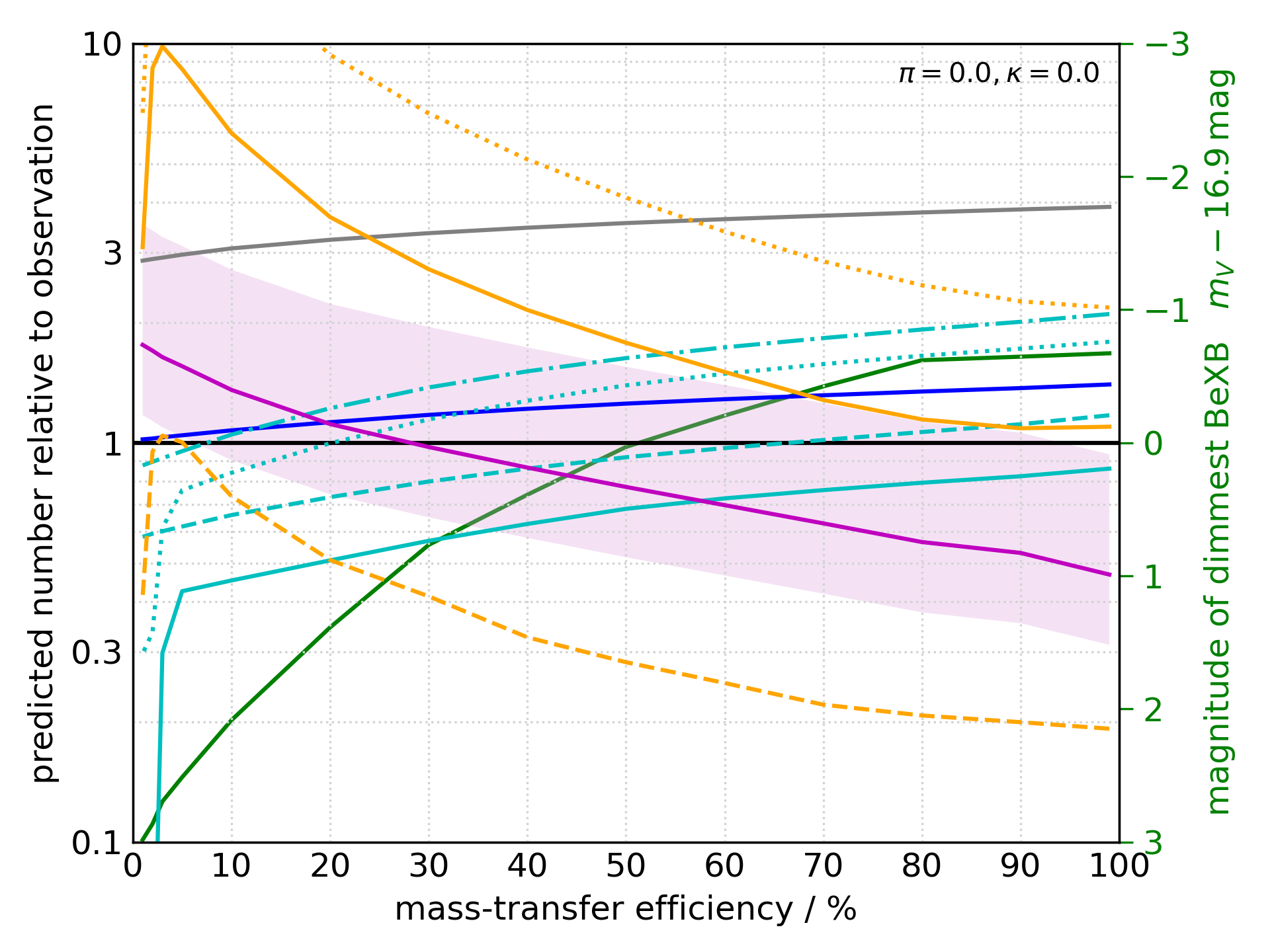}
 \includegraphics[width=\columnwidth]{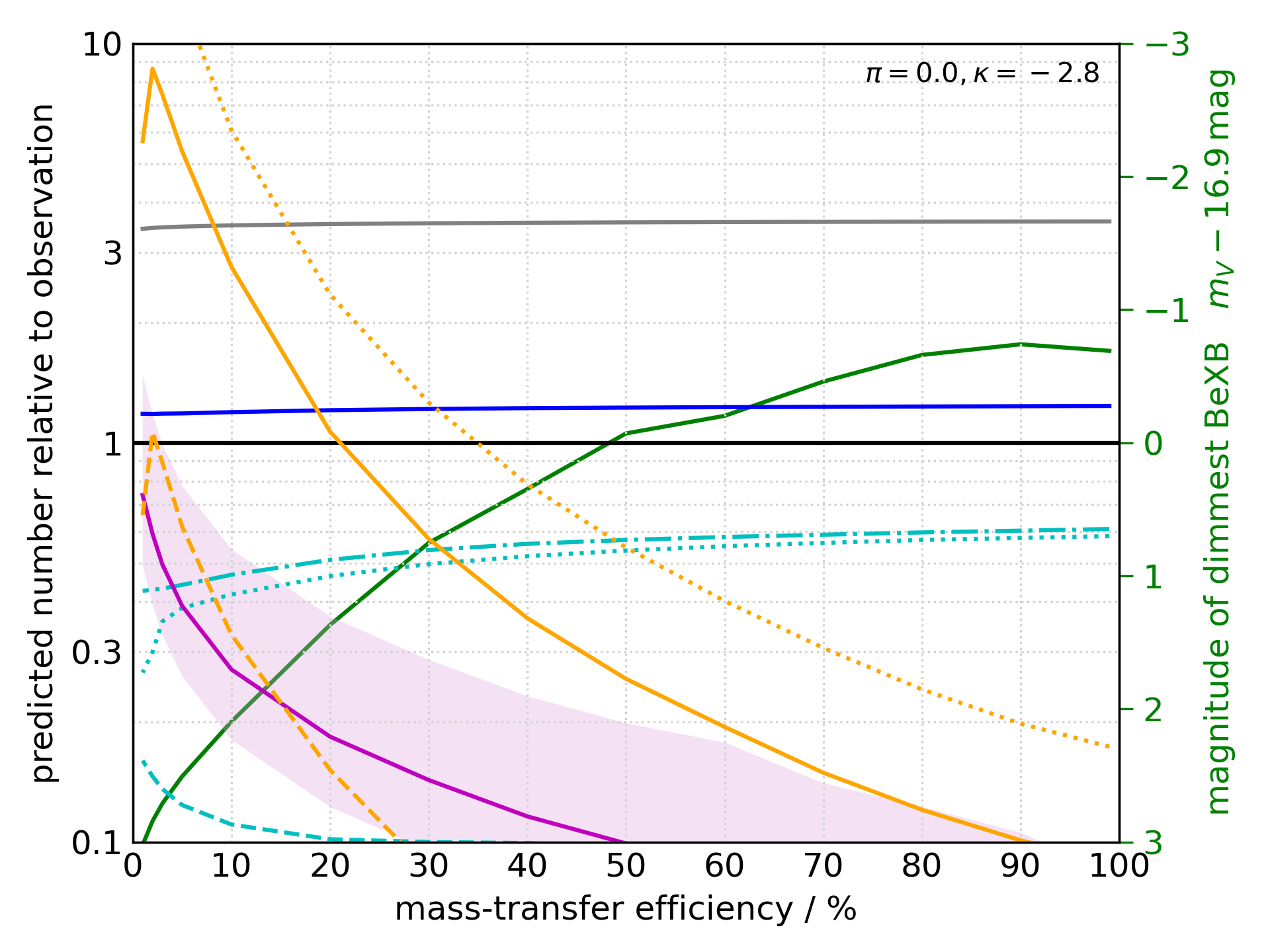}
 \includegraphics[width=\columnwidth]{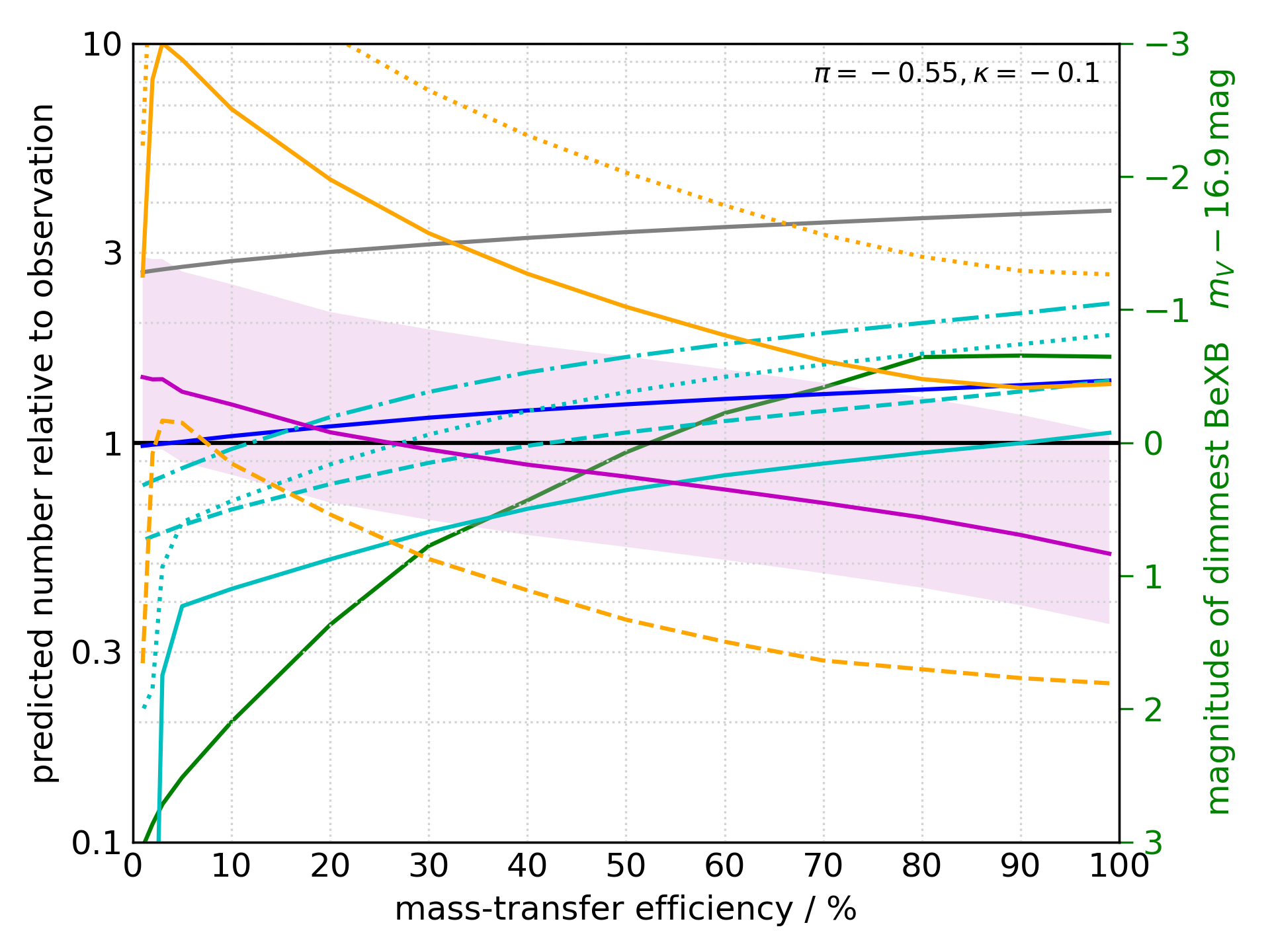}
 \includegraphics[width=\columnwidth]{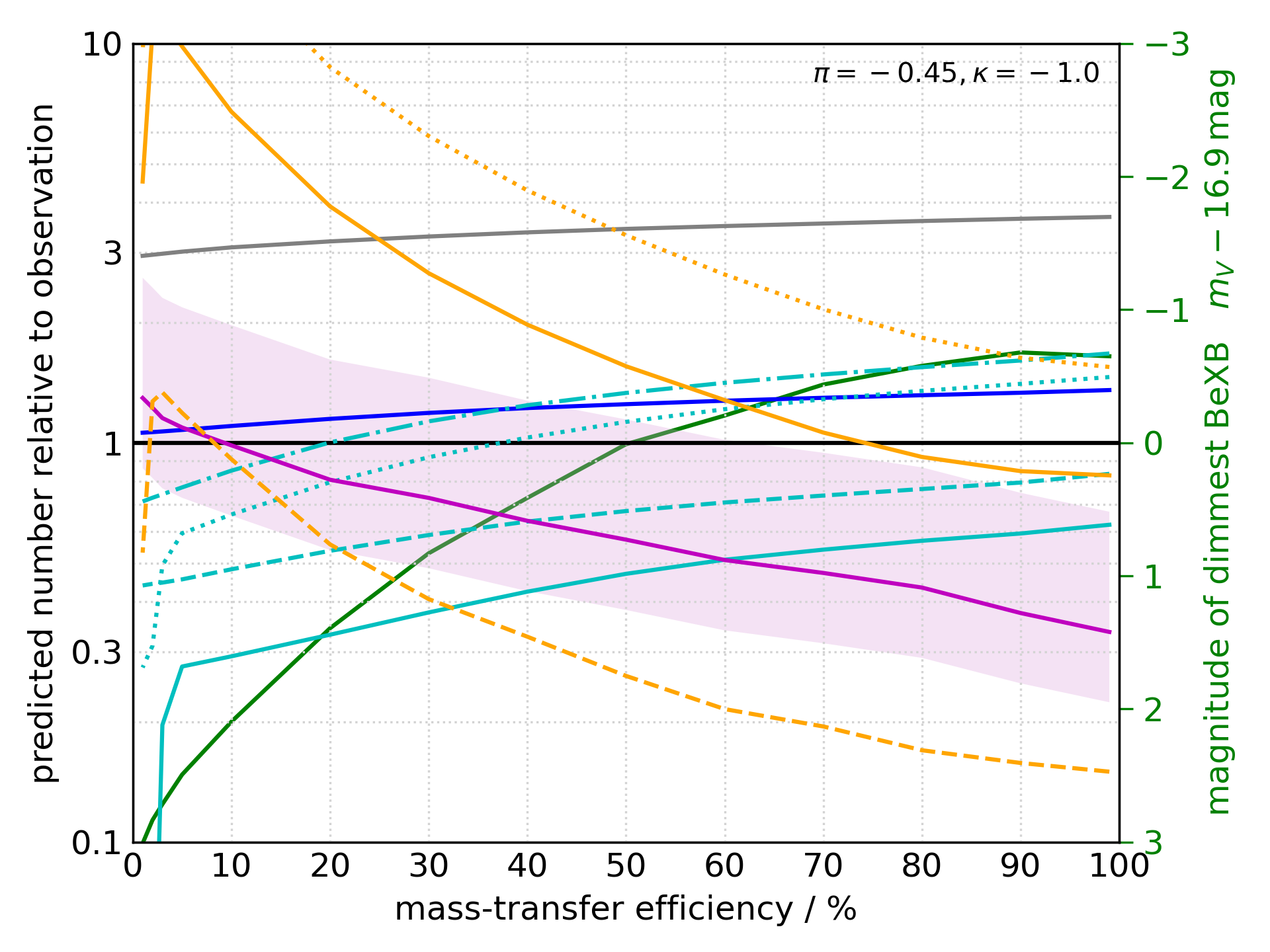}
 \includegraphics[width=\columnwidth, trim={0 5cm 0 4cm},clip]{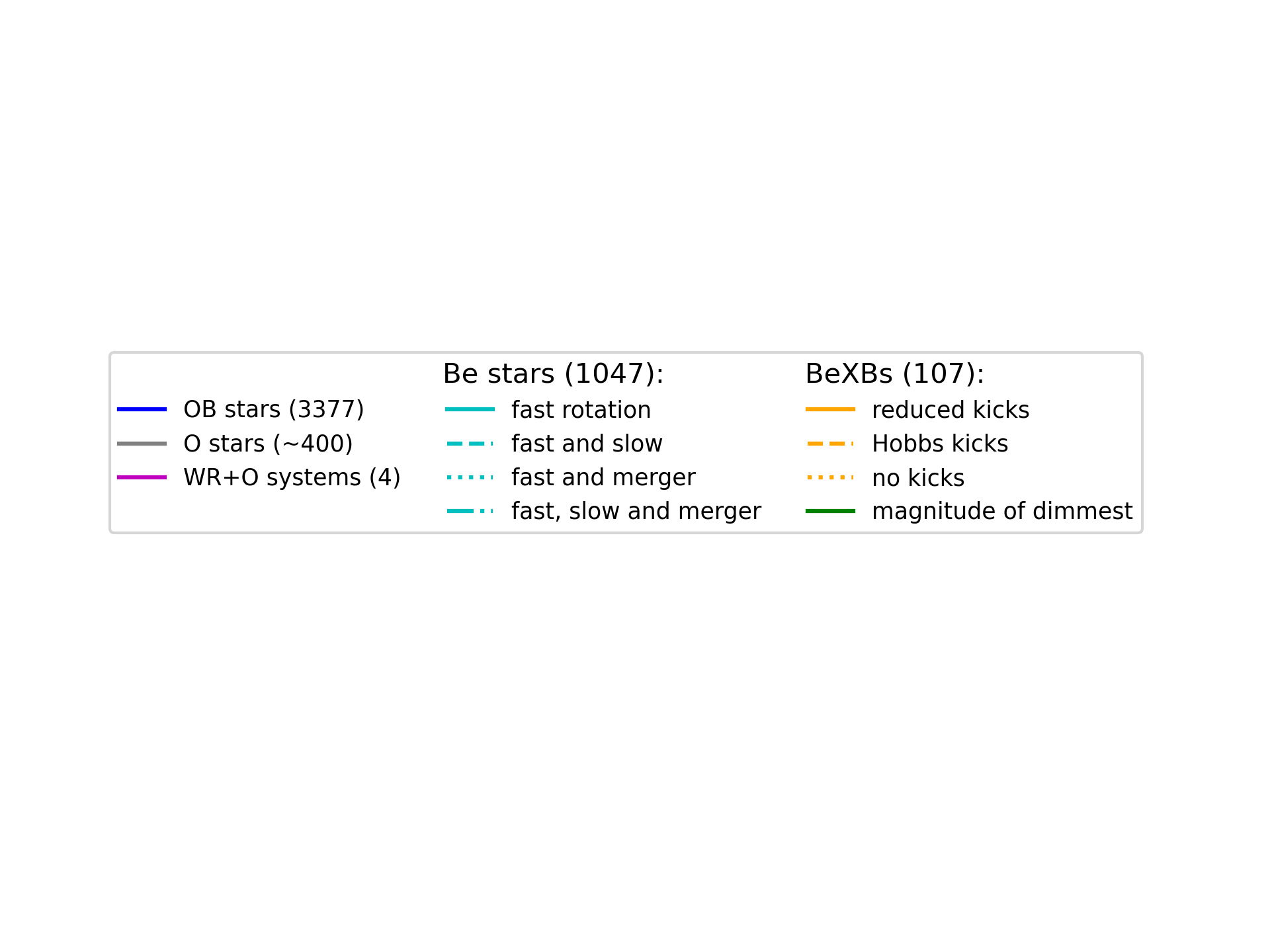}
 \caption{Predicted number of selected types of stars and binary systems relative to the observed number (given in parentheses) as a function of mass-transfer efficiency. We show four different assumptions about the stars that become OBe~stars (continuous line = fast rotating accretors, dashed = fast and slow rotating accretors, dotted = fast rotating accretors and merger products, dash-dotted = fast and slow rotating accretors and merger products) and three different assumptions about the SN kick (continuous = reduced kicks, dashed = Hobbs kicks, dotted = no kicks). The green line with labelling on the right side of the diagram shows the magnitude of the dimmest synthetic BeXB relative to the dimmest observed one. The pink shading indicated the uncertainty due to the low number of WR+O systems. The initial binary parameter distributions are flat (top left), according to \citet[][bottom left]{2012Sci...337..444S}, \citet[][bottom right]{2013A&A...550A.107S}, and \citet[][top right]{2015A&A...580A..93D}.
 %\red{shading needed? Add solid dot to where the lines cross the ``1''-line, in the color of the line.}
 }
 \label{fig:eff}
\end{figure*}

In this section, we generate synthetic binary populations, where the probability to draw systems depends on the initial mass, initial mass ratio and initial orbital period distributions described in Sect.~\ref{method}. We also employ the different kick scenarios and vary the mass-transfer efficiency from 10\% to 100\% in steps of 10\% with additional values of 5\%, 3\%, 2\%, and 1\%. Each simulation contains $10^7$  binary models and is scaled such to a star-formation rate of $0.05\msol/\mathrm{a}$. We extract the number of selected types of stars and binary systems (O~type stars, OB~stars, OBe~stars, BeXBs, WR+O systems) and compare them to the observed numbers in Fig.~\ref{fig:eff}.

The four panels in Fig.~\ref{fig:eff} are ordered in such a way that one can easily identify the impact of the initial binary parameters. The two panels on the left side have rather flat mass ratio distribution ($\kappa\approx0.0$) but those of the panels to the right are skewed towards more unequal systems ($\kappa=-2.8$ and $-1.0$). The upper panels both have a flat period distribution ($\pi=0$), while for the two lower panels, close systems are preferred ($\pi\approx0.5$).

We calculate the number of systems with O~type stars by classifying all models with an effective temperature greater than 31600\,K as such. A margin of error can be derived from varying this number by one spectral class, i.e. 30350\,K and 32900\,K \citep{2021A&A...646A.106S}, which is about 30\%. The observed number of about 400 is the estimate corrected by completeness from \citet{2021A&A...646A.106S}. The numbers of OB (hydrogen-burning star above $10^4\,\mathrm{K}$) and OBe~stars only includes stars with an absolute $G_\mathrm{bp}$ magnitude brighter than $-3$ \citep[$M\simgr8\msol$][]{2021A&A...646A.106S}, which is the completion limit of \citet{2022A&A...667A.100S}. We used a distance modulus of $18.91$ \citep{2005MNRAS.357..304H} and calculated the absolute $G_\mathrm{bp}$ magnitude following \citet{2021A&A...646A.106S}. We classify all hydrogen burning models as OBe~stars if they spin faster than 0.95 times their critical rotational velocity \citep{2004MNRAS.350..189T} and vary this number by including slow rotator and/or merger products. The observed numbers of OB and OBe~stars are from \citet{2022A&A...667A.100S}. We classify a model as WR~star if it is a HeS with a luminosity larger than $10^{5.6}\lsol$ \citep{2020A&A...634A..79S} and compare them to the four WR+O systems in the SMC \citep{2016A&A...591A..22S,2018A&A...616A.103S}. The margin of error is the Poisson counting uncertainty, i.e. $\pm2$. Finally we classify a model as BeXB, if it contains a OBe~star according to the definition from above and a NS. 107 BeXBs are known in the SMC \citep[][living version \url{https://www.mpe.mpg.de/heg/SMC}]{2016A&A...586A..81H} and the dimmest has a magnitude of $16.9$.

For all initial distributions in Fig.~\ref{fig:eff}, we identify a clear overproduction of O~stars. This is not unexpected as this dearth was already discussed by \citet{2021A&A...646A.106S}. The total number of bright OB~stars is consistent with the observations and deviates less than a factor 1.5. Both quantities are almost independent of the mass-transfer efficiency. The predicted number of WR+O stars agrees for the distributions in Fig.~\ref{fig:eff} top left, bottom left, and bottom right in a wide range of mass-transfer efficiencies with the best match at low mass-transfer efficiencies. For the distribution in the top right, only mass-transfer efficiencies below 5\% yield an agreement between simulations and observations. Thus, the number of WR+O systems is unaffected by reasonable variation of the initial period distribution but depends more on the initial mass ratio distribution. More extreme initial mass ratios lead to more mergers reducing the number of systems. For all initial distributions, we find that the WR+O number decreases with increasing mass-transfer efficiencies. This comes from the larger merger area in Fig.~\ref{fig:Pq} and the reduced lifetime of the accretor.

For the same reasons, we find less BeXBs at high mass-transfer efficiencies. Only for very small mass-transfer efficiencies we observe the opposite trend, since in this case not enough mass is transfered to the accretor which then does not spin fast enough to become a OBe~star. We predict most BeXBs for the no-kick scenario and least for the Hobbs scenario, as stronger kicks break up the systems more easily. A initial mass ratio distribution favouring low mass ratios ($\kappa=-0.1\to-1$, lower left to lower right panel in Fig.~\ref{fig:eff}) decreases the number of BeXBs due to more systems merging and a period distribution favouring low periods ($\pi=0\to0.5$, top panels to bottom panels) increases the number of systems especially at high mass-transfer efficiencies as wide systems merge there more often than close systems (Fig.~\ref{fig:Pq}). A higher mass-transfer efficiency causes the magnitude of the dimmest BeXB to be smaller, as expected for a heavier accretor. This behaviour is independent of the initial mass ratio and initial orbital period distribution as it does not consider the number but only the occurrence of such systems.

For OBe~stars we find that their number increases with larger mass-transfer efficiency. While a lifetime effect might be present, here the assumed magnitude limit comes into play. A larger mass-transfer efficiency pushed more binaries with lower initial primary mass, which are preferred by the IMF, over the magnitude threshold than a lower mass-transfer efficiency. We find more OBe~stars for steeper period distributions and flatter mass ratio distributions. For the \citet{2015A&A...580A..93D} distribution (top right) the curve showing the number of OBe~stars is outside the range of the plot, i.e. this distribution underestimates the number of OBe~stars by more than a factor of 10. Only the \citet{2012Sci...337..444S} distribution (bottom left) reproduces the observed number of OBe~stars when only considering fast rotators and only at very high mass-transfer efficiencies. When slow rotators are included, the two left panels of Fig.~\ref{fig:eff} produce the observed number of OBe~stars with intermediate mass-transfer efficiencies. However when doing this one should also consider all OB+NS systems as BeXB, which would lead to an overproduction of BeXBs. Another approach would be to vary the star-formation rate in such a way that it takes a higher value when the OBe~star progenitors are born. However it turns out that this is at the same age when the BeXB progenitors where born, so their number would be lifted, too. Figuratively speaking, both approaches mean that one cannot lift the OBe~star curve without lifting the BeXB curve. Possible solution could be to reduce the number of BeXBs again by assuming less reduced SN kicks or to consider either merger products or single star evolution as a channel for OBe~star production. The above mentioned overproduction of OB~stars simultaneous to the underproduction of OBe~stars supports this finding.

In all but Fig.~\ref{fig:eff} (top right), the line for the number of BeXBs and for the magnitude of the dimmest one meet at an mass-transfer efficiencies of 60\% to 75\%, close to the desired observed level. This suggests that during the formation of BeXBs the mass-transfer efficiency was rather high. The number of WR+O systems is best reproduced with low mass-transfer efficiencies (subject to a large uncertainty), which means that it might be a function of stellar mass. We assume in the following that the mass to be considered is the accretor mass, as we assume that this is the place in the system where the material is ejected. LMC stars are probably a better proxy for SMC stars than MW stars, and so we choose the distributions of \citet[][bottom right]{2013A&A...550A.107S} as our fiducial set. In our model, the median initial mass of the accretor in WR+O systems is $35\msol$, while it is $7\msol$ for BeXBs. So, we vary the mass-transfer efficiency linearly between these two accretor masses $M_\mathrm{A}$, which yields 
\begin{equation} \label{epsofm}
    %\varepsilon = 0.1 + (1.55-\log M_\mathrm{A})\frac{5}{7}, \quad M_\mathrm{d}\in[,]
    \varepsilon = 1.21 - 0.72 \log M_\mathrm{A}, \quad 7\msol \leq M_\mathrm{A}\leq 35\msol
\end{equation}
and take the boundary values of 0.6 and 0.1 outside that range. The median mass of accretors in OB+BH progenitors turns out to lie with $15\msol$ somewhere in between causing an typical mass-transfer efficiency of about 35\% in such systems.

\begin{figure}
    \centering
    \includegraphics[width=\columnwidth]{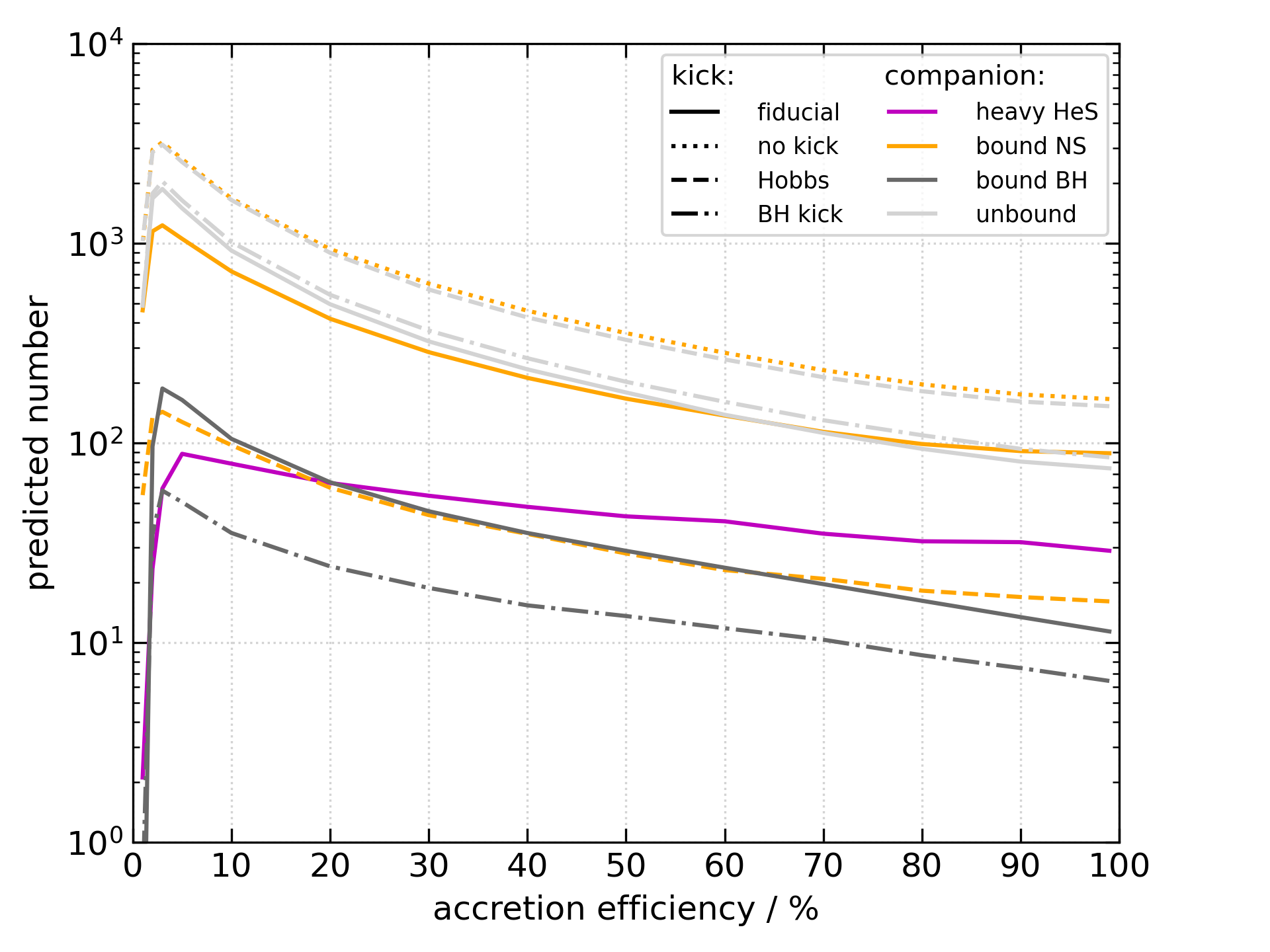}
    \includegraphics[width=\columnwidth]{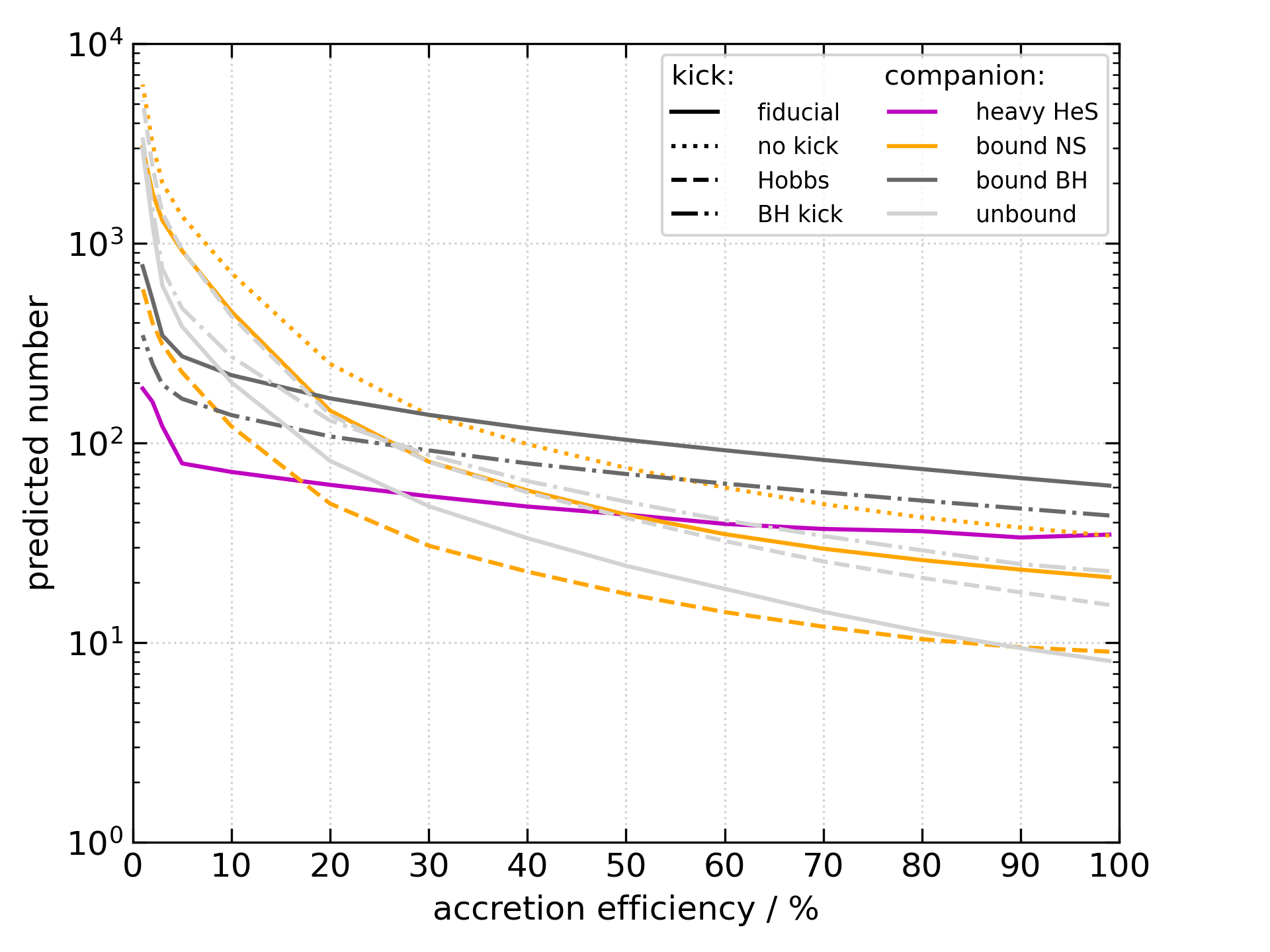}
    \caption{Predicted number of OBe~star (top) and regular OB~star (bottom) companions indicated by colour and different kick scenarios indicated by line style as a function of mass-transfer efficiency for the initial distributions of \citet{2013A&A...550A.107S}.}
    \label{fig:comp}
\end{figure}

In Fig.~\ref{fig:comp} (top) we present the absolute number of OBe stars with HeS, NS and BH companions including those which where unbound from their NS in the SN. We assume the initial distributions from \citet{2013A&A...550A.107S}. (See Fig.~\ref{fig:comp-other} for other initial distributions.) We only consider HeS heavier than $2.55\msol$ which is a rough limit between WD and NS formation. The numbers reach a maximum at mass-transfer efficiencies of 3\%, since for larger values the accretors becomes heavier and thus live shorter and the lowest mass-transfer efficiencies depose not enough angular momentum on the accretor to become an OBe~star. The number of of OBe+HeS does not drop as steeply as the others as it is in general the HeS which evolved faster and ends the OBe+HeS phase. A stronger (weaker) SN kick reduces the number of OBe+NS=BeXB systems, while the number of unbound systems is increased (decreased). We find that the BH kick can unbind a substantial number of systems reducing their number by a factor of two. We predict 10 to 200 OBe+BH systems. For an mass-transfer efficiency of 30\%, which is the typical values one finds with Eq.~\eqref{epsofm}, we predict 50 of them.

Fig.~\ref{fig:comp} (bottom) shows those OB star models which are not rotating fast enough to be classified as emission line stars. Again the numbers shrink with larger mass-transfer efficiency. Towards low mass-transfer efficiencies we find a sharp increase in the number of OB~stars. This is because accretors are no longer spun up enough to become OBe~stars. The OB+BH systems show a different slope than the other species. This may be due to the lower exponent of the mass-lifetime relation for massive stars. Furthermore the SN kicks affect these systems less than the OBe systems since non-OBe stars originate generally from closer and more strongly bound systems. We expect about 150 normal OB+BH systems.

\begin{comment}
\begin{table*}
 \centering
 \caption{Comparison between our work (first column) and the results of Xu~et~al. (in~prep.) (fourth column). In between intermediate models to compare the two approaches (details see text).}
 \begin{tabular}{c||c|c|c|c}
	& \combine & \combine & MESA (Xu) & MESA (Xu) \\
	& Standard & MESA-like & $10\dots50\ms$ & Standard \\ \hline\hline
	SFR*		& 0.05  & 0.05  & 0.05  & 0.05 \\
	O-Stars		& 997   & 1038  & 1470  & 1600 \\
	HeS+OBe		& 611   & 15    & 31    & 223 \\
	HeS+OB		& 219   & 4     & 10    & 12 \\
	WR		    & 4.7   & 0.4   &  4    & 7 \\
	BH+OBe		& 170   & 102   & 160   & 170 \\
	BH+OB		& 193   & 36	& 35    & 41 \\
	NS+OBe		& 999   & 21	& 19    & 23 \\
	NS+OB		& 398   & 2     & 5     & 5 \\
	OBe disr.	& 1502  & 92    & 71    & 72 \\
	OB disr.    & 138   & 1     & 3     & 3 \\ \hline
	NS+BH+disr.	& 3400 & 253    & 293 & 314 \\
	${\frac{\mathrm{NS}}{\mathrm{NS+disr.}}}$    & 46.0\% & 19.7\%	  &	24.5\% & 27.2\% \\
	${\frac{\mathrm{NS+disr.}}{\mathrm{BH+NS+disr.}}}$    & 89.3\% & 45.6\%	  &	33.4\% & 32.8\% \\
 \end{tabular}
 \label{mesa}
\end{table*}
\end{comment}

\section{The fiducial model with mass dependent accretion efficiency}\label{results_fid}
In this Section, we present the properties of our fiducial simulation, i.e. where the mass-transfer efficiency is a function of accretor mass according to Eq.~\eqref{epsofm}, with the initial mass ratio and orbital period distributions of \citet{2013A&A...550A.107S} and the SN kicks as in Table~\ref{kick}. In the following, we will often differentiate between OBe stars, i.e. main-sequence OB~type stars with Balmer emission lines, and regular/normal OB stars, i.e. main-sequence OB~type stars without emission lines. For this analysis we rerun our fiducial model with $10^8$ binary models.

\subsection{Companions of SMC OB stars}

\begin{figure*}
 \centering
 % original size is 162.56mm × 121.92mm, making SNtype square and give 4/7 and 3/7 width.
 %\includegraphics[width=0.57\textwidth]{pic/CompFrac.png}
 %\includegraphics[width=0.42\textwidth, trim={20.32mm 0 20.32mm 0}, clip]{pic/SNtype.png}
 % SNtype gets 1/(original aspect ratio)
 \includegraphics[width=0.64\textwidth]{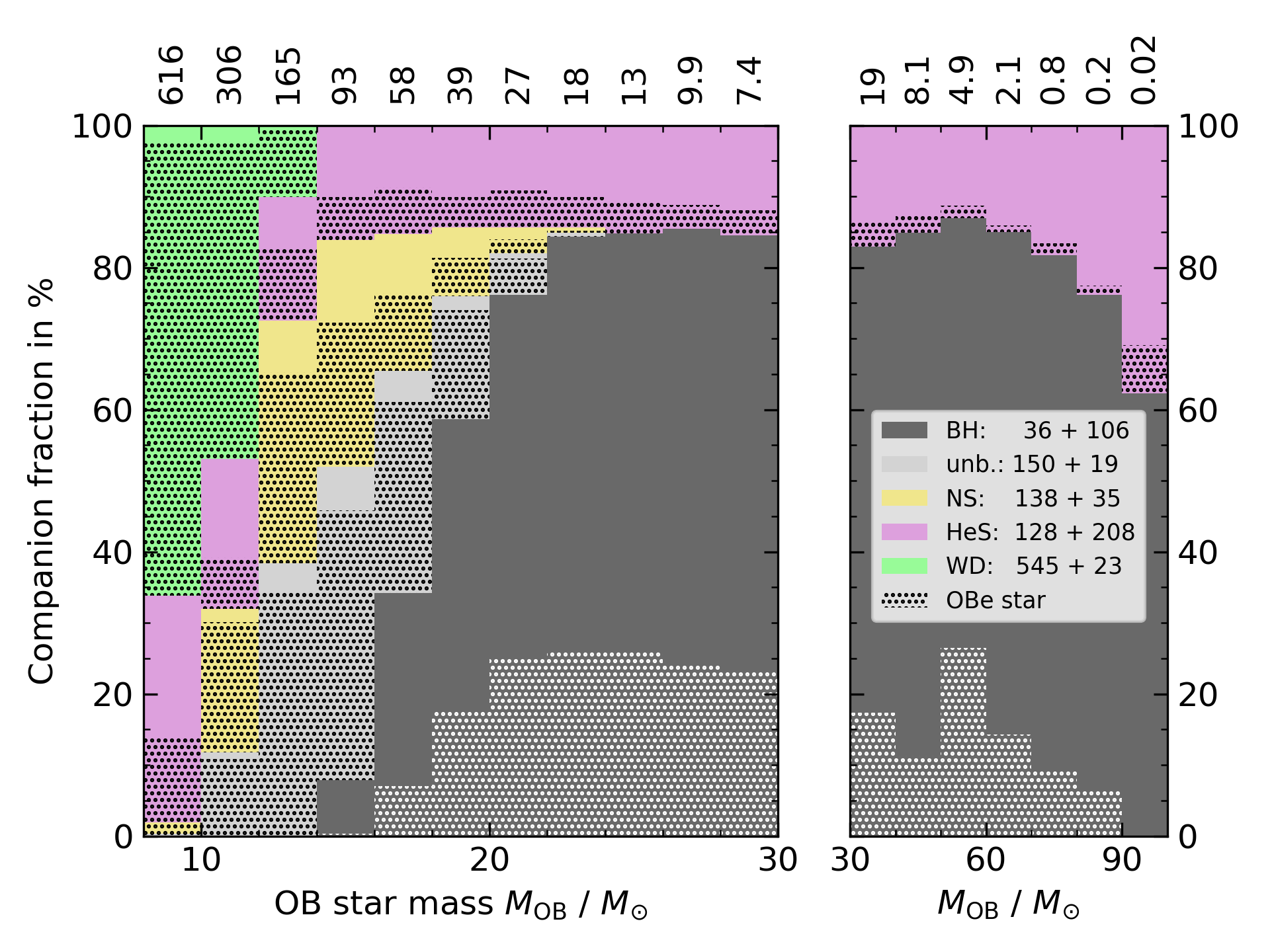}
 \includegraphics[width=0.35\textwidth, trim={35.56mm 0 35.56mm 0}, clip]{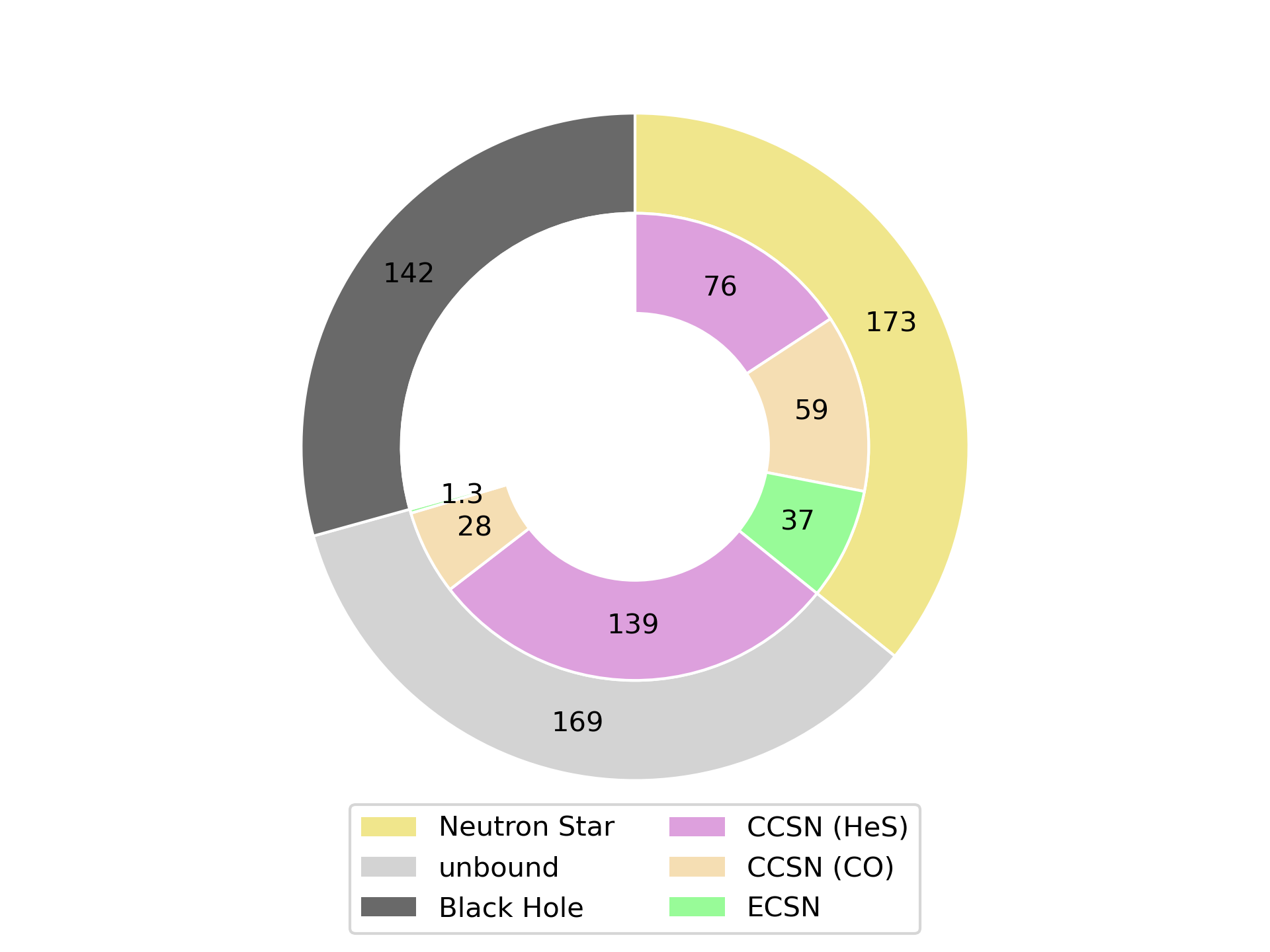}
 \caption{
 Left: Predicted companion fraction of massive ($M>8\msol$) main-sequence SMC stars in post-mass transfer binaries as function of their mass, including system in which the companion (in general a NS) was unbound during the SN, and excluding merger products. Models which we assume to appear as emission line stars are marked with black/white dots. Above each mass bin we give the total number of system in that bin. The total number of each companion type is given in the legend distinguishing between emission-line star (first number) and normal OB star (second number).
 Right: Absolute number of compact companions of OB~stars (outer pie chart), and the assumed SN type of the NS progenitor (inner pie chart), predicted by our fiducial SMC population.}
 \label{CompFrac}
\end{figure*}

In Fig.~\ref{CompFrac} we show the predicted fractions of companion types to our core-hydrogen burning SMC models after mass transfer, as a function of their mass. This figure includes system which become unbound during the SN event, but excludes merger products, single stars and binary stars before interaction. We find four types of companions in such systems as well as systems unbound by the SN explosion, each occurring typically in a certain mass range. HeSs can be found over the whole mass spectrum. This is not surprising, as they are the immediate outcome of stable mass transfer. They make up about 20\% of OB star companions. For the lowest and the highest considered masses, this fraction reaches up to 30\%. Less than half of their OB~star companions are expected to be emission line stars with a shrinking fraction towards higher masses.

Up to an OB star mass of $14\msol$, WDs can be found as companions, where most OB stars are expected to be emission line stars. This is corroborated by the fact that most of these systems undergo a Case~BB/BC RLO after core He-exhaustion of the primary, during which the accretor is spun up a second time.  NSs appear as companions for OB star masses above about $8\msol$ and dominate above about $12\msol$. The NS companion fraction is largest around $15\msol$ and drops to zero above about $24\msol$. The ratio of normal OB+NS systems to OBe+NS systems increases towards higher masses. In contrast to the WDs, we predict a notable number of normal OB stars with NS companions, as the higher mass stripped stars avoid a mass transfer after core helium exhaustion. Many OBe stars spin down during core-He burning of the stripped stars, which can be seen from the synthetic OB+HeS systems, as they are all assumed to rotate critically after mass transfer. Unbound systems are found for the same OB star masses as NS companions, as they share the same evolutionary past. Unbound OB stars have a larger OBe fraction than systems with NSs, which is probably cased caused by a larger probability of systems with high orbital period, where tidal breaking does not play a role, to unbind during the SN explosion. We predict 138 NS+OBe systems which we expect to appear as BeXBs, in good agreement with the observations. Additionally, we find 35 normal OB stars with NS companions.

Lastly, for OB star above about $14\msol$, we expect BH companions to start forming. They become the dominant companion type above around $18\msol$ and reach a nearly constant fraction of around 80\% for OB star masses above $22\msol$. Only for the highest considered OB star masses ($>70\msol$), the fraction of BHs decreases slightly. About $30\%$ of BHs have an OBe companion and this number decreases with OB star mass. The reason are angular momentum loss by stellar wind and the tides imposed by the BH and its HeS progenitor on to the star, which can be seen from their period distribution (see Sect.~\ref{sec-orbit} and \ref{apx-orbit}). At such high masses, also angular momentum loss by wind becomes important, which can be seen by the decreasing number of OBe+BH systems for increasing OB~mass. We predict for the SMC a total number of 142 OB+BH binaries for the scenario without BH kick. 36 of them should have an emission line companions. Additionally we expect 106 normal OB+BH systems. 
%COMPARISON WITH OBSERVATIOSN IS DONE LATER.which may appear as SB1 binaries. NEED BLOEM RESULTS FOR COMPARISON. \citet{2022A&A...658A.129J} predicts that a large fraction of Galatic OB+BH systems should be identifiable with GAIA. 

Figure~\ref{CompFrac} (right) shows that bound OB+NS systems, unbound OB+NS systems, and OB+BH systems occur with roughly the same numbers. The dominant SN mode are CCSNe from a HeS progenitor (pink), which make up about 60\% of all events. About 3/4 of them unbind the NS from the system. Systems with small helium core masses undergo a RLO Case~BB/BC after core helium exhaustion and produce almost naked CO cores. As we assume that they experience smaller kicks, most of them remain bound. ECSN, which have the smallest kicks (Table~\ref{kick}). In general they do not lead to a disruption and make up less then 1\% of the unbound systems. In this scenario without BH kick, no BH is unbound from its companion star and so all unbound systems released a NS. We discuss the effect of BH kicks in Sect.~\ref{sec-BHkick}.

\subsection{Masses of OB stars and their companions}\label{sec-mass}

\begin{figure*}
 \centering
 \includegraphics[width=\textwidth]{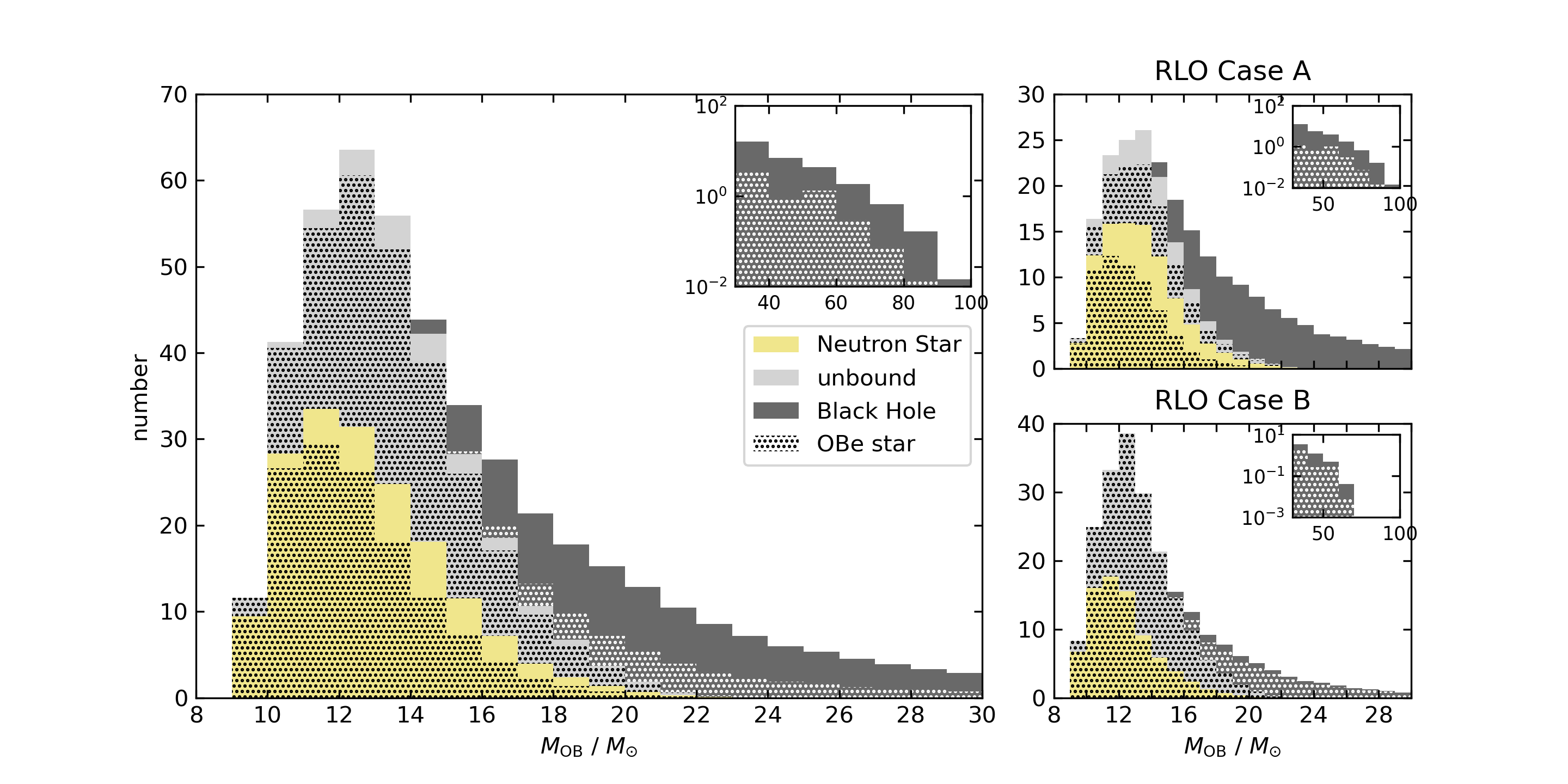}
 \includegraphics[width=\textwidth]{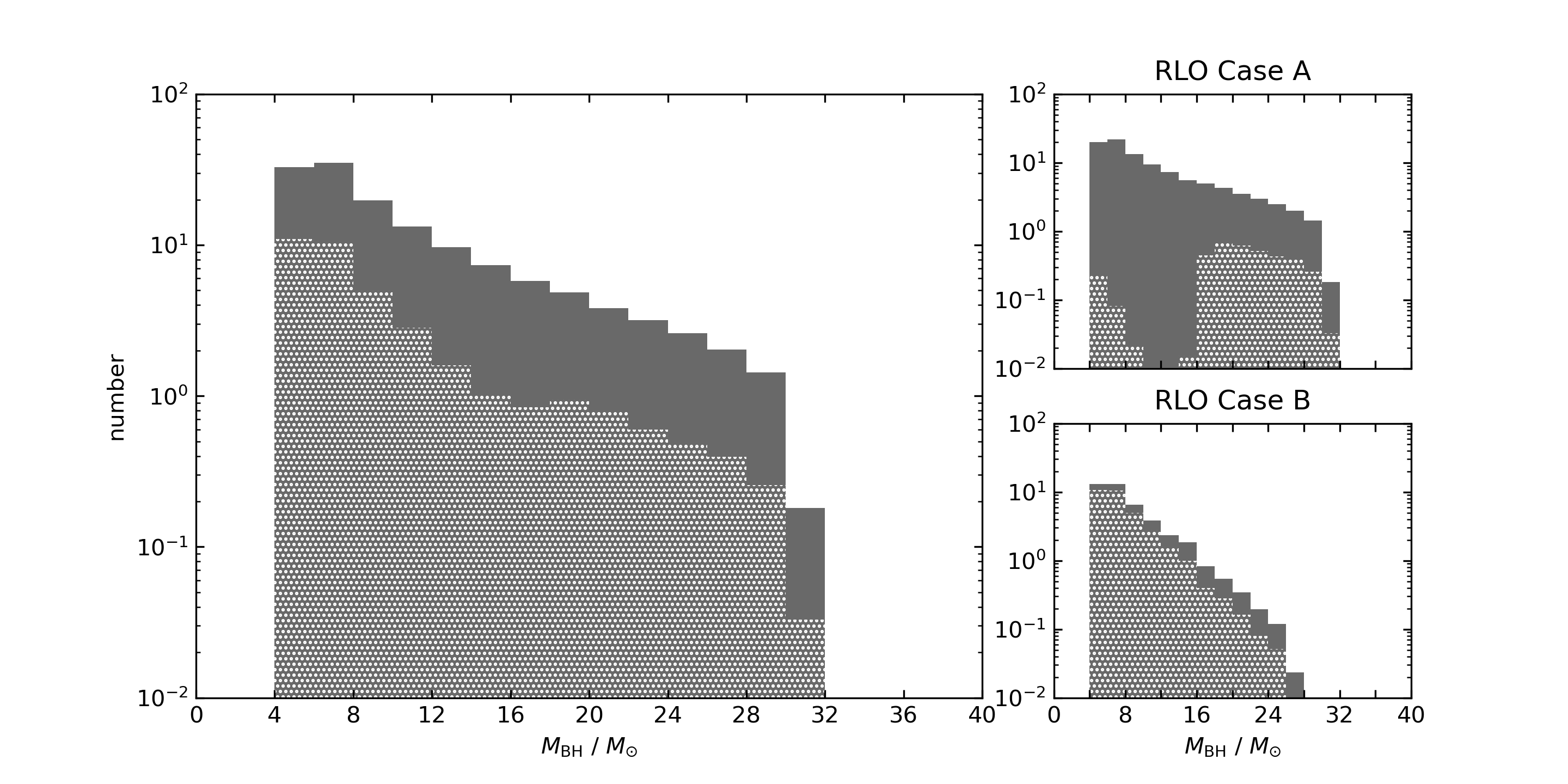}
 \caption{Predicted masses of the components of massive binary systems after RLO and the formation of a SR. Top row: OB~star masses coloured according to their companion type and marked with dots if we expect an emission line star. The inlays in the upper right corners focus on the upper mass end. Bottom row: BH~masses with dots in case of a OBe companion. Note that the vertical axis is logarithmic. Left: Total numbers. Right: Case~A RLO (top) and~B (bottom) separately.}
 \label{mass}
\end{figure*}

In Fig.~\ref{mass}, we show the predicted masses of the components of bound and unbound OB+NS systems and OB+BH systems. We find stellar masses (top row) from 9 to $100\msol$. In different mass ranges different companion types dominate. We find NS companions for stars with 9 to $22\msol$ and BHs from 14 to $100\msol$. Systems which broke up due to the SN kick follow the same patterns as NS systems, with slightly heavier OB stars as the former peak around $13\msol$, while the latter reach their maximum at $12\msol$. OB~stars with BH companions have their mode at $20\msol$.

We can understand the mass distributions considering the assumed binary physics. We find the lightest NS progenitors to have an initial mass of about $10\msol$, corresponding to a $3\msol$ helium core at core hydrogen exhaustion. Thus $7\msol$ are lost from the donor during the mass transfer. The mass transfer efficiency is around 60\% in this mass range and the minimum mass ratio for stable mass transfer is about 0.5 (see Fig.~\ref{fig:Pq}), which yields a lower limit of $5\msol$ for the initial mass of the accretor and therefore a minimum mass of $9\msol$ for the OB star as seen in Fig.~\ref{mass} (top). Similarly the upper limit and the ranges of the BH companions can be understood.

Figure~\ref{mass} (top right) shows Case~A and~B systems separately. We find that almost all regular OB~stars evolved through Case~A since in close orbits tidal forces are more efficient in braking the star. Systems tend to disrupt more frequently in Case~B, which is due to the lower binding energy in wide orbits. In Case~A systems the mass distributions are slightly wider. This is caused by orbital period dependent donor mass after RLO \citep[see][]{2024A&A...690A.282S}. Since the main-sequence models expand stronger for larger mass in our stellar model grid, the preference for BH~systems to evolve through Case~A is not surprising.

The lower panel of Fig.~\ref{mass} show the mass distribution of the BH companions. They range from 4 to $32\msol$ and show a slope $\mathrm{d}\log N/\mathrm{d}M_\mathrm{BH}$ of about $-0.05$. The minimum BH~mass derives from the minimum mass of a HeS to form a BH, which is $6.6\msol$ and the BH formation prescription (Sect.~\ref{method}) and is thus $4.8\msol$. The upper mass limit, about $30\ms$, can be found similarly by taking away 50\% of the initial mass of a $100\ms$ star, our upper limit, to get the mass of the collapsing HeS.

The BH masses depict interesting differences between RLO Case~A and~B (Fig.~\ref{mass}, bottom right). While the mass distribution of Case~A BHs shows the same slope as the total population, for Case~B $\mathrm{d}\log N/\mathrm{d}M_\mathrm{BH}$ is with $-0.1$ twice as large. We attribute this to the widening of the main-sequence for increasing stellar mass. As already mentioned, BHs whose progenitor evolved through Case~B prefer to have OBe companions, while those are rare for Case~A BHs. For those, we find a large dip of OBe companions around $12\msol$. It is unclear to us what causes it but we expect it to have limited effect it would increase the number of emission line companions only by 4 to 5, if one would remove the dip by drawing a straight line over it. BHs that formed after a Case~B RLO do not reach the same maximum mass as those from Case~A because our stellar models above $80\msol$ did not evolve beyond central hydrogen exhaustion. 

% \begin{figure}
%  \centering
%  \includegraphics[width=\columnwidth]{pic/mOB_mCC.png}
%  \caption{Predicted masses of OB+BH systems. The lower left panel shows the number of systems in the mass-mass plane in logarithmic colouring and the other panels are projections of that onto one axis, i.e. the distribution of OB masses (upper left) and BH masses (lower right), where emission lines stars are marked by white dots. We also show the masses of all known BH+OB systems and the masses of the SMC WR+O systems with WR masses as BH masses.}
%  \label{mOB_mCC}
% \end{figure}

\begin{figure}
    \centering
    \includegraphics[width=\columnwidth]{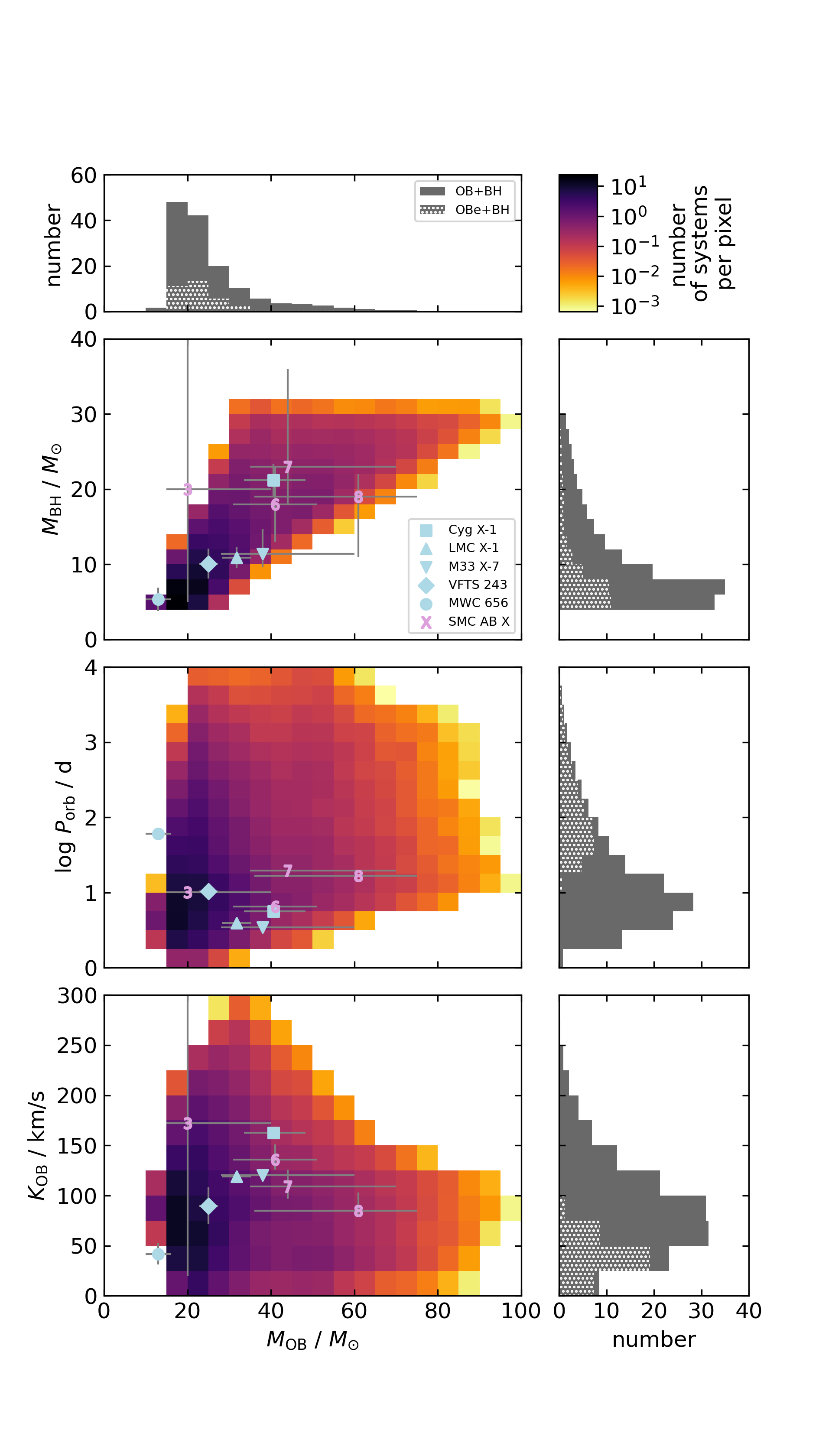}
    \caption{Predicted properties of OB+BH systems. The left panels shows the number of systems in the OB~mass--BH~mass plane (second row), OB~mass--orbital period plane (thrid row), and OB~mass--OB~velocity semi-amplitude plane (fourth row) in logarithmic colouring and the other panels are projections of that onto one axis, i.e. the distribution of OB~masses (first row left), BH~masses (second row right), orbital periods (third row right), and OB~star velocity semi-amplitudes (fourth row right), where emission lines stars are marked by white dots. We also show the properties of all dynamically confirmed OB+BH systems in the local group and of the WR+O systems in the SMC with WR~masses as BH~masses.}
    \label{fig-3er-mass}
\end{figure}

In Fig.~\ref{fig-3er-mass} (second row left), we show the combined mass distributions for OB+BH systems (see also Sect.~\ref{sec-q}). This distribution can be described as being framed by four lines. At first there are the upper and lower BH masses as discussed previously. As we have shown in this Section, one can divide the BH mass by 0.36 to estimate its initial stellar mass. Secondly, the population is limited to the left by a diagonal line following $q\approx1$. These systems stem from those with the initially most extreme mass ratios of about 0.3. Take for instance the upper left corner with $M_\mathrm{BH}\approx30\msol$ and $M_\mathrm{OB}\approx30\msol$. The initial BH progenitor mass was about $100\msol$ and the initial accretor mass about $25\msol$, since the mass-transfer efficiency is small at high masses. The systems in the lower left corner on the other hand ($M_\mathrm{BH}\approx6\msol$ and $M_\mathrm{OB}\approx10\msol$) had initial masses of about $15\msol$ and $6\msol$ due to the higher mass-transfer efficiency.

The systems at the right side of Fig.~\ref{fig-3er-mass} (second row left) are bounded by a line of $q \approx 0.2 \dots 0.3$ and come from systems with a mass ratio initially close to unity. Here, however, effects of Case~A mass transfer come into play. Take for example a system with initial masses of $15\msol$ and $14\msol$. If the donor loses half of its mass of which about a third (Eq.~\eqref{epsofm}) is gained by the accretor, the OB~star would have $16\msol$, but the diagram shows accretors as heavy as $22\msol$ in the lower right corner of the population. This is due to the fact, that donors which evolve though Case~A mass transfer can lose more than half of their mass, especially if a Case~ABC mass transfers strips them even further, reducing the mass of the BH and increasing the mass of the OB~star. For the density of systems per pixel in Fig.~\ref{fig-3er-mass} (second row left) we note, that the number of systems decreases for larger BH masses, but is fairly constant for varying OB masses. The former derives from the initial mass function and the latter from the near flat initial mass ratio distribution. This means it is more likely to find a $30\msol$ OB~star with a $15\msol$ BH than with a $30\msol$ BH, but it is about equally likely for a $20\msol$ BH to have a $30\msol$ and a $50\msol$ OB~companion.

%\red{For BHs we need to argue with the initial mass ratio. We described in the previous paragraph how one can estimate the BH mass, namely multiplying the initial mass by 0.36. From that we can estimate the mass ratio of the OB+BH system as $q \approx 0.72\cdot(q_0+\varepsilon/2)$, where $q_0$ is the initial mass ratio and $\varepsilon$ the mass transfer efficiency and we assumed that the donor lost half of its mass during the mass transfer. Thus we find a maximum mass ratio of 0.9 with $q_0=1$ and $\varepsilon=0.45$. This are the the highest possible values for $q_0$ and $\varepsilon$ as BH form from stars with initial masses larger than about $15\msol$. Similarly, as the highest masses, we take $q_0=0.2$ and $\varepsilon=0.1$ as boundaries and find $q=0.2$.}

%\subsection{Luminosities, temperatures and magnitudes}\label{sec-hrd}

\begin{figure}[h!]
 \centering
 \includegraphics[width=\columnwidth]{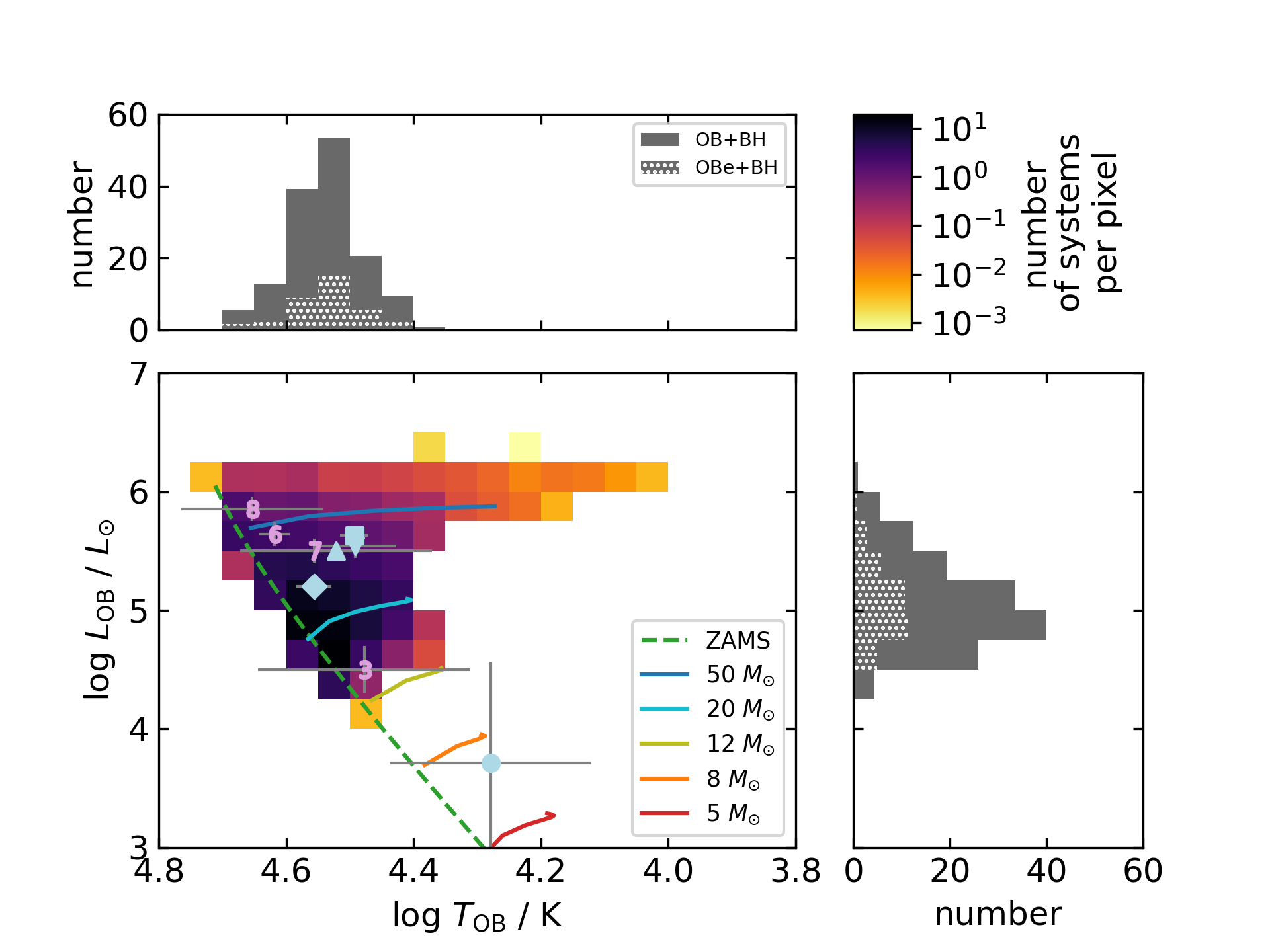}
 \caption{Predicted HRD positions of OB stars with a BH companions in logarithmic colouring together with selected model tracks and the zero-age main-sequence (ZAMS). We include values of all observed OB+BH systems and the O~star of SMC WR+O systems (see Fig.~\ref{fig-3er-mass} for the symbols). The panels on top and on the right show the temperature and luminosity distributions with predicted emission line stars highlighted by dotting.}
 \label{HRDs}
\end{figure}

Fig.~\ref{HRDs} show the predicted OB+BH population in the Hertzsprung-Russell diagram (HRD). It is expected that fast rotating stars redden due to the von Zeipel-theorem. We are not able to include this effect in our analysis as our models are fixed and rather show the non-rotational effective temperature. The OB stars have luminosities from $10^{4.5}\lsol$ to $10^{6.5}\lsol$ and display temperatures from 25\,kK to 50\,kK corresponding to O~and the earliest B~type stars. For the most massive companions, the main-sequence broadens a lot leading to a lowest effective temperature of 10\,kK. Their contribution is, as one can see in the upper panels, negligible. The HRD for the predicted OB+NS systems are provided in Sect.~\ref{apx-hrd}.

\begin{figure}
    \centering
    \includegraphics[width=\columnwidth]{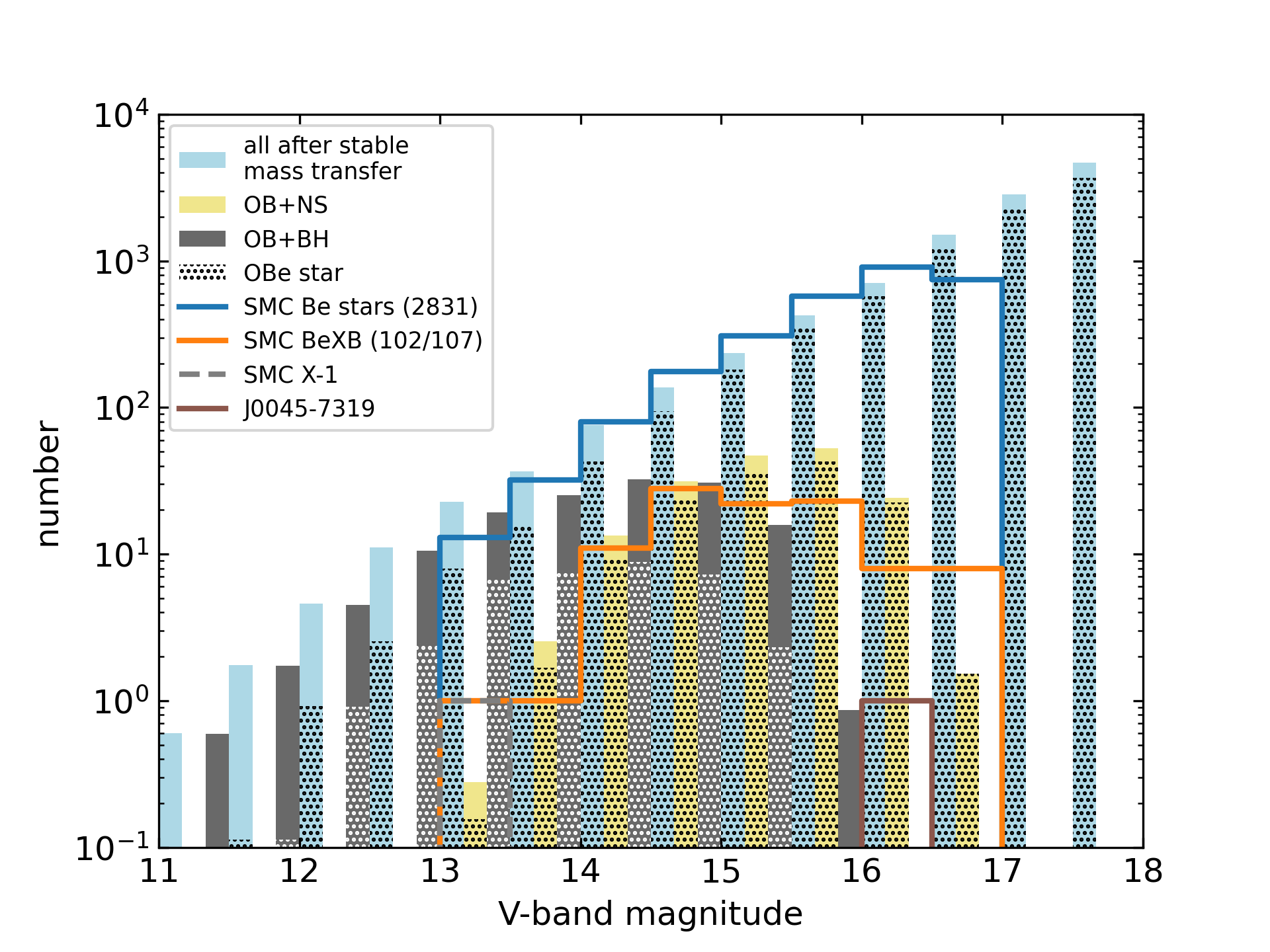}
    \caption{Predicted and observed distribution of V-band magnitudes. The blue bars show the magnitudes of all OB~star models (i.e. with HeS, WD, NS, and BH companion and unbound systems) after stable mass transfer. OBe candidates are marked with dots. As observations we show the SMC OBe magnitudes distribution (blue line) from \citet{2022A&A...667A.100S}, the magnitudes of BeXBs (orange line) from \citet[][living version \url{https://www.mpe.mpg.de/heg/SMC}]{2016A&A...586A..81H}, SMC~X-1 (grey dashed line) and J0045-7319 (brown line).}
    \label{magV}
\end{figure}

To make a meaningful comparison of our simulations with the observations we estimate the V-band magnitudes of our OB star models by using the recipe from \citet{2021A&A...646A.106S}, which is based on \citet{2016ApJS..222....8D} and \citet{2016ApJ...823..102C}, to calculate magnitudes from luminosity and effective temperature. We show the results in Fig.~\ref{magV}. There, the slope of the expected OBe stars $\mathrm{d}\log N/\mathrm{d} m_\mathrm{V}$ of about 0.55 until a magnitude of 12, where their number drops significantly, while the number of all post RLO systems continues to follow the same slope. A difference already appears around $m_\mathrm{V}=14$. The magnitudes we find for OB+NS binaries reach from 13 to 17 with a maximum around 15.5. The distribution is skewed towards dimmer stars. BeXBs are slightly more common at larger magnitudes. BHs can be found around stars brighter than 16th magnitude. Thire distribution is flatter and less skewed than the OB+NS distribution.

\subsection{Orbital parameters}\label{sec-orbit}

% \begin{figure*}
%  \centering
%  \includegraphics[width=\columnwidth]{pic/mOB_kOB.png}
%  \includegraphics[width=\columnwidth]{pic/lgP_kOB.png}
%  \includegraphics[width=\columnwidth]{pic/mOB_lgP.png}
%  \includegraphics[width=\columnwidth]{pic/lgP_ecc_BH.png}
%  \caption{Predicted orbital properties of OB+BH systems with measurements of all known OB+BH systems and of the SMC WR+O systems, in the same manner as Fig.~\ref{mOB_mCC}. Top left: OB~mass--velocity semi-amplitude plane. Top right: orbital period--velocity semi-amplitude plane. Bottom left: OB~mass--orbital period plane. Bottom right: orbital period--eccentricity plane.}
%  \label{orbitBH}
% \end{figure*}

\begin{figure*}
    \centering
    \includegraphics[width=\textwidth]{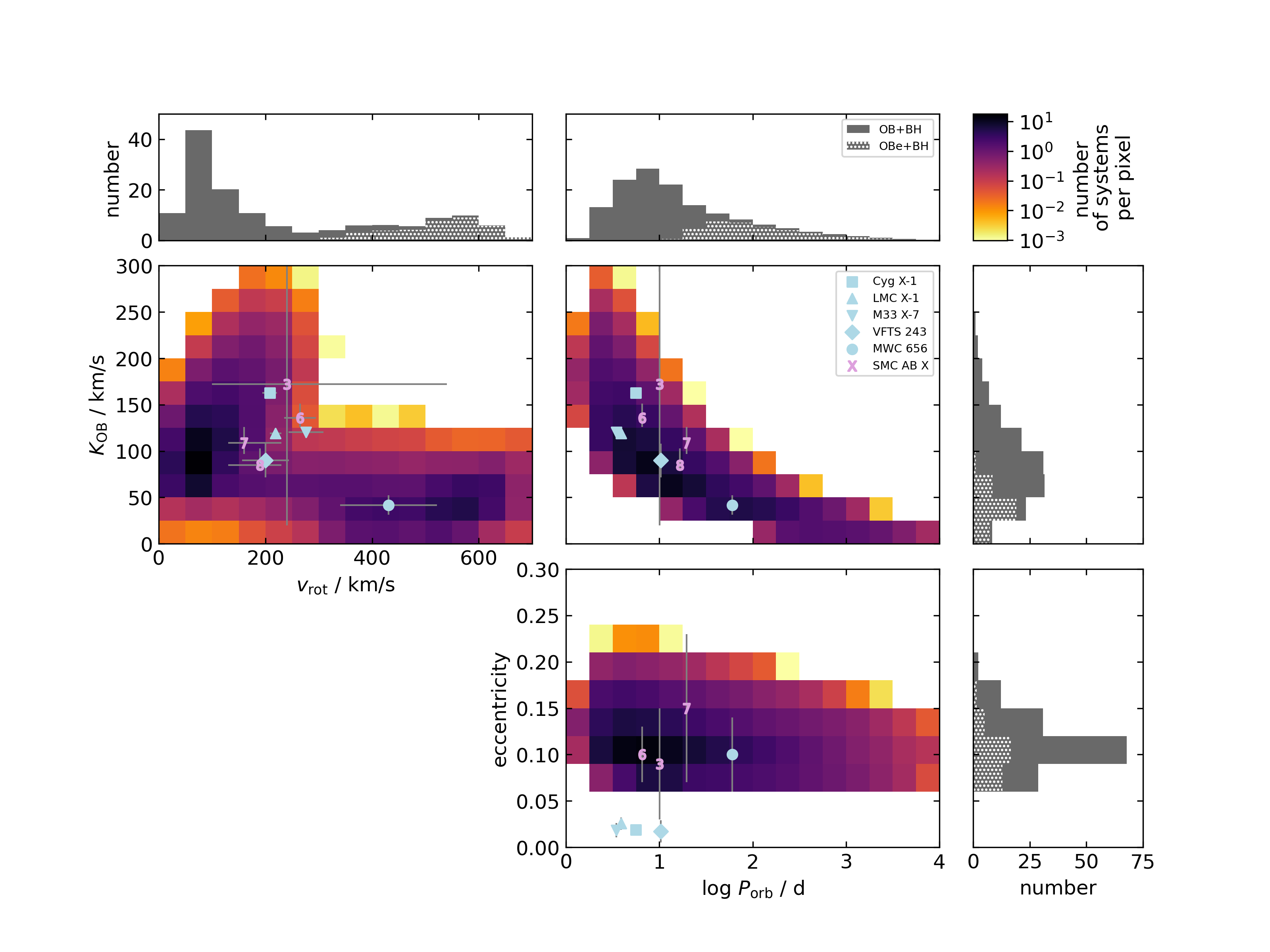}
    \caption{Predicted properties of OB~stars that have BH~companions with measurements of all dynamically confirmed OB+BH systems in the local group and of the WR+O systems in the SMC. 
    The main panels shows the number of systems in the rotational velocity--velocity semi-amplitude plane (first column, second row), orbital period--velocity semi-amplitude plane (second column, second row), and eccentricity--velocity semi-amplitude plane (second column, third row) in logarithmic colouring and the other panels are projections of that onto one axis, i.e. the distribution of OB~rotational velocities (top left), orbital periods (top centre), OB~star velocity semi-amplitudes (centre right), and eccentricities (bottom right), where emission lines stars are marked by white dots.}
    \label{fig-3er-orbit}
\end{figure*}

For the systems that remain bound after the formation of the SR, we can analyse their orbital properties, namely orbital period $P_\mathrm{orb}$, eccentricity $e$, and orbital velocity semi-amplitude of the OB star 
\begin{equation}\label{eq-K}
    K_\mathrm{OB} = \frac{M_\mathrm{SR}}{M_\mathrm{SR}+M_\mathrm{OB}} \sqrt{\frac{G(M_\mathrm{SR}+M_\mathrm{OB})}{a\cdot(1-e^2)}},
\end{equation}
where $a$ is the semi-major axis. Fig.~\ref{fig-3er-mass} and~\ref{fig-3er-orbit} show these orbital properties of our OB+BH binaries. In the bottom left panel of Fig.~\ref{fig-3er-mass} we find that systems with masses around $20\msol$ and velocity semi-amplitudes from 25 to $150\kms$ are preferred. Only the widest systems, those with low orbit velocities, become OBe stars. The upper right corner is empty as their progenitors would have initially overfilled their Roche lobe. No systems can be found at high velocities and low masses because the donor after Case~A being to light to form a BH. The centre right panel of Fig.~\ref{fig-3er-orbit} shows a somewhat narrow relation between orbital period and velocity semi-amplitude, which is due to Kepler's law. In the centre left panel of Fig.~\ref{fig-3er-mass} the combined distribution of orbital period and OB mass is shown. Both quantities are highly skewed and peak at low values. Towards high masses and periods a wide plane can be found. The upper right corner is empty because such system are very rare due to the initial distributions. Systems with low periods and high masses are avoided as such systems would overfill their Roche-lobe initially. Finally, the bottom centre panel of Fig.~\ref{fig-3er-orbit} depicts orbital period and eccentricity. The eccentricity only assumes values between 0.05 and 0.2 and strongly peaks at 0.1. The reason is our BH formation formalism. As 20\% of the helium envelope is lost, the BH progenitor loses some of its momentum, which translates to a non-zero eccentricity \citep{1998A&A...330.1047T}. Varying the mass loss or imposing a kick on the BH would change the resulting eccentricity dramatically. Note however that our prescription does not account for the continuous circularisation of the orbit due to tides after the SN, so the real eccentricity may be lower. Nevertheless the eccentricity distribution of wide OB+BH systems could be a probe for BH kicks.

% \begin{figure}
%  \centering
%  \includegraphics[width=\columnwidth]{pic/rot_kOB.png}
%  \caption{Same as Fig.~\ref{mOB_mCC}, but for the rotational velocity and the orbital velocity semi-amplitude.}
%  \label{rot-orb}
% \end{figure}

We show a combined histogram of rotational and orbital velocity of OB+BH systems in Fig.~\ref{fig-3er-orbit} (centre left). We find that the OB+BH populations divides clearly in two sub-groups. The somewhat larger group can be found at low rotational ($<200\kms$) and medium to high orbital ($>50\kms$) velocities. These systems evolved through Case~A RLO and are normal OB stars. Systems with high rotational ($>300\kms$) and low orbital ($\sim50\kms$) velocities form the second group and are in general emission line stars. This has an important implication for the observational search for BHs, namely that on one hand high radial velocity variations in SB1 systems and on the other hand that OBe stars are a predictor of BH companions.

% \begin{figure}
%  \centering
%  \includegraphics[width=\columnwidth]{pic/lgP_ecc_NS.png}
%  \includegraphics[width=\columnwidth]{pic/periodNS.png}
%  \includegraphics[width=\columnwidth]{pic/ecc.png}
%  \caption{Predicted orbital properties of OB+NS with observations of the SMC BeXBs, SMC X-1, and J0045-7319, in the same manner as Fig.~\ref{mOB_mCC}. Top: OB~mass--velocity semi-amplitude plane. Middle: orbital period coloured by SN-type. Bottom: eccentricity coloured by SN-type.}
%  \label{orbitNS}
% \end{figure}

\begin{figure*}
    \centering
    \includegraphics[width=0.75\textwidth]{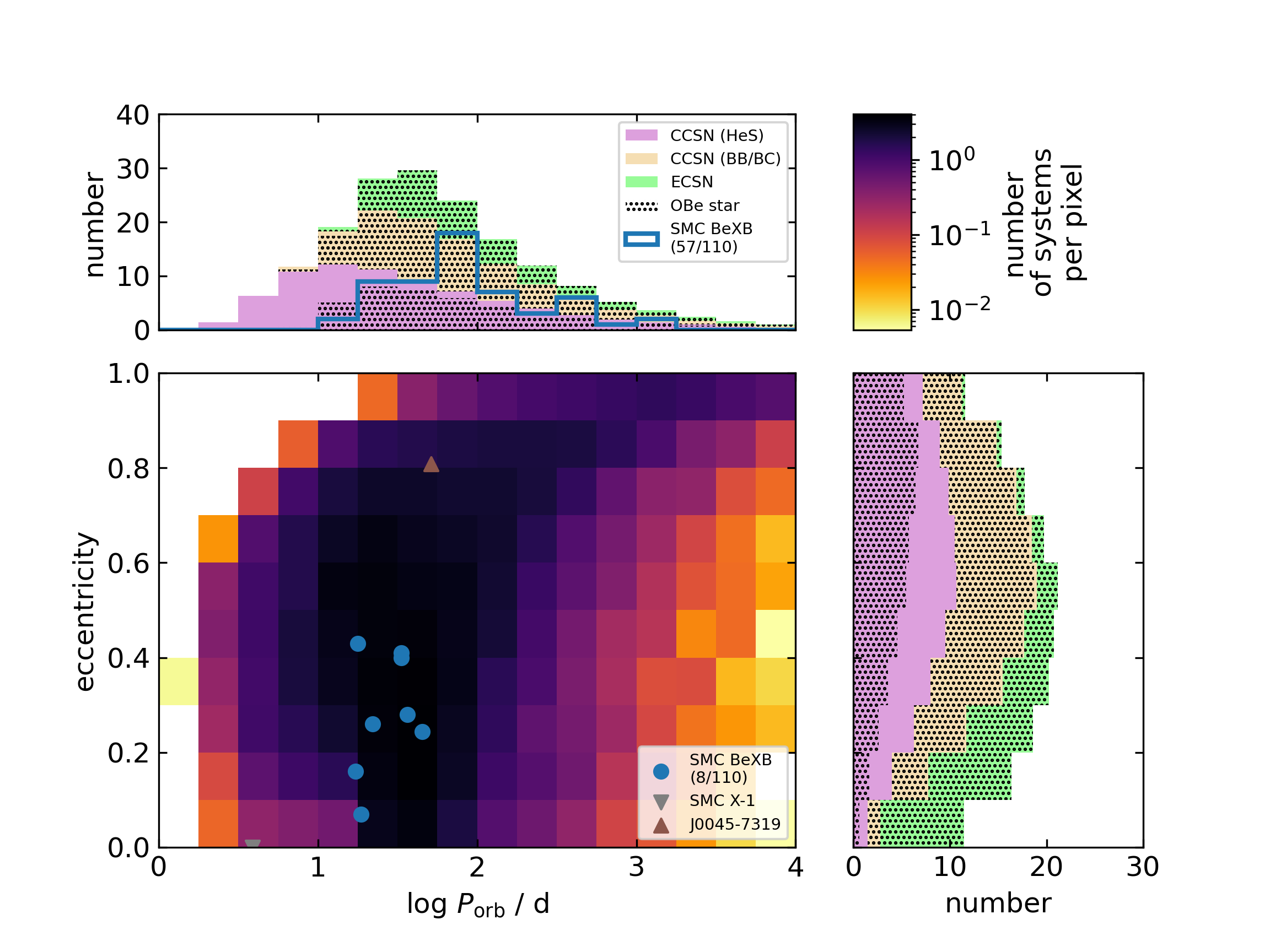}
    \caption{Predicted orbital properties of OB+NS systems with observations of the SMC BeXBs, SMC X-1, and J0045-7319. Centre: orbital period--eccentricity plane. Top: orbital period coloured by SN-type. Right: eccentricity coloured by SN-type.}
    \label{fig-3er-NS}
\end{figure*}

The orbital properties period and eccentricity of OB+NS binaries are shown in in Fig.~\ref{fig-3er-NS}. We do not treat mass ratio and orbital velocity here because the NS mass is strongly confined and the OB stars' velocities are very low due the low NS mass, see however Fig.~\ref{massratio}. The centre panel shows a 2D histogram of orbital period and eccentricity. The eccentricities follow a very broad distribution covering all values from 0 to 1 and the periods' mode is, as discussed above, between 10 and 100\,d. The combined distribution has accordingly a large main feature at these values. A notable exception is a preference for large orbital periods at high eccentricities and an avoidance of high eccentricities at low orbital periods. This is not surprising as this would lead to Roche-lobe overfilling periastron distances. While emission line stars are more frequent at large orbital periods, we find no dependency on eccentricity.

In the top and right panel of Fig.~\ref{fig-3er-NS}, we show the period and eccentricity distributions coloured according to the SN~type. CCSNe from a HeS, CCSNe after a Case~BB/BC RLO and ECSNe have slightly different typical orbital periods, increasing in that order. This means systems with a stronger kick lead to lower orbital periods, which may sound counter-intuitive but can be explained with wider orbits being more likely to unbind than close orbits. Furthermore, CCSNe from a HeS progenitor are the only species that yields a notable amount of regular OB stars as tidal breaking before the SN is only possible here. In case of a CCSN after Case~BB/BC RLO the OB~star was just spun-up again and systems which undergo a ECSN and are close enough for effective tidal braking also experience such a mass transfer. The eccentricities of systems that evolved through ECSNe show the smallest eccentricities and those which experienced a CCSN from a HeS the largest. This also reflects the magnitude of the SN kick as larger kick velocities lead to more eccentric systems. As mentioned above we did not employ continuous tidal induced circularisation of the orbit which may affect the eccentricity distribution.

%\citep{2016A&A...586A..81H}
%\citep{2012A&A...537A..76S, 2020MNRAS.497L..50C}

\subsection{Systemic velocity}

% \begin{figure*}
%  \includegraphics[width=\textwidth]{pic/v3D.png}
%  \includegraphics[width=\columnwidth]{pic/v3D_SN.png}
%  \includegraphics[width=\columnwidth]{pic/v3D_rot.png}
%  \caption{Top: Same as Fig.~\ref{mass} but for the space velocity of the OB~star. Bottom left: Space velocity coloured by SN-type with observation of the transverse velocities of isolated SMC OBe stars (blue line), BeXBs (blue arrows), and SMC X-1 (grey). The part of the histogram within the grey line indicates the unbound systems. Bottom right: Systemic velocity and rotational velocity of OB stars with bound and unbound NS and BH companions as in Fig.~\ref{mOB_mCC} and the same observational data as the left panel. The median error of the transverse velocities is $23\kms$.}
%  \label{v3d}
% \end{figure*}

\begin{figure*}
 \centering
 \includegraphics[width=0.75\textwidth]{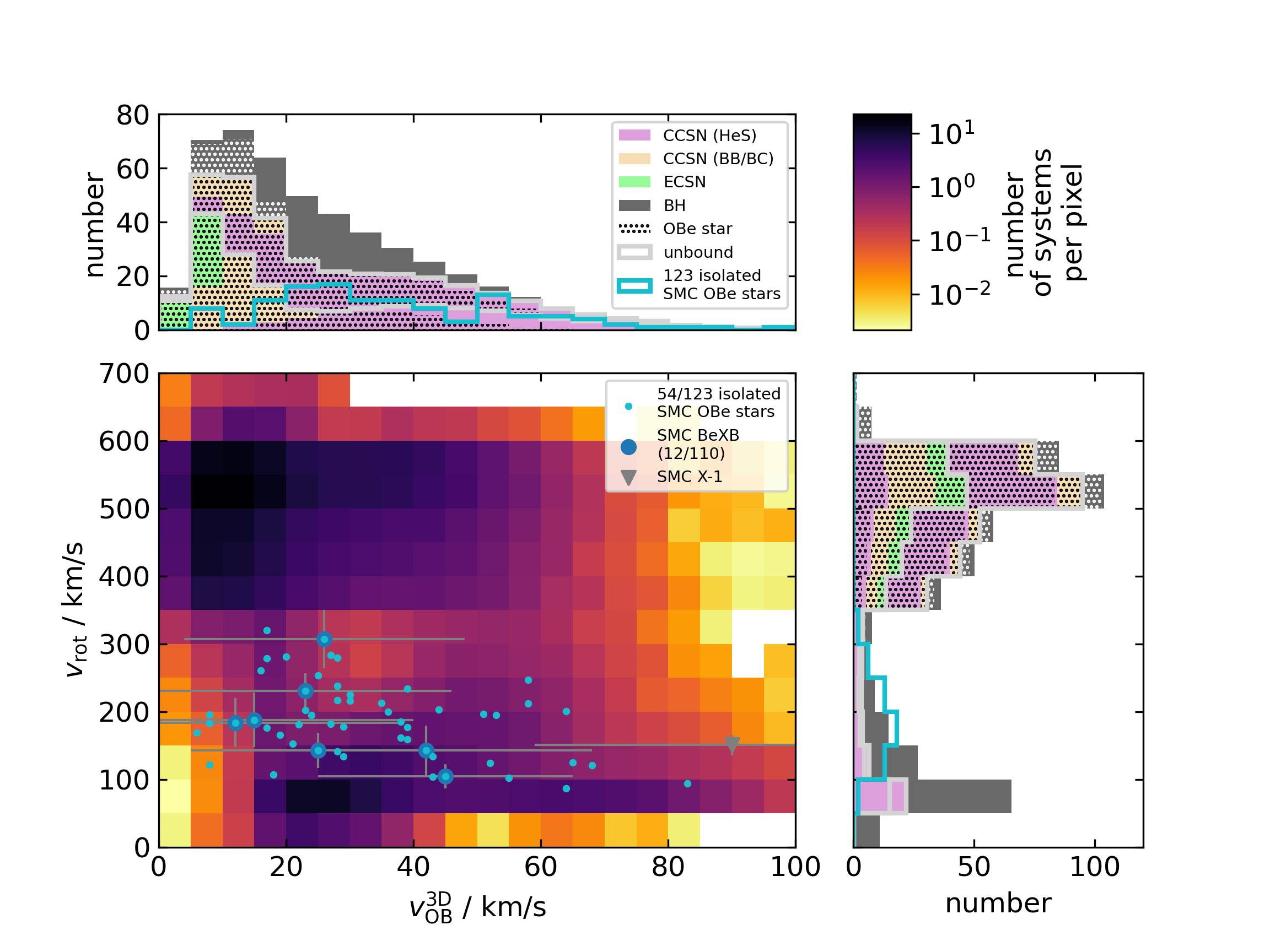}
 \caption{Predicted space velocities and rotational velocities of OB~stars with bound and unbound NS and BH companions with observations of the transverse velocities of isolated SMC OBe stars (blue line and small dots), BeXBs (large blue dots), and SMC X-1 (grey). The median error of the transverse velocities is $23\kms$. Centre: space velocity--rotational velocity plane. Top: space velocities coloured by SN-type. The part of the histogram within the grey line indicates the unbound systems. Right: rotational velocities coloured by SN-type.}
 \label{fig-3er-v3d}
\end{figure*}

Because a SN kick changes the momentum of the system, we are able to predict the space (centre of mass) velocity $\varv_\mathrm{OB}^\mathrm{3D}$ of our OB systems and ejected OB stars as we do in Fig.~\ref{fig-3er-v3d} and~\ref{v3d}. We find that NS systems (yellow) have the broadest distribution from 0 to up to 100\,km/s. Even more striking is that this distribution is bimodal with one narrow maximum at 10\,km/s and a wide one at 40\,km/s. Inspecting the upper left panel we see that the first peak comes from ECSNe and CCSNe of a Case~BB/BC donor. The second peak comes exclusively from CCSN with a HeS progenitor, as they have the largest kick velocities which lead to the large space velocities. CCSN after Case~BB/BC RLO and ECSN lead to lower space velocities in agreement with the strength of their kicks. NS systems with the largest space velocities tend not to be OBe stars. This is due to the fact that the pre-SN orbital velocity leaves an imprint on the space velocity \citep{1998A&A...330.1047T}. Thus we can follow that they come from initially close binaries which explains the absence of fast rotation. The panel on the top right of Fig.~\ref{v3d} support that Case~A produces faster systems.

For unbound systems (grey contour in Fig.~\ref{fig-3er-v3d}) we find the most common space velocity of the OB star just above 10\,km/s. They again reach up to almost 100\,km/s but do not have the bimodality of their bound counterparts. BH systems also show a unimodal distribution, but with lower velocities (peak at 20\,km/s), since we do not assume any BH kicks. Their velocities are, as well as the eccentricities, caused by the mass loss during BH formation. With these assumptions BH systems no to break up.

In the centre of Fig.~\ref{fig-3er-v3d} we show a 2D histogram of spacial and rotational velocities of all OB+SR systems, whether bound or unbound. We find similarly to Fig.~\ref{fig-3er-orbit} (centre left) two populations. First, there are systems with high rotational ($>300\kms$) and relatively low spatial velocity (peak at $10\kms$), which are the Case~B and wide Case~A systems. Then there are systems with low rotational ($\sim100\kms$) and higher spacial velocities (mode at $20\kms$). These are the narrow Case~A systems where tidal locking is relevant which inherited their fast orbital velocity as systemic velocity.

% Reed 2003: v_rad of B(e)
% van den Heuvel: v_tr of BeXB

\section{Discussion}\label{discuss}
In this Section, we compare our results with observations (Sect.~\ref{observ}), with Paper~I (Sect.~\ref{sec-dis-xu}]), and with previous work (Sect.~\ref{lit}). The main uncertainties of our simulations are discussed in Sect.~\ref{uncert}.

\subsection{Comparison with observations}\label{observ}

In this Section we compare our predicted distribution to observations of SMC systems and suitable proxies. We focus here on systems with BH or NS companions and their directly observable properties.

\subsubsection{BH systems}

No OB+BH system was so far observed in the SMC. Therefore we rely on OB+BH systems observed in the local group and WR+OB systems, which are believed to evolve to OB+BH systems \citep{2012ARA&A..50..107L}, as proxies. The OB+BH systems in question are Cyg~X-1 \citep{2011ApJ...742...84O,2021Sci...371.1046M}, LMC X-1 \citep{2009ApJ...697..573O}, M33~X-7 \citep{2007ATel..977....1O,2022A&A...667A..77R}, VFTS~243 \citep{2022NatAs...6.1085S}, and MWC~656 \citep[][the true nature is under debate, see \citet{2022arXiv220812315R} and \citet{2023A&A...677L...9J}]{2014Natur.505..378C}. For the WR+O systems, we rely on \citet{2016A&A...591A..22S} and \citet{2018A&A...616A.103S} and references therein. We align all analysed quantities by considering their inclination. %For all analysed quantities, we calculate and discuss the un-inclined values. 
The BH~systems exists in environments with higher metallicities than the SMC, which has effects on the mass limit for BH formation, the BH mass, and the OB~star mass and rotation by wind mass loss and wind angular momentum loss \citep{2012ARA&A..50..107L,2023pbse.book.....T}.

In Fig.~\ref{fig-3er-mass} we can see that our predictions and the observed masses cover the same ranges. We treat the WR mass \citep{2018A&A...611A..75S} as BH mass assuming the WR collapses completely to a BH. It is possible that some material is ejected and the total mass reduced by gravitational binding energy, see Sect.~\ref{method}. MWC~656 is at the extreme low mass end of both components, assuming it is a BH system. However we find that the observations are roughly equally distributed over the predicted area. This is not surprising as the selected observations are biased towards higher masses. First, massive stars are brighter and more likely to observe, and second WR+O systems are the OB+BH progenitors with the highest masses and far more likely to identify observationally than HeS+OB systems, where the HeS is not a WR.

In Fig.~\ref{fig-3er-mass} (bottom) we find the observed systems to cluster around a velocity semi-amplitude of 100 to $150\kms$. The reason may be the difficulty in observing systems with lower values. In the top centre panel of Fig.~\ref{fig-3er-orbit} we find that the observations follow the shape of the predicted population. Only MWC~656 lies in the region where we expect OBe stars (high period, low orbital velocity) and it is the only known BH+OBe candidate. As predicted, we find a typical orbital period of around 10\,d.

The bottom centre panel of Fig.~\ref{fig-3er-orbit} shows the predicted the OB+BH systems' eccentricities to range from 0.05 to 0.20 but the observed OB+BH systems have values clearly below this, with the sole exception of MWC~656. The reason may be tidal circularisation in close orbits, no mass and energy loss during the BH formation, a kick, as post-SN eccentricities below about 0.06 can only be produced by a kick \citep{1998A&A...330.1047T}, or our BH formation scheme overestimates the mass ejection during BH formation process. Even more interesting is that the WR+O systems from the SMC have a non-zero eccentricities while we would expect a perfectly circular orbit after RLO.

All observed BH/WR+OB systems (but MWC~656) are observed to have relatively similar orbital ($\sim 120\kms$) and rotational velocities ($\sim 200\kms$) of the OB star (Fig.~\ref{fig-3er-orbit}, centre left). These observations are close the our predicted subpopulation of BH+OB systems with slow rotation. However our synthetic population rather has with about $100\kms$ a most likely rotational velocity that is a half as large as observed. This may be due to an overestimate of tidal, wind braking in our models or due to a different stellar structure. We can conclude that these systems likely evolved through Case~A mass transfer. MWC~656 however seems to belong to the OB+BH sub-population at high rotation. Indeed this system harbours an emission-line star matching the predictions. Thus MWC~656 might be the tip of the iceberg of yet undiscovered BH+OBe systems.

Fig.~\ref{HRDs} offers an important counterargument to the BH nature of MWC~656. While all other BH/WR+OB systems can be observed with temperatures and luminosities in well agreement with observations, MWC~656 lies clearly outside our population of OB+BH systems. This may be due to uncertainties in our BH formation criterion or a different mass transfer efficiency, but we rather see as a support for the claim of \citet{2022arXiv220812315R} and \citet{2023A&A...677L...9J} that MWC~656 harbours a subdwarf instead of a BH. %\red{The HRD position is from an older paper. Maybe we can do better by using Gaia measurements, but I am not able/Abel.}

\subsubsection{BeXBs, SMC X-1 and J0045--7319}

\citet[][living version \url{https://www.mpe.mpg.de/heg/SMC}]{2016A&A...586A..81H} report 107 BeXBs in the SMC. They give V-band magnitudes for 102 and orbital periods for 54 of them. \citet{2011MNRAS.416.1556T} and \citet{2015MNRAS.452..969C} list eccentricities for seven BeXBs. Next to the BeXBs two other systems with a NS are found in the SMC. These are the super-giant X-ray binary SMC X-1 \citep{2007A&A...473..523V} and the pulsar/B-type star binary J0045--7319 \citep{1995ApJ...447L.117B, 1996Natur.381..584K, 2005AJ....129.1993M}. Magnitude, orbital period and eccentricity are well known for them, which we show in Fig.~\ref{magV} and~\ref{fig-3er-NS}.

The V-band magnitudes of BeXBs covers the same range as our simulation (Fig.~\ref{magV}), but the distributions have slightly different shapes. SMC X-1 lies at the bright end of our distribution and J0045--7319 at the dim end, while we would have rather expected such systems without emission lines over the whole range of magnitudes.

The orbital period distribution of BeXBs behaves with values above 10\,d like the synthetic one (Fig.~\ref{fig-3er-NS}). Both samples spread out over several decades and show a clear maximum. The differences, especially the large number of BeXBs with periods between 56 and 100\,d may be due to sampling. The period of J0045--7319 is close to the mode of the model's, however it is not a OBe~star like almost all of our systems at such a period. SMC X-1 has with 3.89\,d the shortest orbital period. At such short periods we predict that the OB star does not show emission lines. Indeed, SMC X-1 is a super-giant X-ray binary.

If we consider eccentricities we find a notable difference between our work and the observations. While we predict an almost flat distribution for the BeXB, the observations range only up to 0.5. This may be due to an observational bias against high eccentricities. The high eccentricity of J0045--7319 and the near zero one of SMC X-1 fit to our model. In Fig.~\ref{fig-3er-NS} (centre) we find that all observations lie where we predict the largest density of systems. The only exception is SMC X-1, which may be explained by our ignorance towards tidal circularisation. The mass ratios of SMC X-1 ($0.068\pm0.010$) and J0045--7319 ($0.16\pm0.03$) are at the edges of our predictions (Fig.~\ref{massratio}). The deprojected rotational velocity of the optical counterpart of J0045--7319 is $163\pm21\kms$ and fits to the predictions for normal OB~star in Fig.~\ref{OBrot}. The recently discovered LMC BeXB system A0538-66 \citep[$P_\mathrm{orb} = 16.6\days, e=0.72$][]{2017MNRAS.464.4133R,2022A&A...661A..22D} is covered by or synthetic systems in Fig.~\ref{fig-3er-NS}.

% \citep{2016A&A...586A..81H} \citep{2011MNRAS.416.1556T, 2015MNRAS.452..969C}

\subsubsection{Spatial velocities}

In Fig.~\ref{fig-3er-v3d} we have added local transverse velocities of 123 isolated SMC OBe stars from \citet{2020ApJ...903...43D}. For most of these observations the binary status is unknown, but since \citet{2020A&A...641A..42B} found evidence that such systems are post-interaction systems, we assume this and compare them with our OB+CC systems (including unbound accretors). The observations likely also contain OB+HeS systems for which we expect no local spatial velocity as they did not undergo a SN. Twelve of them have a counterpart in the BeXB table of \citet{2016A&A...586A..81H} wherefore we highlight them as well as SMC X-1. \citet{2020ApJ...903...43D} also provide rotational velocities for 54 of their systems. Notice that the observations are projected velocities but our simulations are not.

The velocity range of observed and simulated systems agree well, while the shape of the distributions are different (Fig.~\ref{fig-3er-v3d} top). Our data are more strongly skewed towards lower velocities, which one would rather expect from the observations due to projection effects. A possible explanation is an observational bias against slow moving systems and the relatively large uncertainties. (The median error of the local transverse velocity is $32\kms$.) The BeXBs on the other hand seem to follow the simulated distribution more closely as they only reach $45\kms$. However the system with the largest observed projected velocity is already moving as fast as the fastest (not projected) synthetic BeXBs. SMC X-1 is with $90\kms$ in agreement with our OBs+NS systems.

In the centre panel of Fig.~\ref{fig-3er-v3d} we find that the bulk of the observations shows lower rotational velocities than we predict. All of the observed OBe star lie in the part of the diagram where we predict normal OB stars. The not considered inclination of the observed system is not able to account for this. However, the observations may underestimate the rotational velocity at the equator as this region cools by the von~Zeipel-theorem and thus dims \citep{2004MNRAS.350..189T}. We also may have underestimated the effects of tidal and wind braking.

\subsection{Comparison with detailed binary models}\label{sec-dis-xu}

In this Section we compare our results to Paper~I, which is a companion study with the same aims but a different approach. While we use a rapid binary population synthesis code to generate a large range of model populations, Paper~I use a given set of detailed MESA binary models. Besides that, the major differences between the two works are, first, that our models accretion during RLO is not limited to the achievement of critical rotation of the accretor but is normalised to the observed number of BeXBs and WR+O systems. Second, Paper~I uses a luminosity criterion to decide whether a RLO is stable or leads to a stellar merger, while we rely on the occurrence of a L\textsubscript{2} overflow due to the swelling of the accretor. The major consequence is that the $q_\mathrm{i}$-$\log P_\mathrm{i}$-diagram of the two approaches look very different (our Fig.~\ref{fig:Pq} and Sect.~\ref{app:Pq} and their Figs.~2, A.1 and A.2). While in Paper~I the region where donor stripping occurs has the shape of two triangles with stable mass transfer up to smaller mass ratios at high orbital periods, in our work the transition between Case~A and~B is continuous and if systems merge they tend to be those at high orbital periods and mass ratios away from unity. These two mass transfer criteria leads to very different OB+BH populations. Papr~I predict a large fraction of long-period binaries associated with fast-rotating main-sequence companions but our predicted OB+BH binaries are dominated by close systems with normal OB~stars

By direct comparison of our Fig.~\ref{CompFrac} and Fig.~6 of Paper~I, we find that they predict an overall smaller number of OB+SR systems. Also their number in each mass bin is smaller than ours. The different companion types are also differently distributed over the OB star masses. Their OB stars up to $8\msol$ and $30\msol$ have WD and NS companions, respectively. The NS with such massive companions evolved through Case~A mass transfer which may cause difference in final fate as MESA models the internal structure of the stripped star detailed. On the other hand, stars as light as $6\msol$ harbour NSs and BHs in their study, while our lowest masses are $8\msol$ and $14\msol$ due to the difference in mass-transfer efficiency and mass transfer stability.\footnote{In Paper~I, this $6\msol$ companion corresponds to an initial primary star of $10\msol$ and an initial mass ratio of 0.6 with zero mass-transfer efficiency.} The ratio of OB+NS to OB+BH systems of Paper~I is smaller than ours, because their merger fraction for NS-forming systems is much higher than ours. Furthermore, we predict more Case B mergers and less Case A mergers, resulting in a lower ratio of OBe to normal OB stars than that in Paper~I, and a weaker dependence of the OBe-to-OB fraction on OB~mass for OB+BH and OB+HeS systems, which is probably due to the treatment of wind angular momentum loss (see Sect.~\ref{uncert}).

%The impact of the different assumptions for mass transfer efficiency and stable mass transfer as a function of mass ratio and orbital period can be explain the major differences between our results and those of \citet{xu}. While \citet{xu} predicts more BH systems than we do, we predict more NS systems than \citet{xu}, because under the assumptions of \citet{xu} more low mass systems merge during RLO while we have slightly more merger for BH progenitors as well as a shorter lifetime due to the more massive companions caused by a larger mass transfer efficiency.

A important consequence of the merger criterion of Paper~I is the bi-modality of the orbital period (their Fig.~8), which is not present in our Fig.~\ref{fig-3er-NS}. The range of orbital period is similar nevertheless. While the systems at low orbital periods may not appear as BeXB and are rather X-ray quiet (Paper~I), this prediction could be tested. The current data seem to support our approach. 

In our model, the mass-transfer efficiency is only a function of accretor mass, but in Paper~I, it can depend on both masses and the orbital period. This may be the cause of several differences between our results, such as the the larger spread of OB masses for a fixed BH mass (our Fig.~\ref{fig-3er-mass} and their Fig.~11) and the larger range of luminosities of NS and BH companions and stronger overlap of the populations (our Fig.~\ref{HRDs} and their Fig.~4) in Paper~I. Overall Paper~I predicts less massive stars in OB+BH systems due to their lower mass-transfer efficiency. This also causes that our predicted mass ratios of OB+BH systems to be smaller than in Paper~I.

An interesting and testable difference can be found in our Fig.~\ref{fig-3er-mass} and Fig.~11 of Paper~I. We predict the lower left corner in the $M_\mathrm{OB}$-$P_\mathrm{orb}$ diagram much stronger populated than Paper~I does. In our Fig.~\ref{orbit} and Fig.~E.3 of Paper~I one can find accordingly the lack of close normal OB+BH systems. It may be explained with the fact that our merger criterion lets more systems with initially short periods and extreme mass ratios survive at low ($\sim15\msol$, the lowest BH progenitor initial mass) initial donor masses. If these close and light OB+BH binaries exist, they may produce observable X-ray emission \citep{2024A&A...690A.256S}.

There are also differences in the rotational velocities of OB~stars predicted by the two studies as can be seen in our Fig.~\ref{OBrot} and Fig.~G.7 of Paper~I. The population at low rotational velocities is missing in Paper~I, which could either be due to the different implementation of stellar rotation and angular momentum loss by tides and winds (see also Sect.~\ref{uncert}) or due to the merger criteria. We find also that the high rotation peak is broader in Paper~I, which is probably due to the implementation of rotational evolution. In our work this peak is only very weak for the BH companions. Paper~I has also has a smaller number of normal OB+NS systems and Case~B systems do not have $\vrot/\vcr$ below 0.8 in general.

For the OB+NS systems, we find that Paper~I predicts a slightly flatter eccentricity distribution but the tendency to have higher eccentricities for the largest periods remains (our Fig.~\ref{fig-3er-NS} and their Fig.~9). For the SN types, we find about the same tendencies with less CCSN after Case~BB/BC in Paper~I, which may be caused by the ability of the HeS to expand due to the presence of a hydrogen shell. It is also interesting to note that Paper~I finds much less OB+NS systems with velocity semi-amplitudes above $40\kms$ (our Fig.~\ref{orbit} and their Fig.~G.6).

\subsection{Comparison with earlier work}\label{lit}

In the last three decades many population synthesis studies of X-ray binaries and/or OBe~stars \citep[e.g.][]{1989A&A...226...88M,1989A&A...220L...1W,1991A&A...241..419P,1993ARep...37..411T,1995ApJS..100..217I,1995ApJ...440..280D,1995A&A...296..691P,1996A&A...309..179P,1997A&A...322..116V,1998MNRAS.293..113T,1998A&A...340...85R,2000A&A...358..462V,2001A&A...367..848R,2011RAA....11..327L,2014ApJ...796...37S,2014MNRAS.437.1187Z,2019A&A...624A..66R,2020MNRAS.498.4705V,2022A&A...667A.100S,2023A&A...672A..99M,2024MNRAS.527.5023L,2024ApJ...971..133R} have been conducted. While many of them focused on the X-ray output, we discuss here those who focus on aspects of binary evolution.

A popular approach in the past was to use a mass ratio dependent mass transfer efficiency, as \citet{1991A&A...241..419P}, \citet{1995A&A...296..691P} or \citet{1996A&A...309..179P} did. Typical assumptions are a high efficiency for systems with mass ratios close to unity and low values for rather unequal systems with a narrow transition zone. This description is in general not able to reproduce the mass (or spectral type) distribution of BeXBs, as B~stars as light as $2\msol$ are predicted. SN kicks are not able to regulate that sufficiently \citep{1995A&A...296..691P}. Rephrasing of this criterion into a thermally limited accretion \citep{1996A&A...309..179P} leads to the same problem. Many studies \citep[e.g.][]{1991A&A...241..419P} introduce an ad hoc minimal mass ratio for stable mass transfer of about 0.3 to 0.5 to remove systems with light companions. \citet{1995A&A...296..691P} identified with the specific angular momentum of the ejected material a further key parameter, which they used to regulate which binaries evolve to BeXBs. They found that the larger the specific orbital angular momentum of the ejected material the larger is the minimal mass ratio for donor stripping. This was confirmed by \citet{1997A&A...322..116V} and \citet{1998MNRAS.293..113T}. More recently, \citet{2014ApJ...796...37S} tested three models within the stable RLO channel to examine the formation of OBe~stars by binary interaction. They could best explain the observed  mass and orbital period distribution of OB+NS systems neither with the rotation limited accretion model nor with the thermally limited accretion model but only with the model with 50\% mass transfer efficiency, for which the authors found in contrast to the other two models no physical motivation. Lastly, \citet{2020MNRAS.498.4705V} studies a combination of three fixed mass transfer efficiencies and three models for the loss of angular momentum. Each combination resulted in a certain minimal mass ratio for successful donor stripping. With that, they confirmed the earlier findings about the minimum mass ratio, the need for a moderate mass transfer efficiency and notable angular momentum loss from the system. All these findings are in line with the results of our study and furthermore we are able to connect the minimal mass ratio for stable mass transfer with the mass-transfer efficiency by a physical mechanism. In general our results for distribution functions of parameters of BeXBs agree with previous work and differences can easily be explained by the underlying physics.

Several studies address the number of BeXBs. \citet{2020MNRAS.498.4705V} found that their mechanism under-predicts their number and need to invoke a time dependent star formation rate. They need to do this since many of their systems disrupt during the SN because the authors do not use reduced SN kicks for stripped stars as we do. \citet{2014ApJ...796...37S} and \citet{2019A&A...624A..66R} report for the same reason a larger ratio between unbound and bound NS than we do. That SN kicks regulate the number of BeXBs was also found by \citet{1996A&A...309..179P}, \citet{1998A&A...340...85R} and \citet{2019A&A...624A..66R}. \citet{1998A&A...340...85R} furthermore showed that the magnitude of the kick does not change the shape of the orbital period distribution and only slightly the shape of the eccentricity distribution.

In Sect.~\ref{results_var} we found evidence that binary evolution alone can not account for the large number of OBe~stars in the SMC. The literature shows no consensus on the question how important this channel is for their formation. While e.g. \citet{1991A&A...241..419P} find that no more than 60\% of them can come from the binary channel, \citet{2014ApJ...796...37S} can explain all OBe~stars by binary evolution. \citet{2020A&A...633A.165H} found that single star evolution is not able to explain the observed number of OBe~stars in open clusters, but binary evolution might be able to \citep{2021A&A...653A.144H}. At this point, neither single star evolution nor binary evolution alone can explain the observed amount of OBe~stars.

We identified further characteristics shared by our and previous studies: In general 20\% of mass gainers of all masses are predicted to have a HeS companion \citep{1995A&A...296..691P,1997A&A...322..116V}. Several further studies \citep[e.g.]{1993ARep...37..411T,1998MNRAS.293..113T,2014ApJ...796...37S} found that the common envelope channel is not relevant for the formation of BeXBs. Tides are the reason why OB+NS systems with small orbital periods do not evolve to BeXBs \citep{1998A&A...340...85R} and circularise close systems which we neglected \citep{1998MNRAS.293..113T,2014ApJ...796...37S}. \citet{2019A&A...624A..66R} identified that the minimum mass ratio for successful donor stripping and the mass transfer efficiency impact the velocity distribution of unbound stars. % linden2009 boubert2018

Several studies make predictions about the yet unobserved OBe+BH population. \citet{1999A&A...349..505R} predicted orbital period and eccentricities in agreement with our results, even though they require the BH progenitors have initial masses of at least $50\msol$. \citet{2014ApJ...796...37S} predicts OB(e)+BH binaries for their rotationally limited and their semi-conservative model but not for their thermally limited accretion model as in the latter the OBe stars become too massive. In the former two models the OBe+BH populations have different OB masses. They compare well to our model (e.g. Fig.~\ref{fig-3er-mass}) as our effective mass-transfer efficiency for BHs lies in between their two models. Their ratio of OBe+NS and OBe+BH systems is larger than ours as they do not consider O~stars as emission line stars. \citet{2019ApJ...885..151S} predict the Galactic population of BHs with normal star companions assuming a rotationally limited mass transfer efficiency. Their model~B, which is similar to our scenario with BH kicks, yields relatively similar results, while differences in the companions masses are smaller by a factor of 2 (probably due to the lower mass-transfer efficiency) and the period distribution appears to be flatter than ours. While they did not consider whether the accretor becomes a OBe~star, they predict a large number of OB+BH systems. Their follow-up study \citep{2020ApJ...898..143S} treats the possibility of these systems to become X-ray binaries. They predict a large number of OBe+BH systems in the Galaxy. Their orbital periods are as in our results roughly larger than 10\,d and the OB masses are lower for the mentioned reasons. Their ratio of normal OB+BH to OBe+BH systems is about 1:4 in contrast with our result of 4:1. \citet{2020A&A...638A..39L} predicted the LMC's OB+BH population with a MESA binary grid similar to Paper~I with rotation limited accretion and the same luminosity limit for stable mass transfer. Thus their OB~stars are lighter than in our study and the period distribution is bimodal due to the same reason as for Paper~I. Therefore, we refer to our discussion about the work of Paper~I in Sect.~\ref{sec-dis-xu}.

On the other hand, the lack of BeXBs with BH accretors is an open question in the literature, and several mechanisms are proposed, why such systems are not observed. \citet{2001A&A...367..848R} proposed that a luminous blue variable phase occurs to the BH progenitor before the RLO. By the strong winds of that phase the BH progenitor loses so much mass before collapse that a RLO is prevented and its companions does not become an emission line star. This idea breaks apart if the companion can evolve to a OBe~star by single star evolution, which however is difficult \citep{2020A&A...633A.165H}. \citet{2004ApJ...603..663Z} argue that these BHs would not accrete material from the disk \citep[see also][]{2024A&A...690A.256S}. Their argument is based on the results from \citet{2003MNRAS.341..385P} that BH binaries prefer short orbital periods, which is challenged by our and other results. \citet{2009ApJ...707..870B} find, in addition to a suppression by the IMF, evolutionary arguments, why BeXBs with BHs are suppressed. This study however relies heavily on a common envelope ejection, which was found to be unrealistic for the formation of BeXBs \citep{2014ApJ...796...37S}. \citet{2020A&A...638A..39L} suggest that the BH progenitor is a WR star whose wind has removed the OBe~star disk. Finally, we compare our WR+O population to those of \citet{2022A&A...667A..58P}, even though their study focuses on the LMC. Compared to their Fig.~5 to~7, we find a larger period range but our mass ratio distribution is narrower. We share the preference for systems with a period of 10\,d and O~stars as massive or up to 50\% heavier than the WR~star with them. %1998NewA....3..443V

In Sect.~\ref{results_var} we reported evidence for a mass-dependent mass-transfer efficiency with more massive accretors accreting a smaller fraction of the lost donor mass than less massive accretors. Throughout the literature we found evidence supporting this claim, both from theoretical and observational studies. \citet{2018A&A...618A.110B} found evidence for non-conservative evolution through infrared nebulae for O to A~typ stars. \citet{1993MNRAS.262..534S} reports a mass transfer efficiency for the late B~type star \textbeta~Per of 60\% and \citet{2018MNRAS.481.5660D} find that the late B~type star \textdelta~Lib must have gone though near-conservative mass transfer. For the early B~type star \textphi~Per, \citet{2001A&A...367..848R} states that the mass-transfer efficiency during its formation could not have been non-conservative. Similarly, \citet{2007ASPC..367..387P} and \citet{2018A&A...615A..30S} report an mass-transfer efficiency for that object of more than 70\% and more than 75\%, respectively. For LB-1, an object at a similar stellar mass, \citet{2021ApJ...908...67S} and \citet{2022A&A...667A.122S} report a moderately non-conservative evolution and for \textbeta~Lyr~A \citet{2021A&A...645A..51B} found close to conservative mass transfer. \citet{2021AJ....161..248W} states that the mass transfer efficiency in their Be+sdOB systems was close to 100\%. For G0 to B1 type Algols, \citet{2001ApJ...552..664N} find conservative mass transfer fitting and hints that OB~type Algols may bee non-conservative. \citet{1994A&A...283..144F}, \citet{2007A&A...467.1181D}, \citet{2006A&A...446.1071V} and \citet{2008A&A...487.1129V} confirmed that. For HMXBs, \citet{1995ApJ...440..280D} found an mass transfer efficiency of about 30\%, and for BeXBs, \citet{2020MNRAS.498.4705V} found an mass transfer efficiency of at least 30\% while \citet{2014ApJ...796...37S} found an mass-transfer efficiency of 50\% to agree with observations. On the other hand, \citet{2021A&A...653A.144H} derived from OBe~stars in coeval populations a highly non-conservative mass transfer. At higher masses, \citet{2021ApJ...923..277R} find an mass-transfer efficiency for the late O~type star \textzeta~Oph of about 30\%. For even more massive stars, namely O+WR systems, in agreement with each other, \citet{2005A&A...435.1013P}, \citet{2016ApJ...833..108S} and \citet{2024A&A...691A.174S} identified these systems to be highly non-conservative (mass-transfer efficiency of 10...25\%). While certainly a period dependency for the mass transfer efficiency is reported \citep{2007A&A...467.1181D,2021arXiv211103329S} which occults the picture, a trend for lower mass-transfer efficiency at higher mass seems to emerge. This is in agreement with our findings from Sect.~\ref{sec-pq}. We can only speculate about its origin. Maybe winds or luminosity play a role. %2010Natur.468...77V
As a last remark, our mass dependent mass-transfer efficiency implies a mass dependent minimal mass ratio for donor stripping. Such an effect was also found by \citet{2021A&A...653A.144H} from OBe~stars in star clusters. %pols1994

%(StarTrack, e.g. \citet{2008ApJS..174..223B}, binary\_c, e.g. \citet{2004MNRAS.350..407I}, COMPASS e.g. \citet{2017NatCo...814906S})

%An extensive amount of work has been done on the evolution of binary stars. Key questions were the properties of binary distributions, the outcome of the RLO and in which kinds of binaries to find black holes.

%Over the past 40 years an immense amount of studies has been published. The following discussion therefore cannot be complete.

%On the market exists a lot of models to describe RLO: 

%Estimates for eps

%estimates about pq plane

%BH binaries

%The predictions about the populations of post MT binaries are numerous and of couse depent on the imput physikcs. ... who did what ...

%Previous work on population synthesis codes and especially Be population synthesis was done e.g. by \citet{1994A&A...288..475P, 1999MNRAS.305..763B, 2002ApJ...572..407B, 2002MNRAS.329..897H, 2003MNRAS.342.1169V, 2004MNRAS.350..407I, 2006A&A...460..565I, 2008ApJS..174..223B, 2008MNRAS.384.1109E, 2009A&A...508.1359I, 2012ApJ...759...52D, 2016MNRAS.462.3302E, 2018MNRAS.481.4009V, 2018MNRAS.473.2984I}. [More Be pop syn paper], StarTrack, binary-c, BYPASS, Compass.

%These systems may appear as $\gamma$~Cas stars, as speculated by \citet{2020A&A...633A..40L}, or OB+WR binaries. \citet{2016ApJ...819...55G} reported 29 Oe stars in the SMC. This is in agreement with our numbers and we predict that $\sim$ 20 of them to have a BH companion. The OBe+BH systems should not be visible in X-rays, see Quast+ in prep. Their mass distribution is similar to (dieses Paper mit den BeXB massen)

\section{Conclusions}\label{concl}

We used the rapid binary population synthesis code \combine\ to investigate the properties of evolved massive binary systems in the SMC. Compared to an analogous effort using detailed binary evolution calculations (Paper~I), our approach allows us to vary mass-transfer efficiency and merger fraction during the first mass transfer. We base our merger criterion on the condition that contact is avoided \citep{2024A&A...691A.174S}, and we do not consider the energy requirement for removing the non-accreted matter from the binary systems as in Paper~I. This leads to the coupling of both properties, such that a higher mass-transfer efficiency leads to more mergers, due to the stronger radius increase of the accretors (Figs.~\ref{fig-pq8} to~\ref{fig-pq100}). This trend is in agreement with several earlier studies (Sect.~\ref{lit}).

With our self-adjusting merger condition, the mass-transfer efficiency remains as the key parameter in our approach. We computed synthetic SMC binary populations with various fixed mass-transfer efficiencies, and compared the result with the observed different populations of evolved massive binaries. We found that in the lower mass range considered here, which produces most of the observed OBe stars and Be/X-ray binaries, a rather high mass-transfer efficiency ($\sim50\%$) is required to reproduce these populations. For the higher mass systems, which evolve into  WR+O star and possible into WR+BH systems, a low mass-transfer efficiency ($\sim10\%$) gives a better agreement. Based on these results, we developed a heuristic relation between accretion efficiency and accretor mass (Eq.~\ref{epsofm}), which we used to build our fiducial synthetic SMC binary population (Fig.~\ref{CompFrac}). 

This method yields a good agreement with the observed number of WR+O systems, and with the number of observed BeXBs, without assuming an enhanced recent star-formation rate \citep[e.g.][]{2020MNRAS.498.4705V}, as well as their magnitude and orbital period distribution. To reproduce the number of observed OBe~stars appears possible, but is affected by uncertainties in the wind-induced spin-down (Sect.~\ref{umod})

We find that the SMC may harbour about 140 OB+BH systems, that is roughly the same number as OB+NS systems. The OB~star in about 40 of these systems may appear as OBe star. We predict two sub-populations of OB+BH systems, namely close systems with small OB~star rotation velocities but medium to large orbital velocities, which may appear as single lined spectroscopic binaries, and wide systems with small orbital velocities but large rotational velocities, which may appear as OBe~stars. %We argue that the eccentricities of OB+BH systems reveal the occurrence of a BH formation kick at BH, and conclude from current observational data that a weak or no kick is more likely.

Our fiducial population contains less OB+WR and OB+BH binaries with orbital periods above $20$\,d than predicted by Paper~I. However, the long-period tail of the period distributions still contains about 1/3\textsuperscript{rd} of the expected population. While their absence in recent surveys \citep{2016A&A...591A..22S,2018A&A...616A.103S,2018ApJ...863..181N,2024A&A...689A.157S} remains unexplained by our models, the properties of the shorter-period systems agree well with the observations. 

%but struggles to describe the rotational velocities of O~stars with WR companions. 

The prediction of a large number of undiscovered OB+BH systems is shared by Paper~I. Even given the uncertainties in the physics of NS and BH formation, detecting these OB+BH systems, or excluding their existence, either as spectroscopic binaries or OBe~systems, e.g. by multi-epoch spectroscopy, will greatly help to reduce the uncertainties in massive binary evolution. Future work based on this will be able to more accurately predict the contribution of binary evolution to the properties of pulsars in binary systems, WR+SR systems and gravitational wave mergers.

%ecsn? -> He massenfenster, tauris, pods.
%belzinski BH kick, bexb-chinesen über ob+bh -> bh kick
%Pottschmidt, Wilms

%\begin{acknowledgements}
% ???
%\end{acknowledgements}

\bibliographystyle{aa}
\bibliography{bib}

\appendix
\section{Binary evolution for mass ratios close to unity}\label{app-q1}

Systems with mass ratios close to unity show a large variety of fates depending on the precise value of the mass ratio as visible in Fig.~\ref{fig:Pq} and \ref{fig-pq8} to \ref{fig-pq100}. Those closest to $q=1$ merge due to mass transfer on a post main-sequence star as for Case~A mentioned in the previous paragraph (orange). Here, the accretor ends hydrogen burning while accreting. For slightly lower mass ratios the system merges due to mass transfer onto a HeS (pink). In this case, the donor was able to loose its complete envelope but the former accretor initiates a reverse RLO shortly after that and before the HeS can end its life. This fate is only present for orbital periods below about $10^{1.7}\days \sim 50\days$ as for wider orbits the initially less massive star cannot fill its Roche lobe any more. This initially more massive star rather becomes a SR while its companion is beyond hydrogen burning ($\times$-like orange hatching). In some models (light grey) both SN happen at the same time (due to the finite temporal resolution of the code). For even lower mass ratios we find mergers due to mass transfer on hydrogen-rich core hydrogen exhausted models again as now the accretor has ended hydrogen burning when a Case~BB/BC mass transfer happens. This fate happens only for periods below about $10^{1.3}\days \sim 20\days$ as above the HeS cannot fill its Roche lobe so the companion of the SR will be a hydrogen exhausted model again. This behaviour is most prominent for low mass transfer efficiencies as rejuvenation is weak.

The region with mergers due to accretion on a HeS is very interesting as there may be the possibility of stable mass transfer. If the accretion efficiency is high, the initially less massive star transfers its envelope onto the initially more massive one, swapping the two stars' roles of OB~star and HeS. However \citet{2012sse..book.....K} predicts that HeS with a hydrogen layer with mass fraction 0.1 ... 0.5 should expand to become red giants. So, this channel would probably lead to a merger. However, if the accretion efficiency is low, the initially less massive model looses its envelope becoming a HeS while the initially more massive one does not accrete and expand remaining being a HeS. So a wide system of two HeS can be formed. Even if the mass transfer is unstable, we found that it is possible that a common envelope is ejected by the HeS yielding a system consisting of two close HeSs. The analysis of these evolutionary paths is unfortunately beyond the scope of this work, but may be the origin of the double WR star SMC AB~5. If we assume a current mass of $50\msol$ for both components \citep[][their acual mass ratio is 0.935]{2018A&A...611A..75S} and initial masses of $80\msol$ \citep[using the models from][]{2019A&A...625A.132S}, then we have an initial mass ratio of almost 1, after the first mass transfer $q\approx1.6$ and after the second one $q\approx1$ (completely non-conservative mass transfer). Following \citet{1997A&A...327..620S}, the first mass transfer increases the period by a factor of 2.014 and the second one by 1.144. Thus with the current orbital period of $19.6\days$, the initial period would have been $8.51\days = 10^{0.93}\days$, which is a value at the lower end of this behaviour in our models (Fig.~\ref{fig-pq80}).

\section{More predicted numbers for different initial binary distributions}\label{app:ini}

\begin{figure*}
    \centering
    \includegraphics[width=\columnwidth]{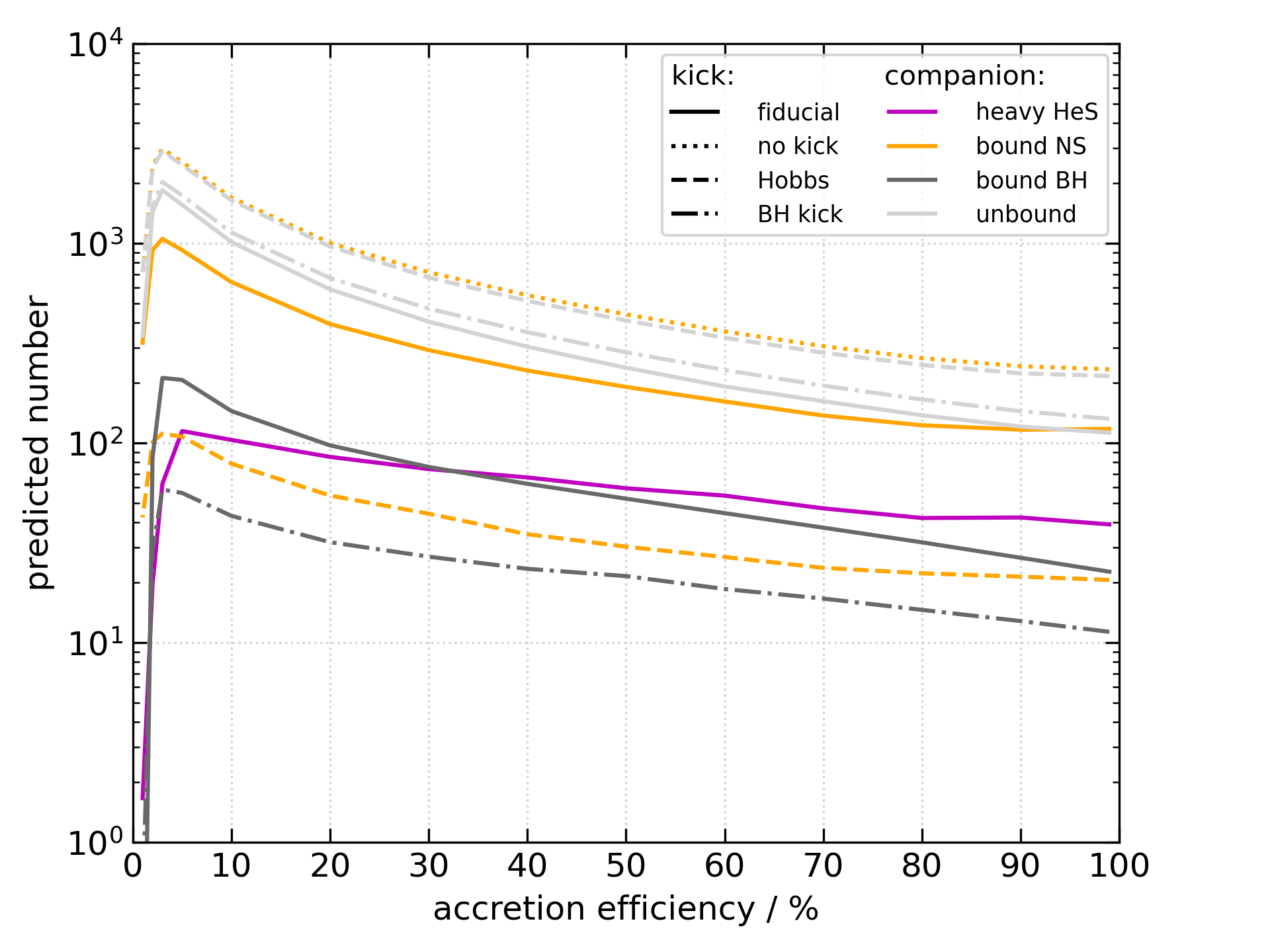}
    \includegraphics[width=\columnwidth]{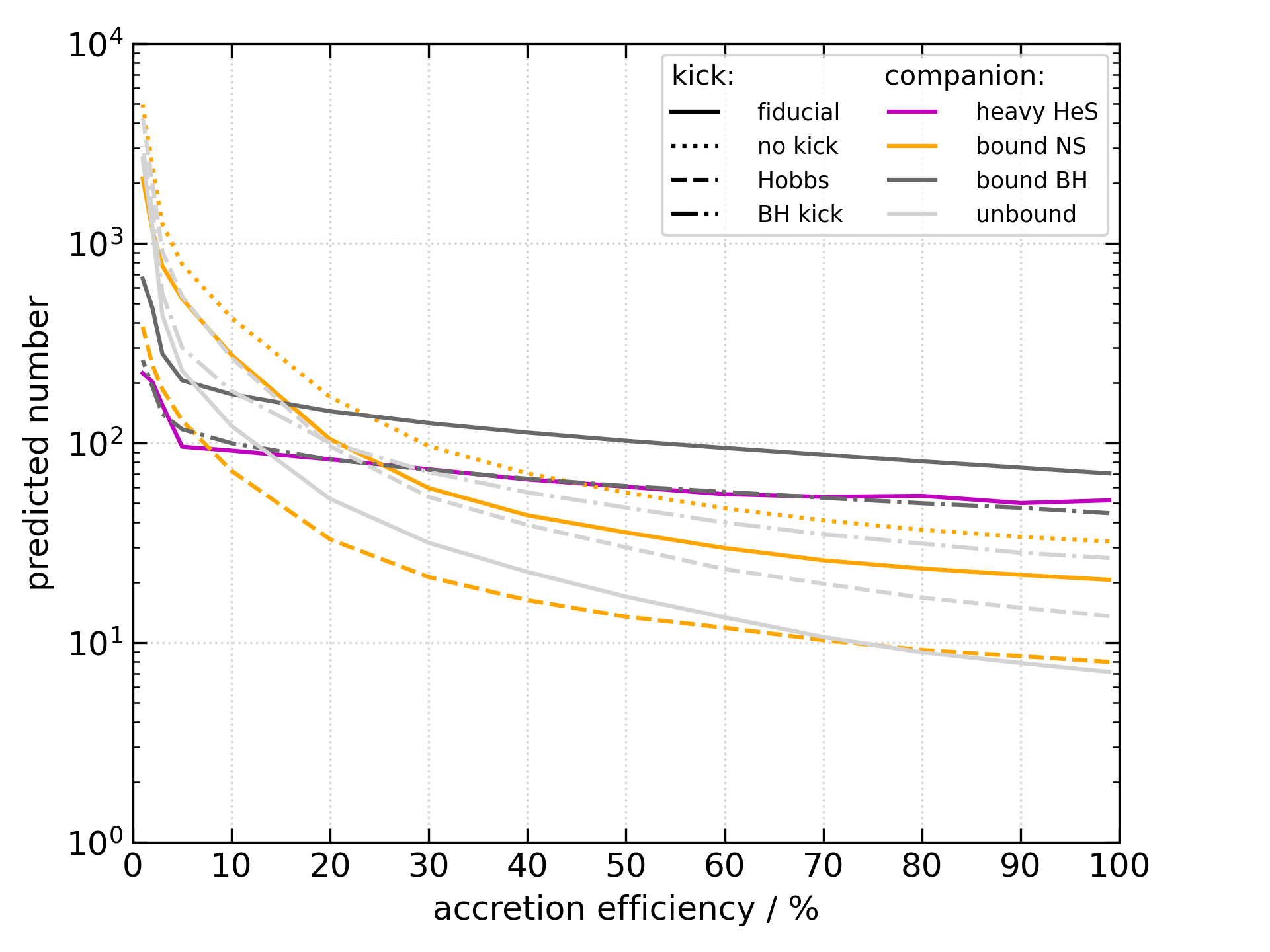}
    \includegraphics[width=\columnwidth]{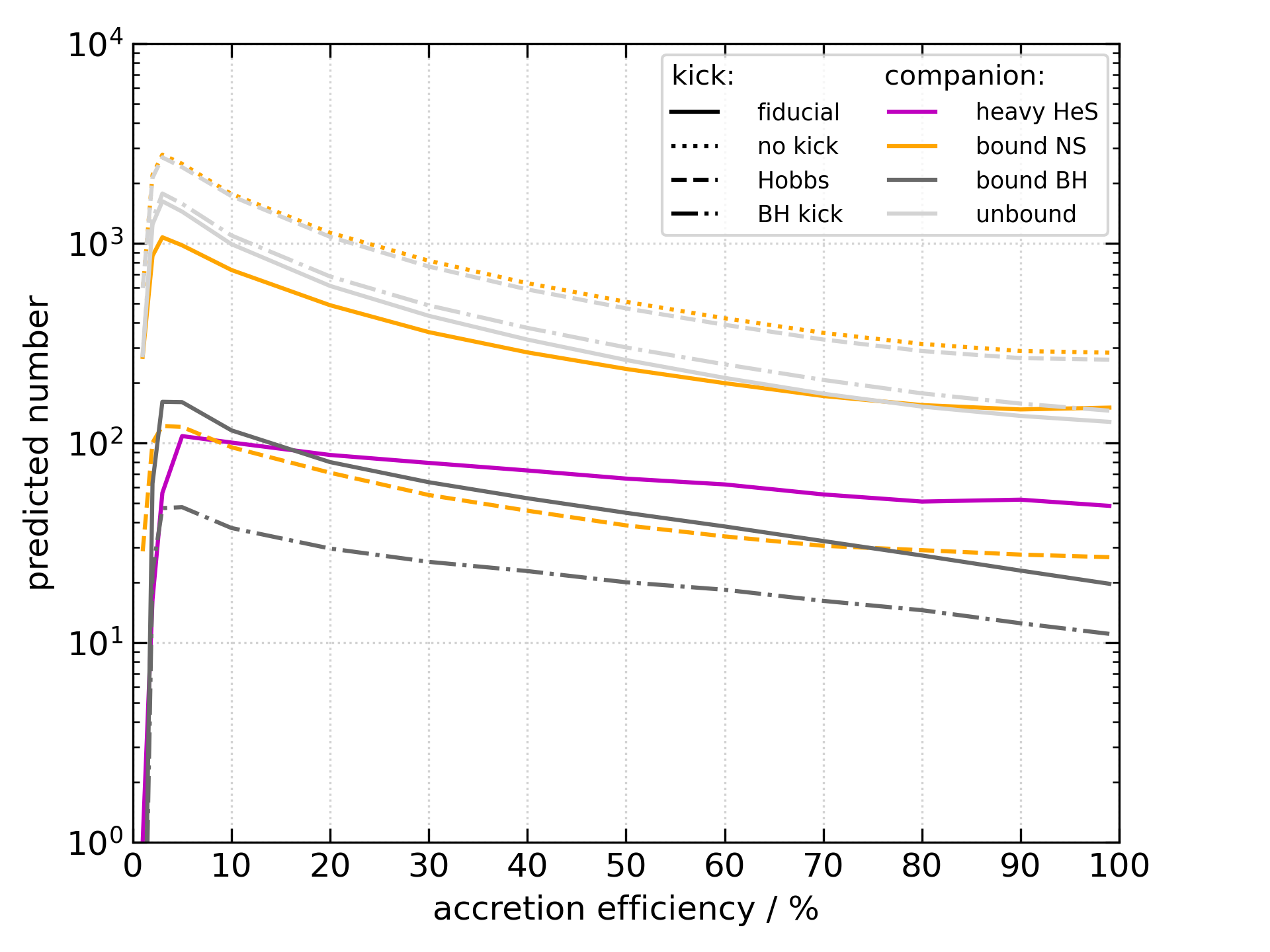}
    \includegraphics[width=\columnwidth]{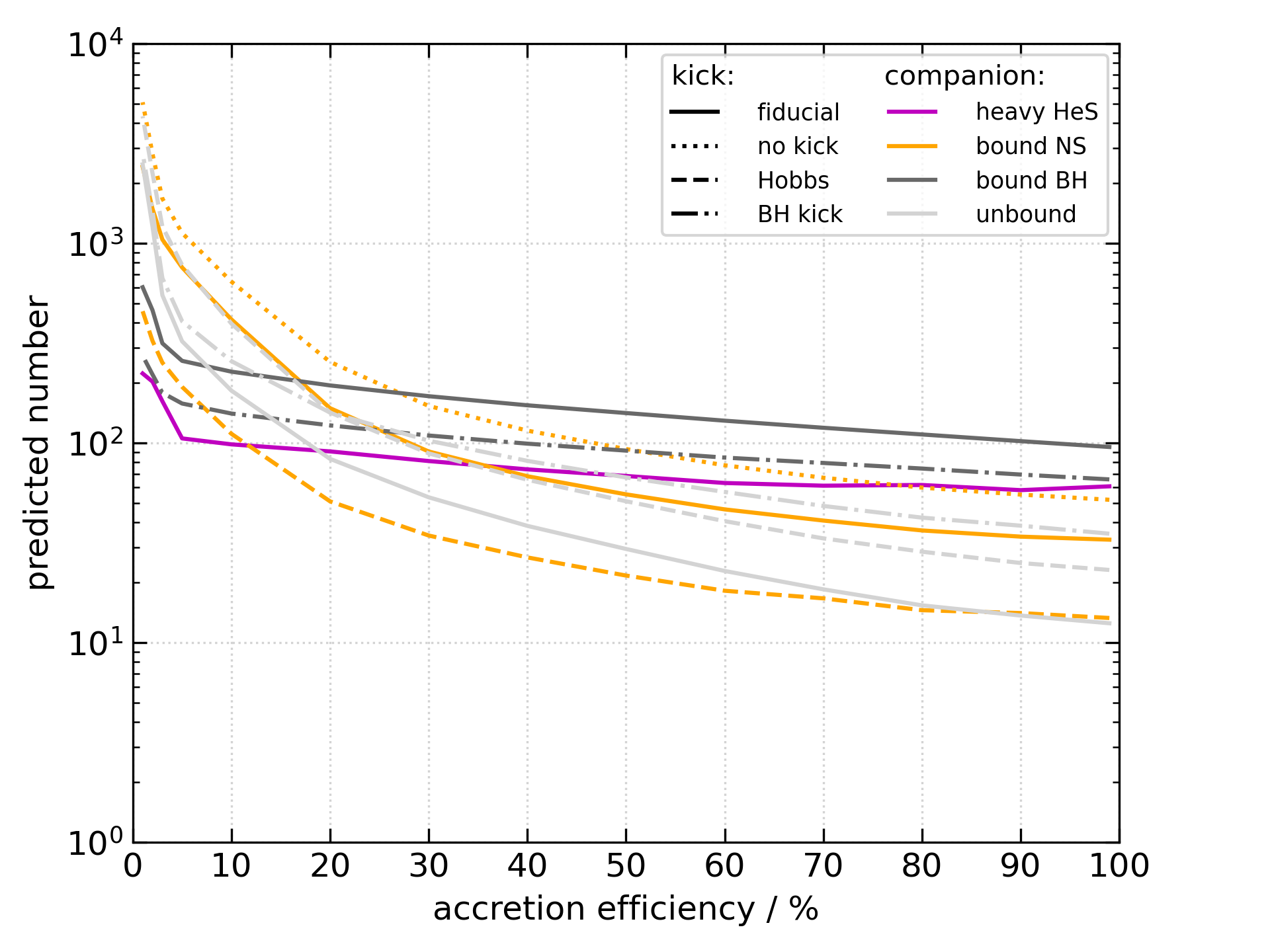}
    \includegraphics[width=\columnwidth]{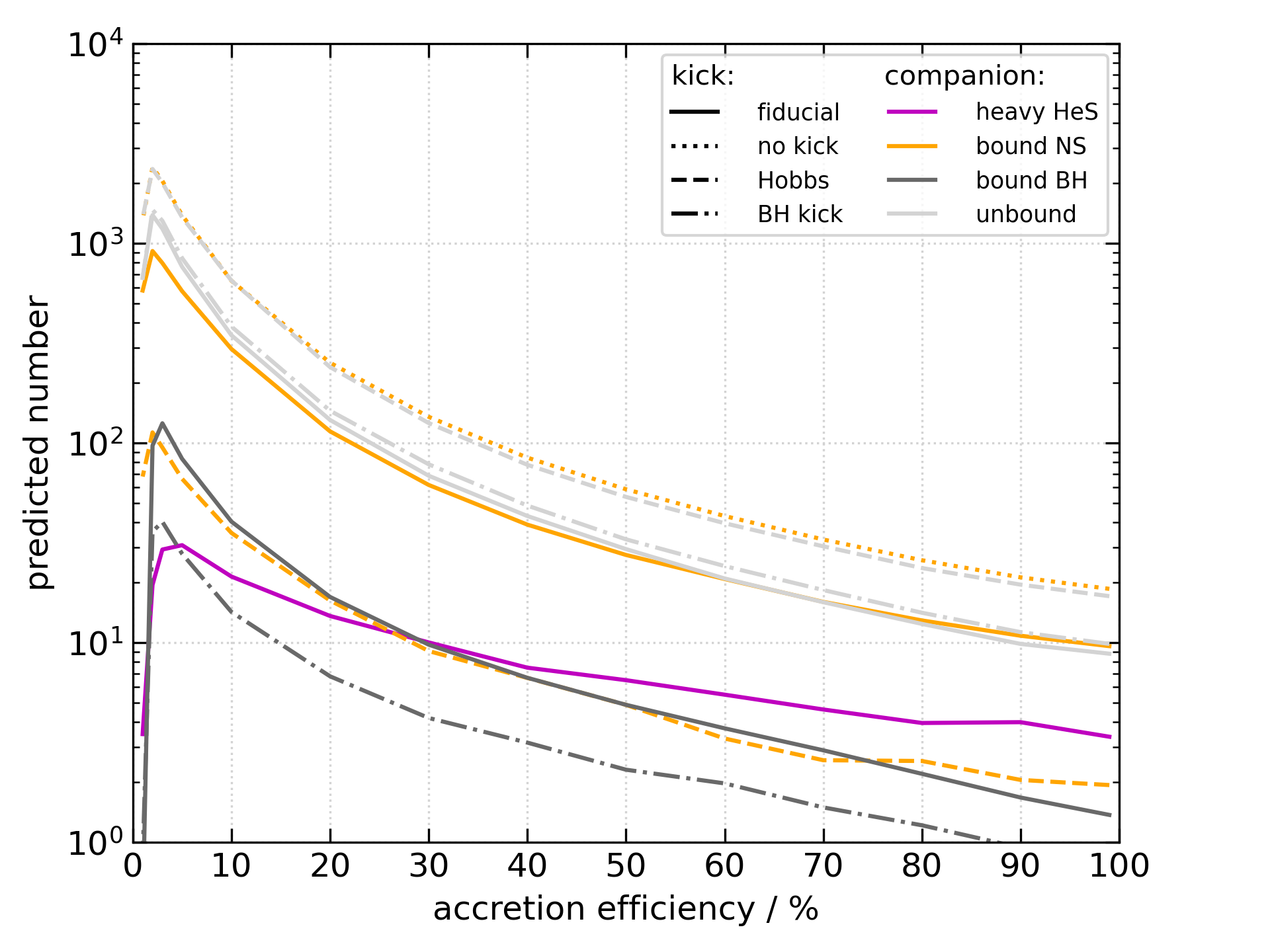}
    \includegraphics[width=\columnwidth]{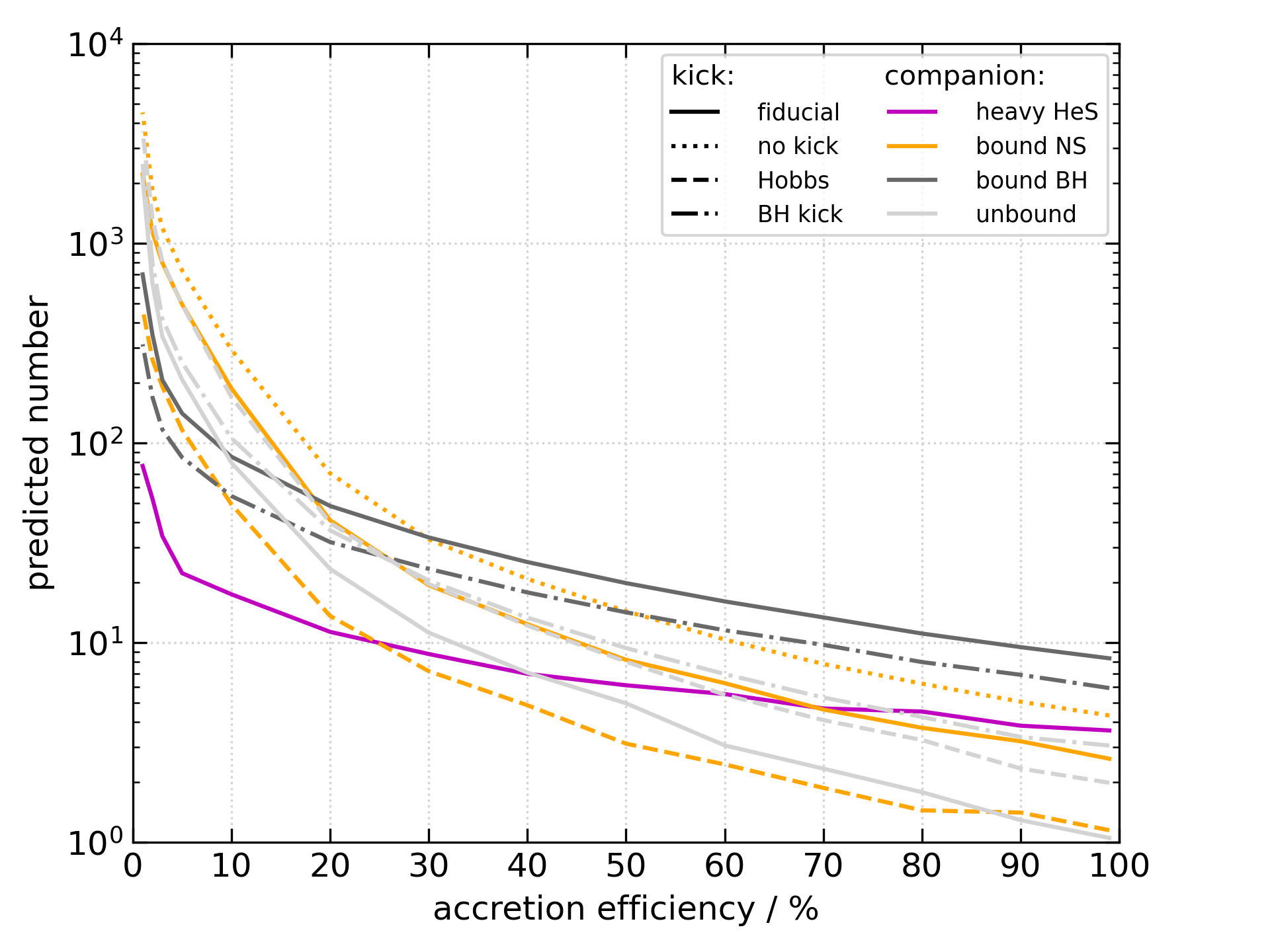}
    \caption{Predicted number of OBe and normal OB~star companions (colour) and different kick scenarios (line style) as a function of mass-transfer efficiency for initially flat binary distributions (top), \citet[][middle]{2012Sci...337..444S}, and \citet[][bottom]{2015A&A...580A..93D}.}
    \label{fig:comp-other}
\end{figure*}

Fig.~\ref{fig:comp-other} shows the number of OB(e) stars for two further initial distributions similar to Fig.~\ref{fig:comp}.

\section{More details of the fiducial model}

\subsection{Mass ratios}\label{sec-q}

\begin{figure*}
 \centering
 \includegraphics[width=\textwidth]{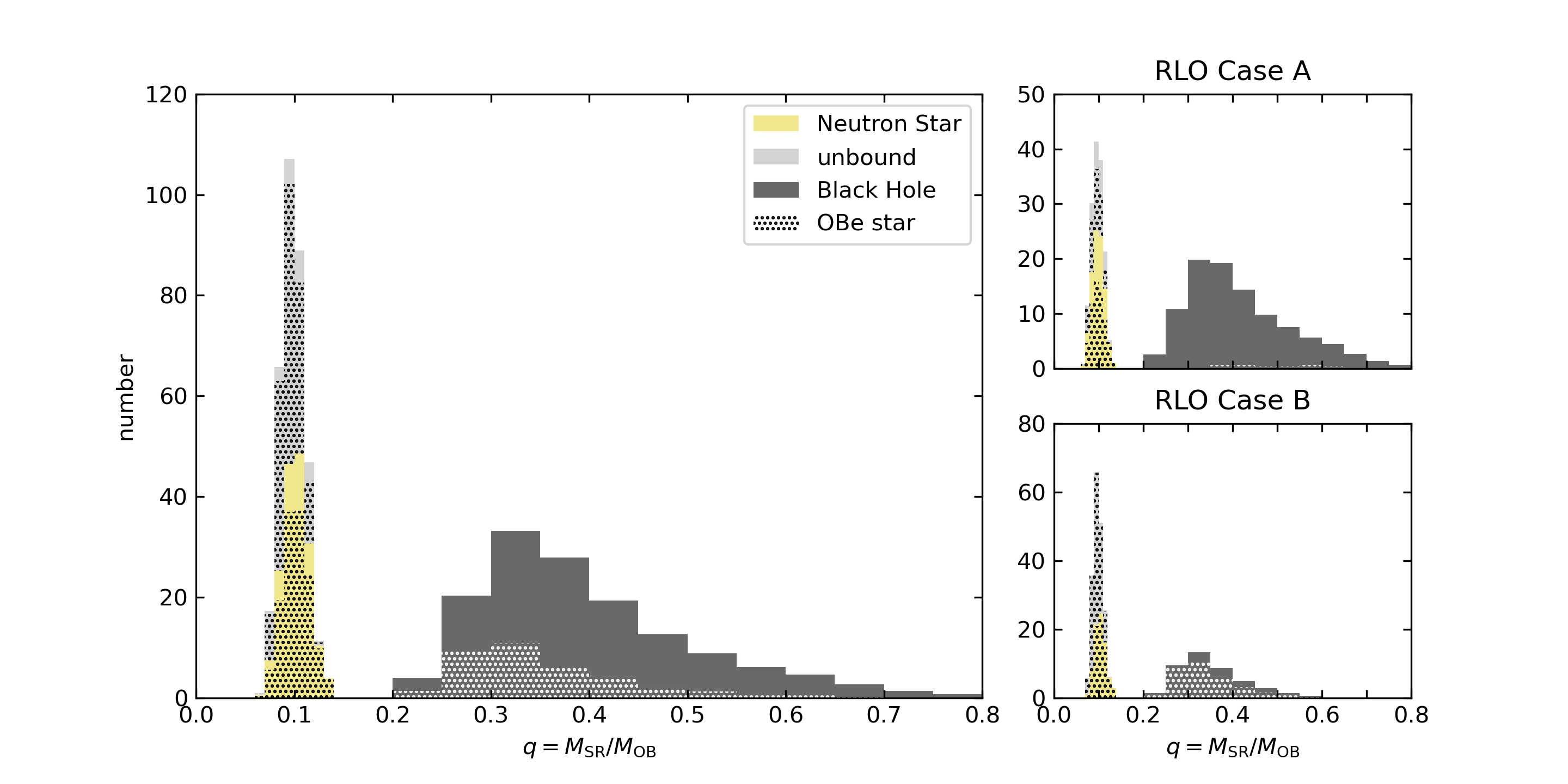}
 \caption{Predicted mass ratios ($M_\mathrm{SR}/M_\mathrm{OB}$) of massive binary systems after RLO and SN, including systems which became unbound due to the SN. (Step size for NSs is 0.01 and for BHs it is 0.05.) Left: Total numbers. Right: Case~A RLO (top) and B (bottom) separately.}
 \label{massratio}
\end{figure*}

In Fig.~\ref{massratio} we show the mass ratio $q = M_\mathrm{SR}/M_\mathrm{OB}$ of our OB+SR systems. We find values from 0.05 to 0.15 for NS companions and 0.2 to 0.8 for BH companions. As expected, NSs dominate the lower mass ratio regime and BHs the higher mass ratios. BHs whose progenitors evolved through Case~A mass transfer reach masses almost as great as their companions star. The distributions for NS can be understood with the values of Fig.~\ref{mass} and a typical NS mass of $1.3\msol$. For OB+BH systems this is more complex, as the BH mass is a function of initial stellar mass and the mass of the OB star depends on the accretion efficiency, which is a function of the accretor mass (Eq.~\eqref{epsofm}). Most features can be derived from Fig.~\ref{fig-3er-mass}.

\subsection{Rotation}

\begin{figure*}
 \centering
 \includegraphics[width=\textwidth]{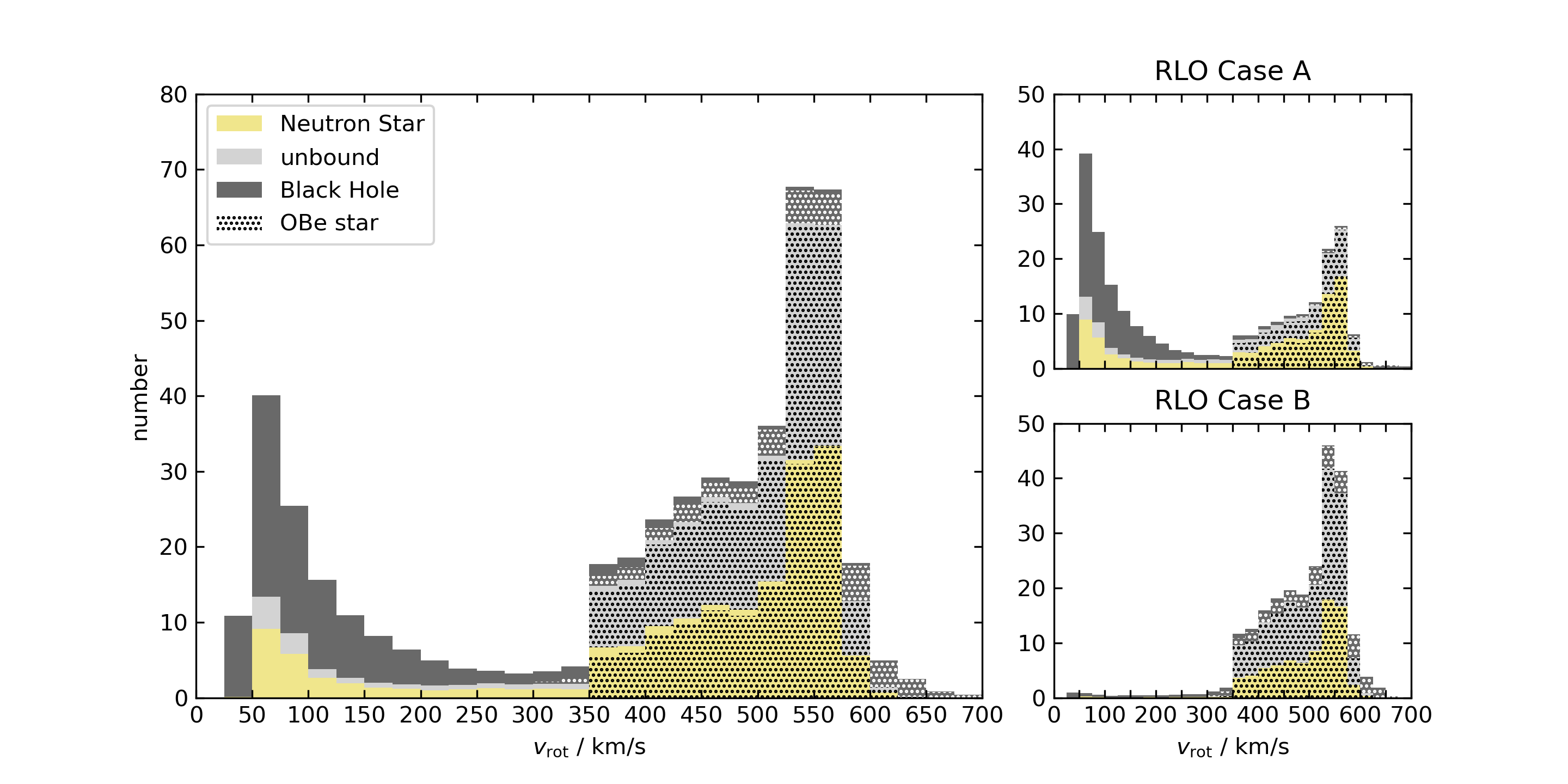}
 \includegraphics[width=\textwidth]{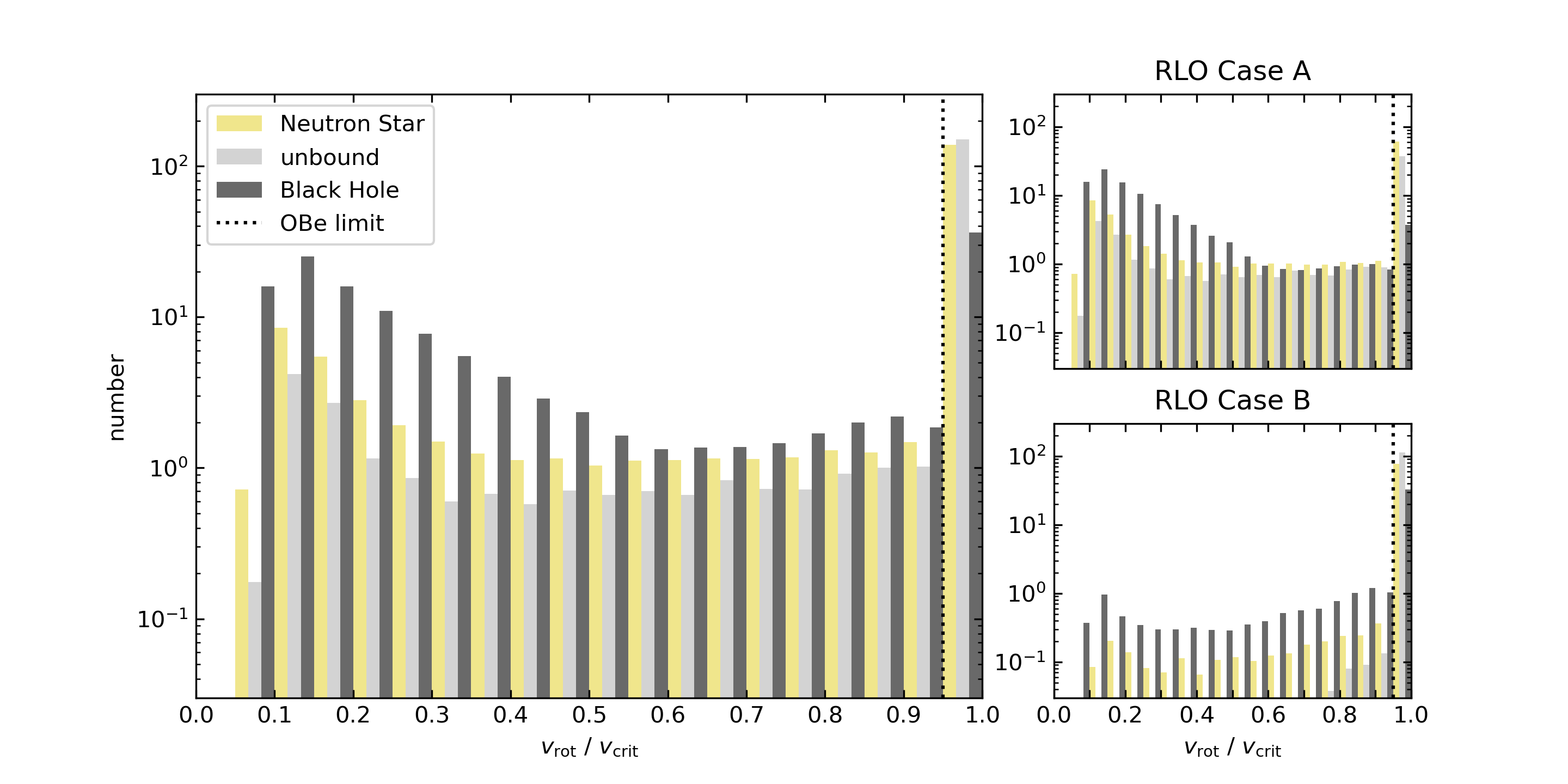}
 \caption{Same as Fig.~\ref{mass} but for the equatorial rotational velocity of the OB~star (top row) and the ratio of rotational and critical velocity of the OB~star (bottom row). Our criterion for an emission line star is indicated by the dotted line.}
 \label{OBrot}
\end{figure*}

Fig.~\ref{OBrot} shows two different measures for the rotation of the OB~star. The upper panel shows the equatorial rotational velocity $\vrot$ in km/s and the lower panel that relative to the critical velocity $\vcr$. Form the former we find velocities from 25 up to almost 700\,km/s in a bimodal distribution. The first peak is near 50\,km/s and the second one is around 550\,km/s, which is much broader for BH companions than for NS and unbound systems. Similarly the low $\vrot/\vcr$-peak is stronger for BHs than the high velocity peak while for the other two species the high velocity peak is clearly the dominant one. At 350\,km/s the velocity distribution has a discontinuity, especially for NS and unbound systems, coinciding with the occurrence of the formation of emission line stars. This becomes much clearer in the bottom panel. All potential OBe star lie on the right side of $\vrot/\vcr=0.95$ and form an extremely strong peak. Due to the variance of the critical velocity it was broadened in the upper diagram. The low velocity peak remains broad and also a clear gap between the two modes is present (0.6 to 0.95).

The panels on the right of Fig.~\ref{OBrot} indicate Case~A and~B. Systems which evolved through Case~B only belong to the high velocity peak and are OBe stars. Only a negligible fraction of them has $\vrot/\vcr<0.95$. Thus we can understand this peak as those stars which where spun up by accretion unaffected by tides. The Case~A systems are more complex. While one group exisists at high velocities, which corresponds to wide Case~A evaluations similar to Case~B, the low rotational velocity peak is only produces through Case~A RLO. It is due to synchronising tides that the peak is so broad as the range of orbital periods after RLO leads to a range of synchronous rotational velocities. BH systems are more strongly affected by this as BHs are more massive and thus induces stronger tides than NSs.

\subsection{Luminosities and temperatures}\label{apx-hrd}

\begin{figure}[h!]
 \centering
 \includegraphics[width=\columnwidth]{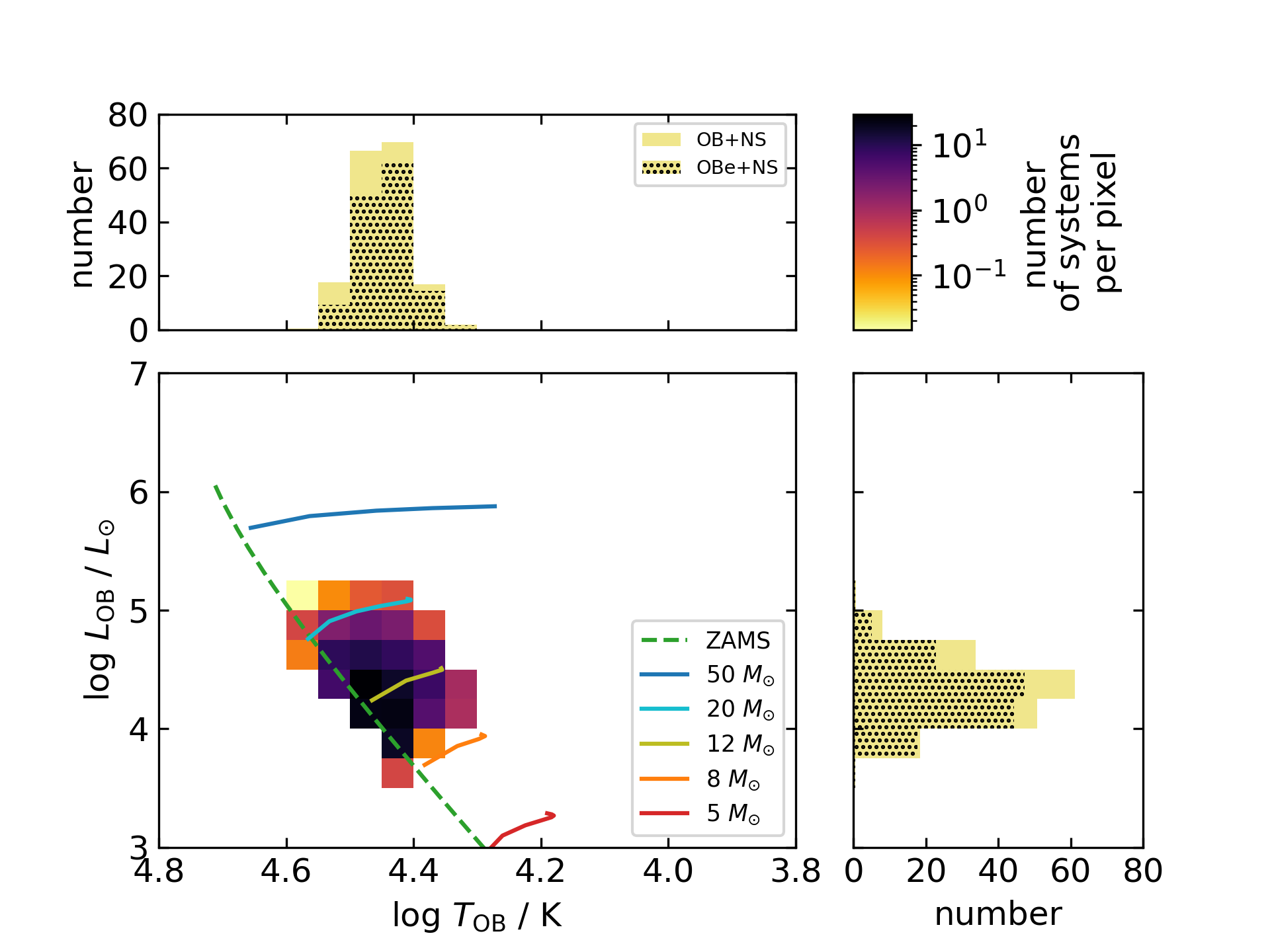}
 \includegraphics[width=\columnwidth]{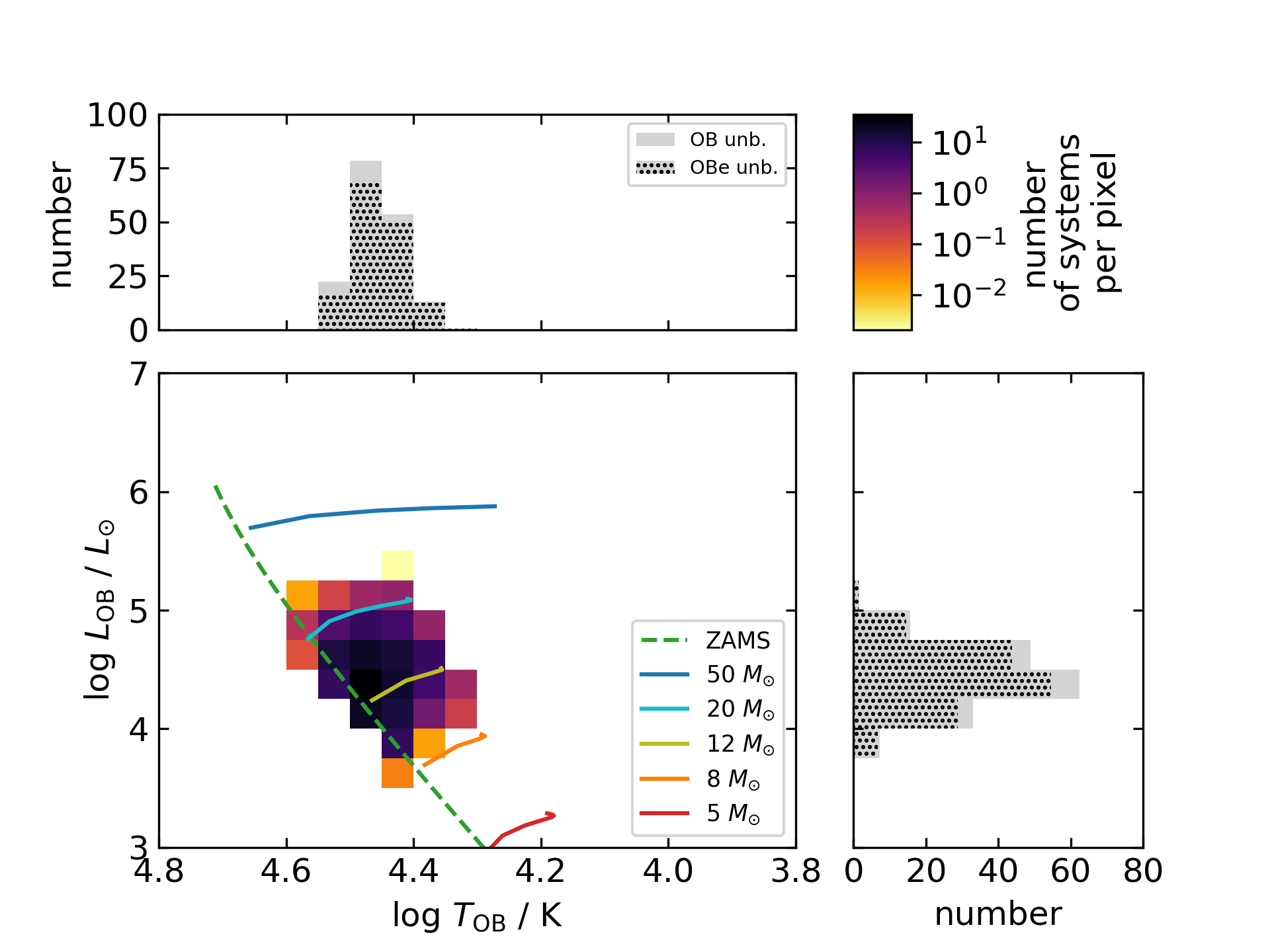}
 \caption{Predicted HRD positions of OB stars with a NS companions (top) and unbound companions (bottom) in logarithmic colouring together with selected model tracks and the zero-age main-sequence (ZAMS). The panels on top and on the right show the temperature and luminosity distributions with predicted emission line stars highlighted by dotting.}
 \label{HRDs-2}
\end{figure}

The HRDs for the companions of NSs (Fig.~\ref{HRDs-2} top) and those stars which became unbound during the SN (bottom) are very similar as expected. Their luminosities are between $10^4\lsol$ and $10^5\lsol$ and their temperatures range from 20\,kK to 40\,kK which may appear as late O~and early B~type stars. Comparing the population the the model tracks, one can see that they lie between the $8\msol$- and the $20\msol$-track in agreement with the results of Sect.~\ref{sec-mass}.

\subsection{Orbital periods and velocities}\label{apx-orbit}

\begin{figure*}
 \centering
 \includegraphics[width=\textwidth]{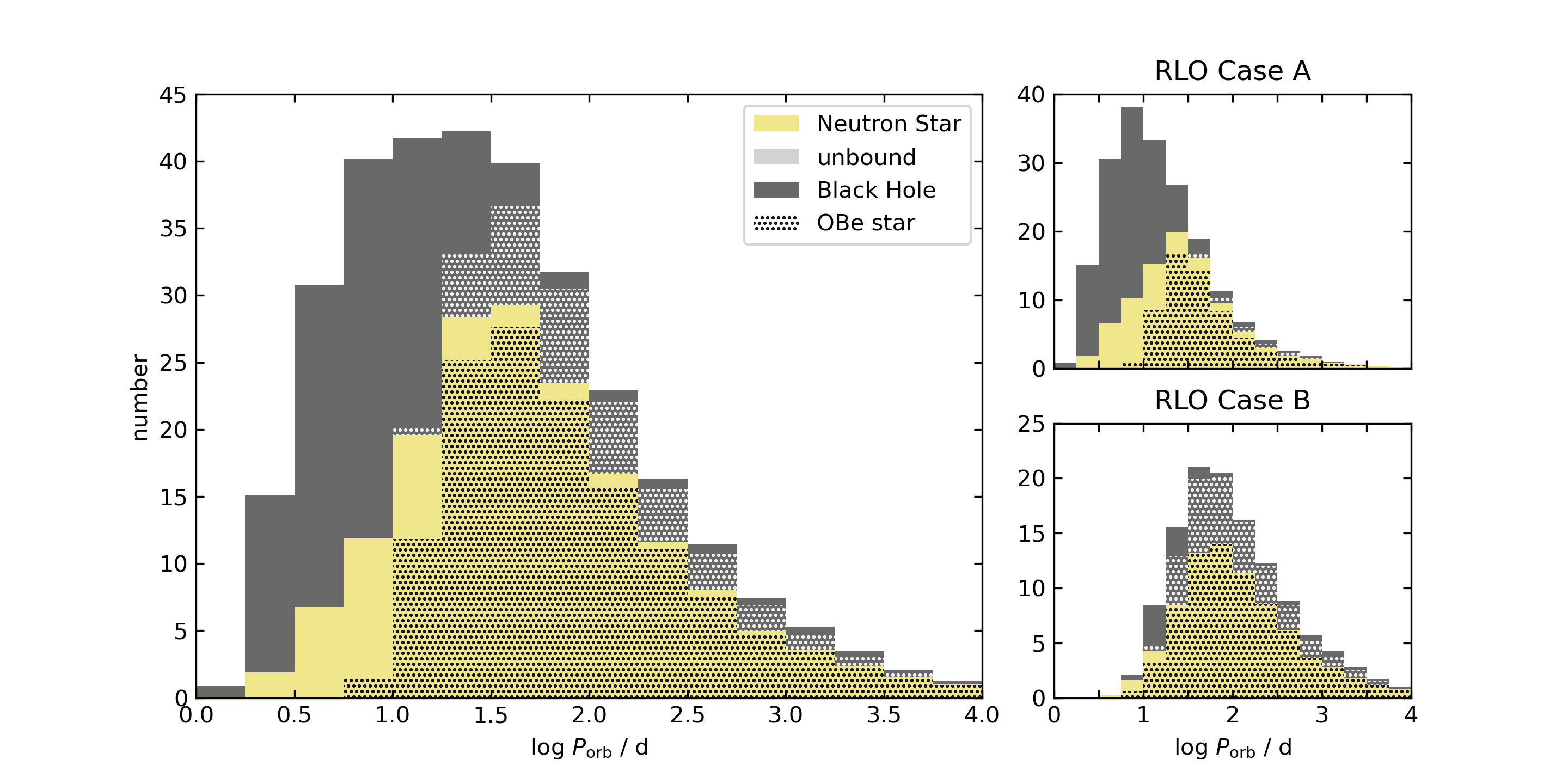}
 \includegraphics[width=\textwidth]{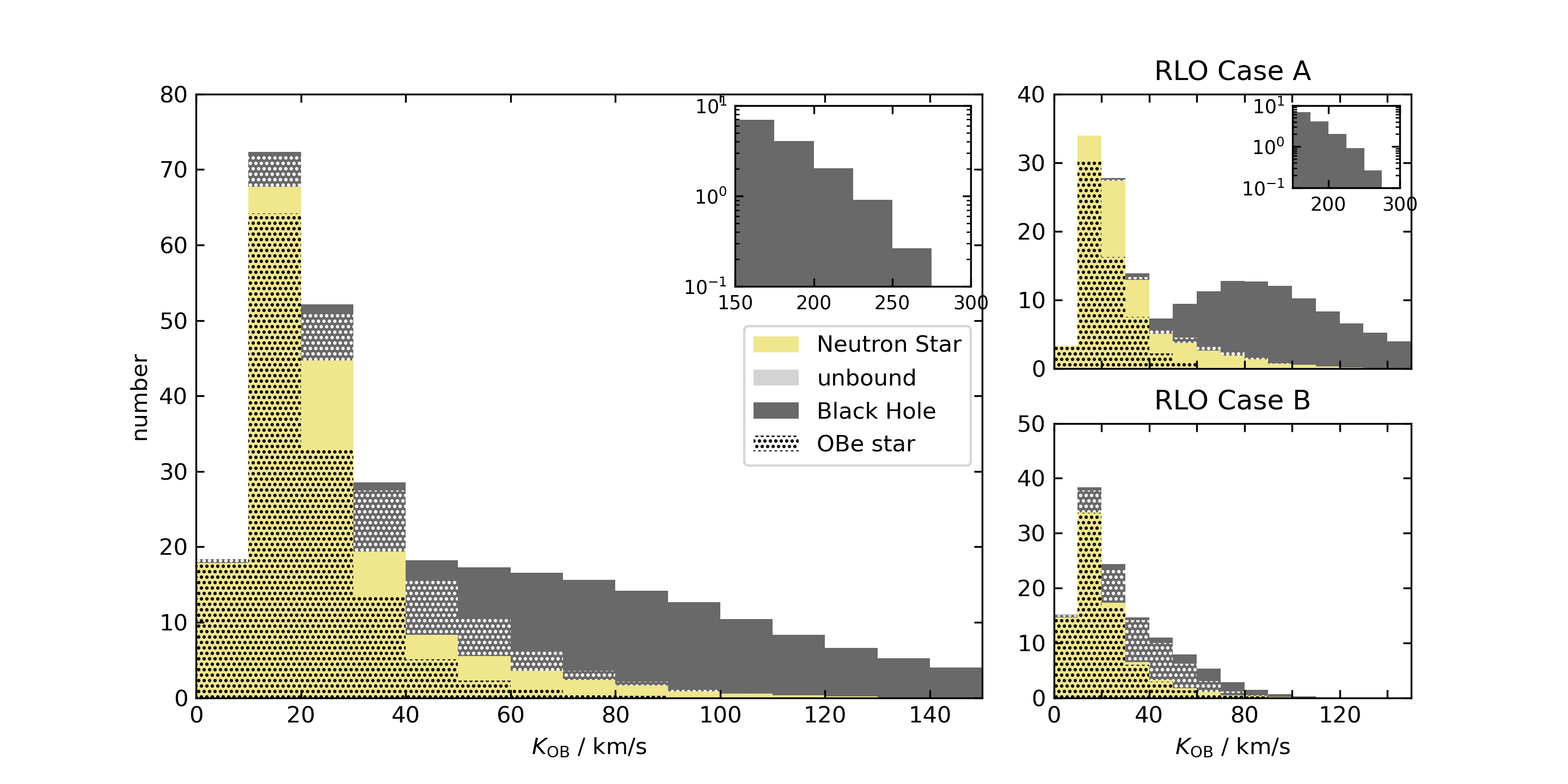}
 \caption{Same as Fig.~\ref{mass} but for the orbital period (top row) and orbital velocity semi-amplitude of the OB~star (bottom row)}
 \label{orbit}
\end{figure*}

Fig.~\ref{orbit} (top) shows the orbital period. For both BH and NS systems we find orbital periods ranging from 1\,d up to 1000\,d. Most NS systems have orbital periods of about 30\,d while BH systems orbit slightly faster with the mode at 10\,d. Both populations' numbers decrease towards higher periods because of the initial period distribution which prefers close systems and, in case for the NS systems where the accretion efficiency was higher, the reduced stability of mass transfer at greater periods. Main-sequence mass gainers are rapidly spun up during mass transfer. Whether they can appear as OBe stars depends on tidal interaction, which is significant in close binaries. Hence in our synthetic population, normal OB stars dominate the low-period systems. The transition from normal OB to OBe stars occurs around 10\,d for NS and 30\,d for BH due to difference in masses. As expected, Case~A systems tend to show lower and Case~B system tend to show larger orbital periods.

The orbital velocity semi-amplitude of the OB star $K_\mathrm{OB}$ is given by Eq.~\eqref{eq-K}. We show a histogram in Fig.~\ref{orbit} (bottom). Note that we show the maximally possible semi-amplitude and not the projection onto the sky-plane, i.e. an inclination of $90^\circ$ and with an argument of periastron of $90^\circ$. OB stars with NS companions reach velocities up to $100\kms$, but peak at $20\kms$. The companions of BHs show a much broader distribution between $10\kms$ and $250\kms$. OB stars in BH systems are typically faster than in NS systems, because of the high BH mass compared to NSs, and are so diverse in velocity semi-amplitude because of the broad mass distribution of BHs. Again, OB stars with a large semi-amplitude tend to be normal OB stars as they are orbits is close and tides are braking the stellar rotation. While the distributions for Case~A and~B look similar for the NS, they clearly differ for the BHs, as the high velocity contribution comes from the Case~A systems.
%While BHs should be easily observable, NS systems can have high eccentricities leading to high velocities for only a short time. 

\begin{figure*}
 \includegraphics[width=\textwidth]{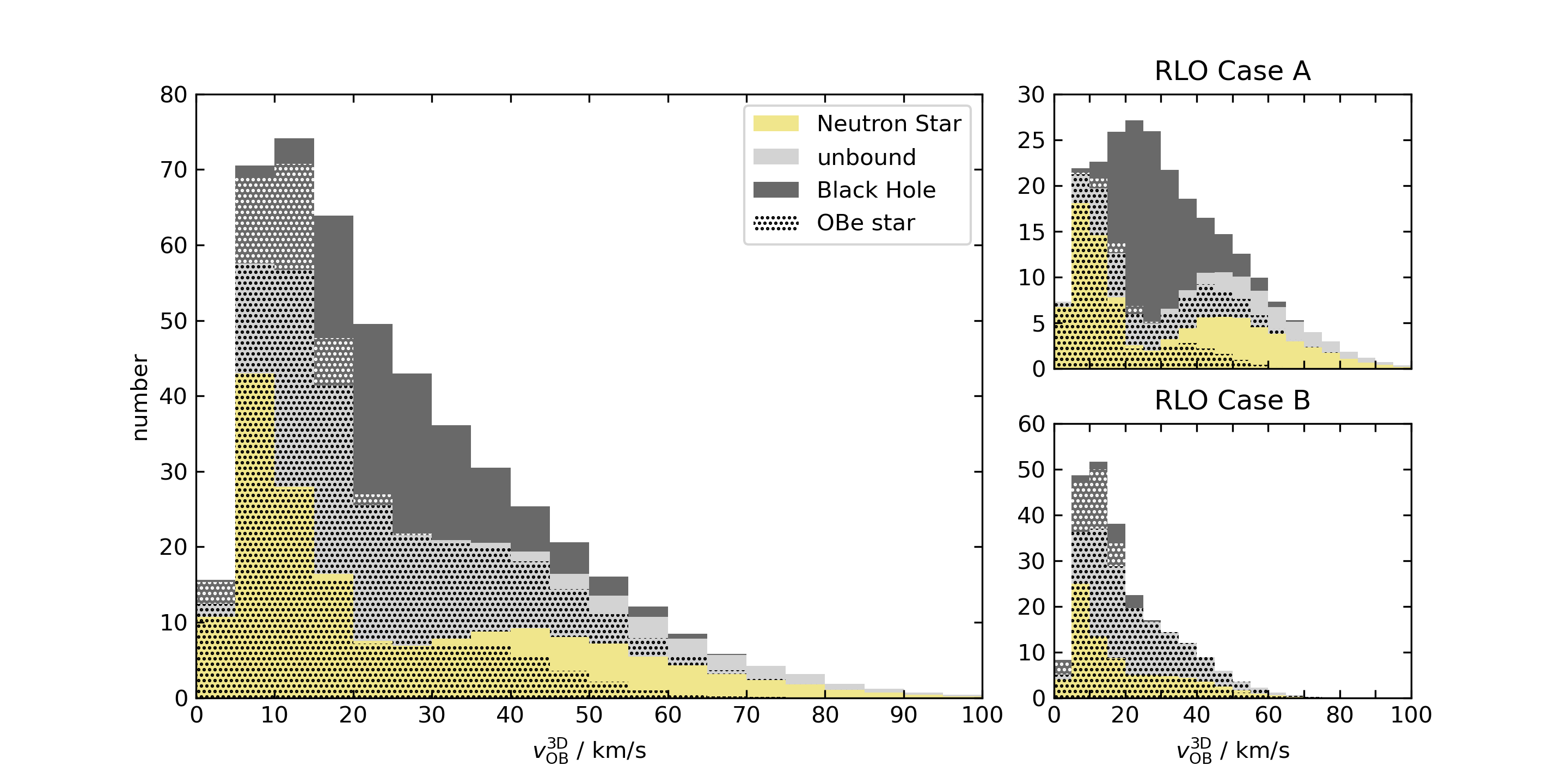}
 \caption{Same as Fig.~\ref{mass} but for the space velocity of the OB~star.}
 \label{v3d}
\end{figure*}

\subsection{Wolf-Rayet stars} \label{sec-wr}

\begin{figure*}
 \centering
 \includegraphics[width=\columnwidth]{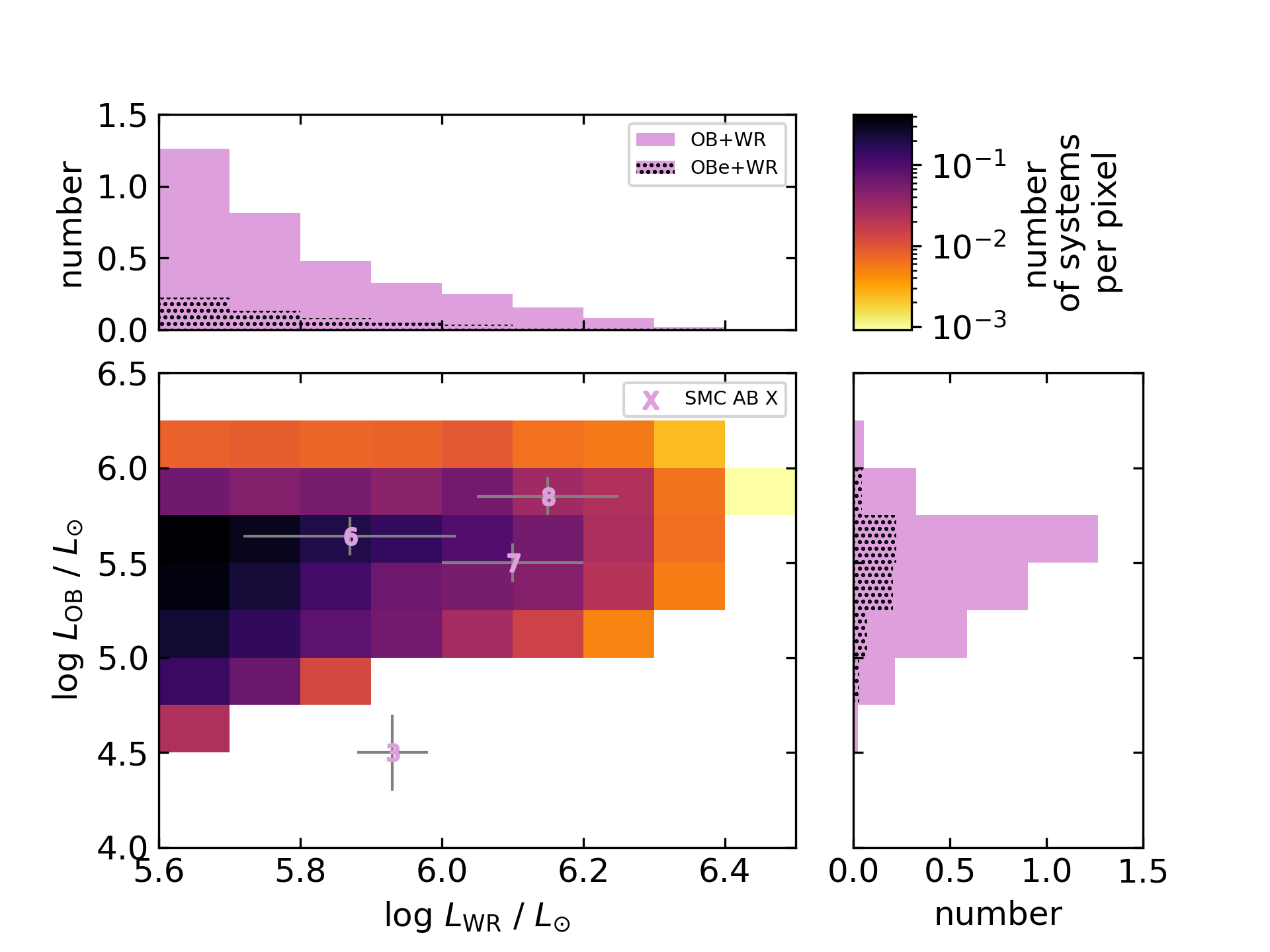}
 \includegraphics[width=\columnwidth]{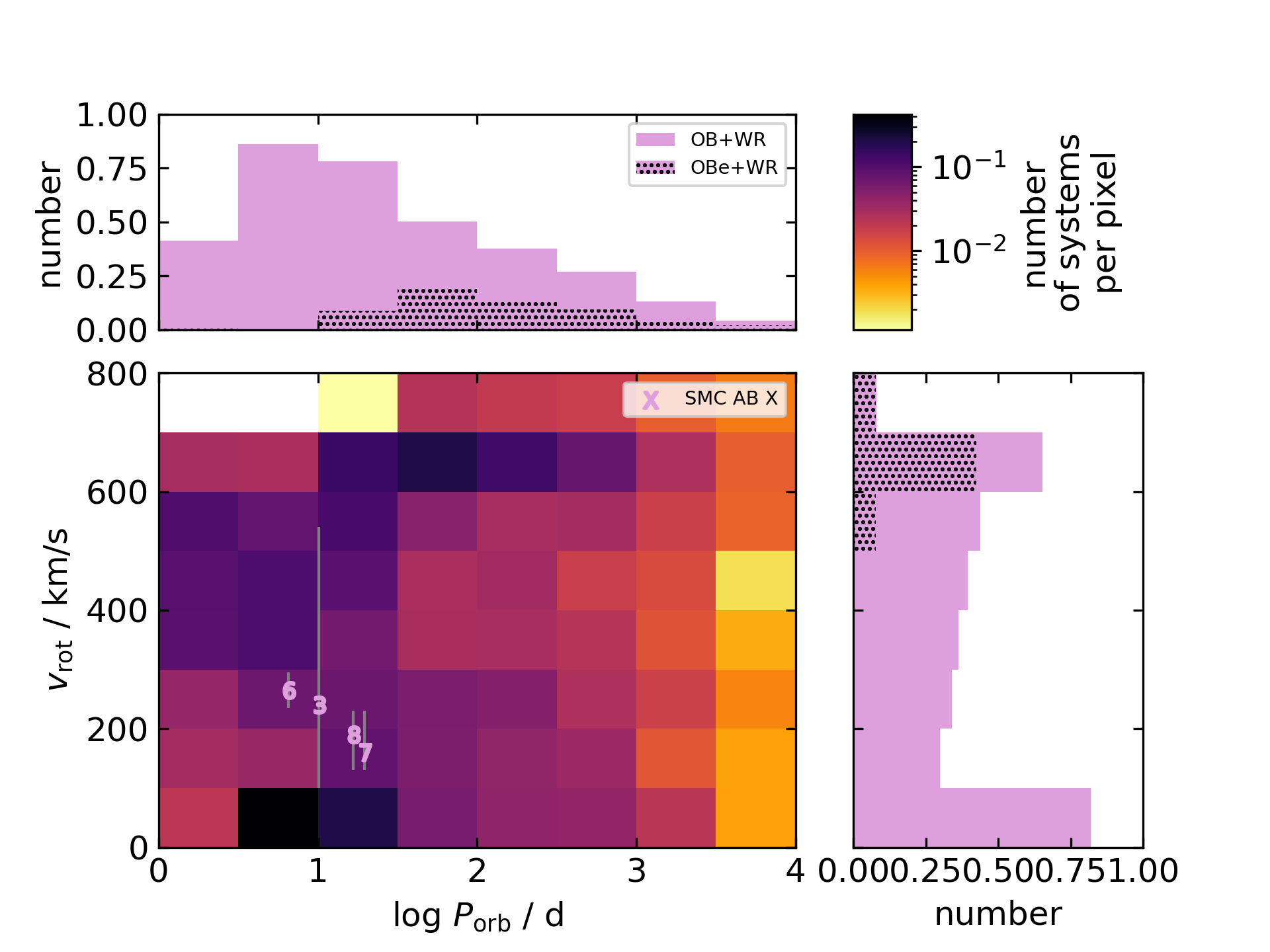}
 \includegraphics[width=\columnwidth]{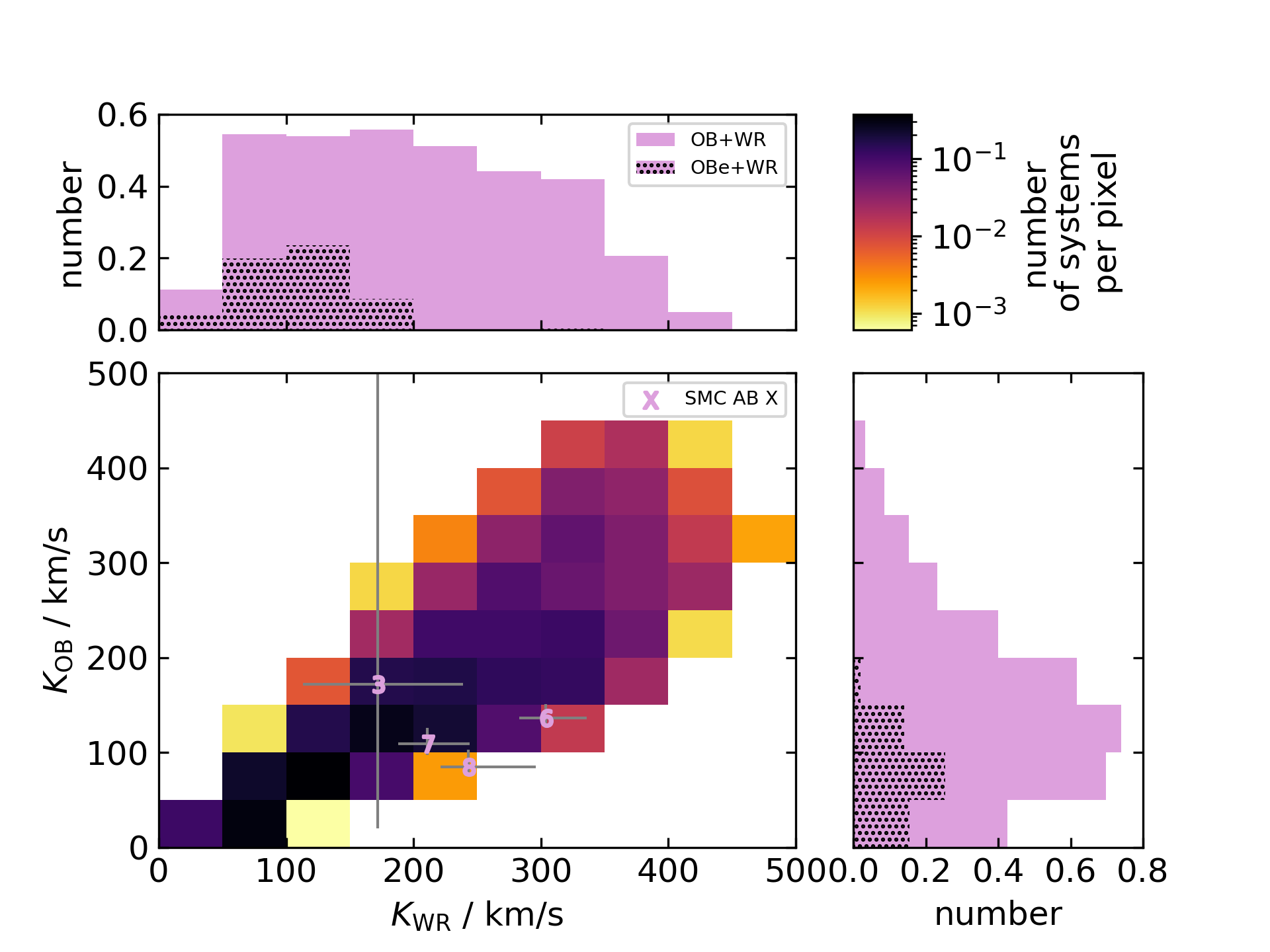}
 \includegraphics[width=\columnwidth]{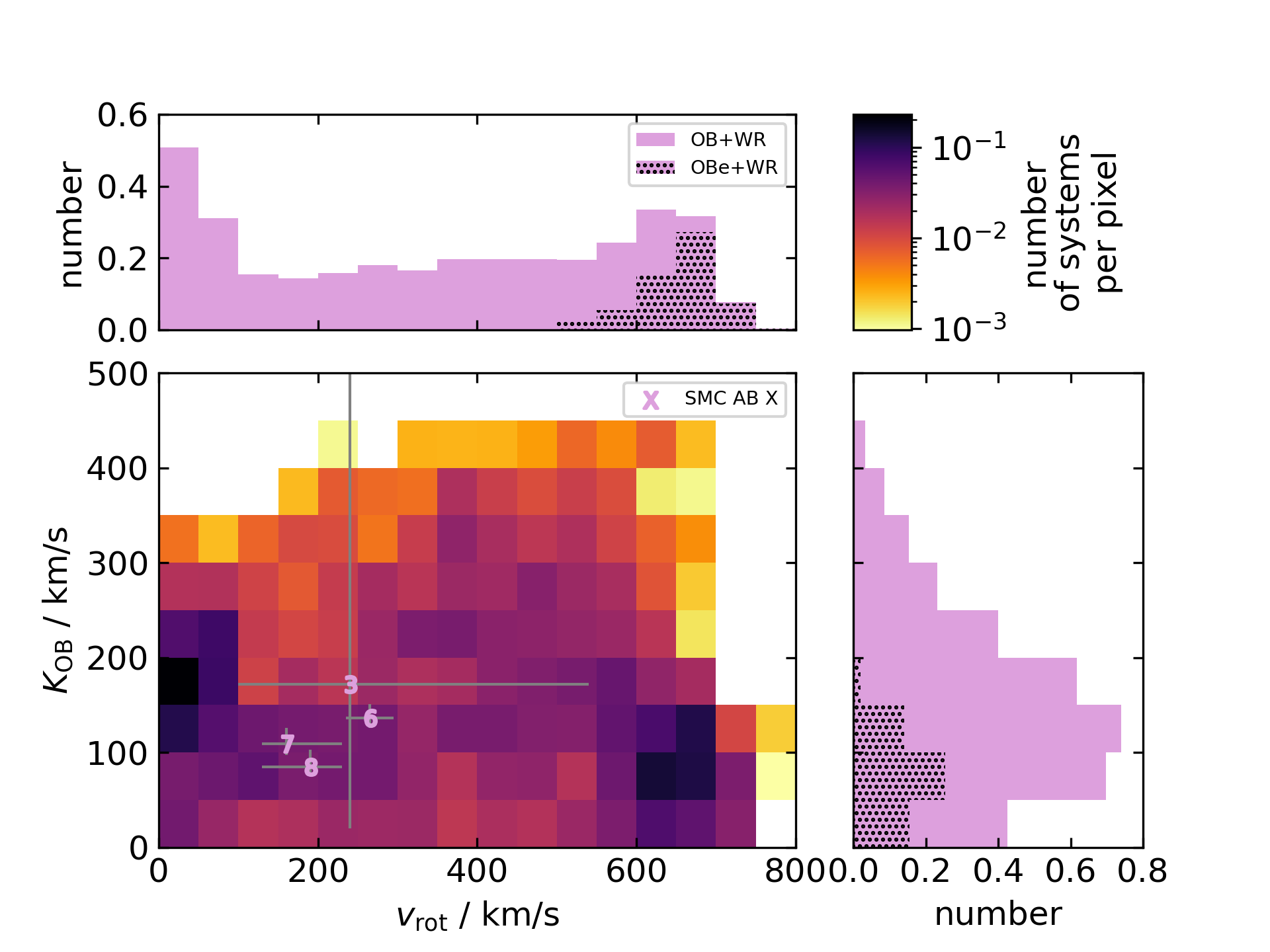}
 \includegraphics[width=\columnwidth]{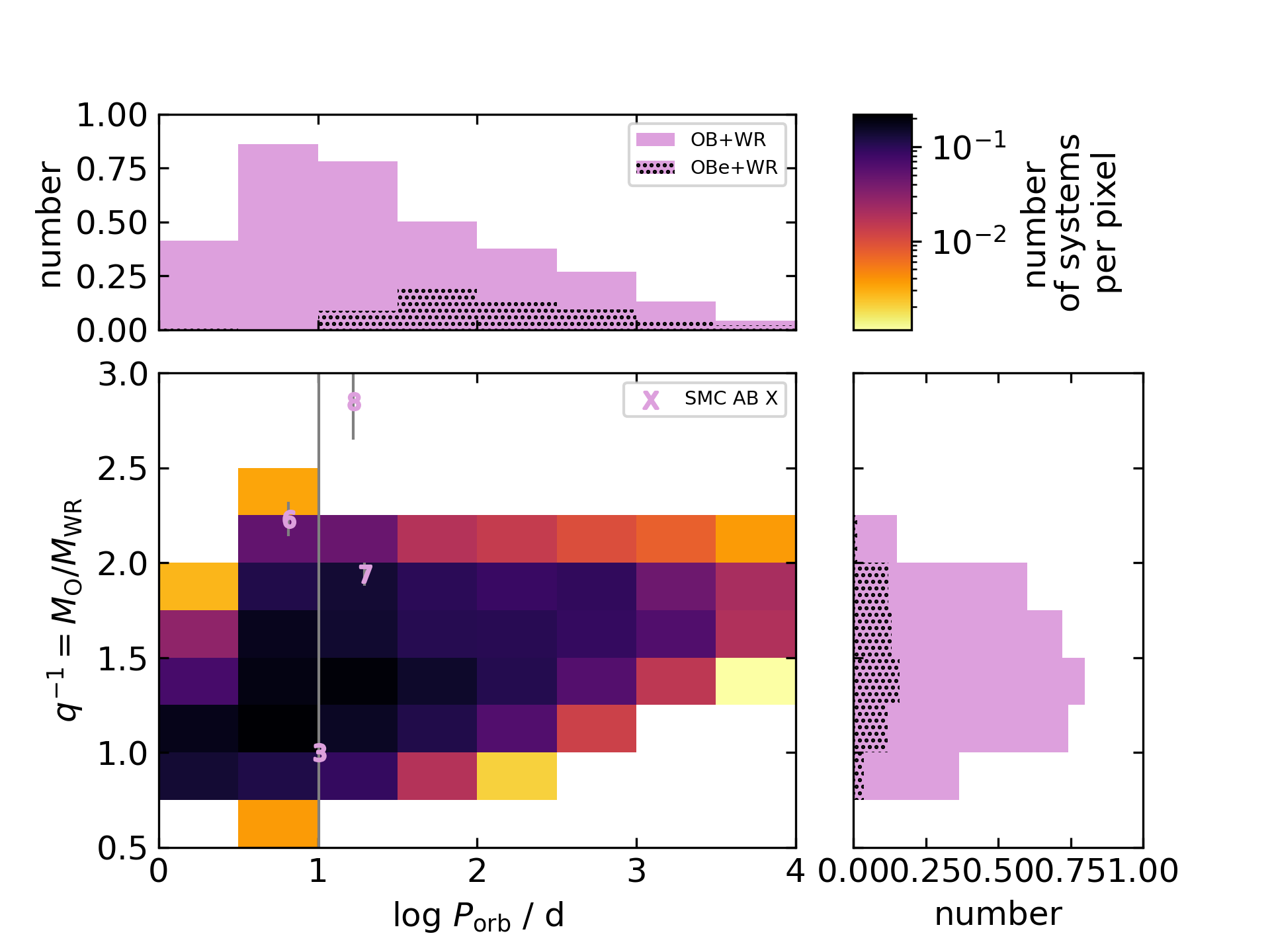}
 \caption{Predicted properties of WR+O systems with observations of the SMC WR+O systems \citep{2003MNRAS.338..360F,2016A&A...591A..22S,2018A&A...616A.103S}. Top left: luminosity-luminosity plane. Top right: orbital period--rotational velocity plane. Middle left: orbital velocity semi-amplitude plane. Middle right: orbital velocity semi-amplitude--rotational velocity plane. Bottom: orbital period--mass ratio plane.}
 % P-K_WR, P-q_WR
 \label{WR}
\end{figure*}

We assume, following \citet{2020A&A...634A..79S}, that a HeS with a luminosity above $10^{5.6}\lsol$ appears as a WR-star. We show the properties of our simulated O+WR systems in Fig.~\ref{WR}. As usual we label O~star which rotate faster than 95\% their critical rotation as OBe stars, even though the WR or O~star wind may prevent the formation of a disk or slow the O~star's rotation down. WR+WR and WR+SR systems are beyond the scope of this work.

The upper left panel shows the luminosities of the two components. While the number of WRs decreases towards higher values as expected from the initial mass function, the luminosities of the O~stars are more symmetrical, but still skewed. In the 2d histogram, the population forms a rough triangle since more luminous and thus more massive WRs have more massive and luminous companions. The $K_\mathrm{WR}$-$K_\mathrm{OB}$ diagram (lower left panel of Fig.~\ref{WR}) shows that the orbital velocity semi-amplitudes correlate tightly. This is not surprising as both stars have mass ratios between 0.5 and 2. Low velocities, especially for the OB star are somewhat preferred but a notable amount of stars still shows values above $200\kms$.

In the two panels on the right side of Fig.~\ref{WR} we investigate the relation between the OB star's rotational velocity to the orbital properties. We expect orbital periods from 1\,d to 1000\,d with a preference for short periods around 10\,d. Velocity semi-amplitudes range up to $450\kms$, but most common are values of about $100\kms$. As for the OB+BH systems we find a bimodality in the rotation rate of the OB star, but a less pronounced one. The two 2d histograms show again a distribution with two peaks, one for the close systems where tidal braking slowed the OB star down and one for the wide systems where the OB star could keep most of its spin angular momentum.

%\red{\textbf{NOT UPDATED YET!!!}}

%As we predict a large number of OB+HeS binaries it is straight forward to check whether the HeS appears as a Wolf-Rayet star. We use the technique proposed by \citet{1989A&A...210...93L} and used e.g. by \red{David} to calculate the optical depth $\tau$ of the He star's wind. If $\tau>1$ we call the HeS a WR star. Our standard simulation predicts 1-2 WR+OB systems.

%We calculate the periods for WR+OB systems and show them in Figure~\ref{wr-p}. It mirrows Fig.~\ref{per}. Short periods are preferred due to the initial period distribution. Very close systems do not reach this stage as they merge. Figure~\ref{wr-k} gives the orbital velocity for the WR and the OB component. The values here are much higher than in Fig.~\ref{OBorb} since WRs and their companions come from the very high mass range. Surprisingly there are two maxima in the WR's distribtion. This reflects the masses of the OB star. If the systems went through Case~A the OB star accreted more matter and became heavier than in Case~B. In contrast to Fig.~\ref{OBorb} this is visible because WR stars probe a small range of initial masses.

To compare the synthetic WR+O system to observations, we consider the WR+O~binares in the SMC of which four are known \citep{2016A&A...591A..22S,2018A&A...616A.103S}. Additionally, there are seven single WR stars. Recently, observations suggest they are truly single \citep{2024A&A...689A.157S}, which have strong implications on massive binary evolution \citep{2024A&A...689A.157S,xu} as well as the WR+WR system AB~5 (see however Sect.~\ref{sec-pq}). The WR population in the SMC is thought to be complete \citep{2016A&A...591A..22S,2018ApJ...863..181N}. We use the values given by \citet{2016A&A...591A..22S} for AB~3, AB~7, and AB~8 and by \citet{2018A&A...616A.103S} for AB~6.

Fig.~\ref{WR} (top left) shows that AB~6, 7, and 8 have luminosities as predicted, while the OB-luminosity of AB~3 does not match. In the other three panels we find the observations only cover a certain subspace of our simulations. While the orbital periods align with the peak of the simulations (about 10\,d), the observations are not distributed as broadly as the models. The same is true for the velocity semi-amplitude distribution (bottom left panel). We find also a mismatch in the rotational velocities of the OB~stars. The observations lie closely together in the brought trough between the two expected maxima (right panels in Fig.~\ref{WR}). This mismatch could have a similar origin as for the OB+BH systems, because perhaps the synthetic subpopulation at low rotation velocities corresponds to the observations, but our model overestimates the braking by tides, winds or structural evolution. The bottom panel shows that our synthetic population underestimates the mass of the O~star compared to the WR~star, which indicates that the mass-transfer efficiency might be higher than assumed.

Lastly, we compare our synthetic WR+O systems to Paper~I. Both studies find a similar WR luminosity function, however differences in the orbital properties emerge. Our WR velocity semi-amplitude distribution is almost flat, but Paper~I prefers low values. Similarly, our period distribution shows a clear maximum while theirs is rather flat even though it has a maximum at low periods. These differences are unexpected since at such high masses we have assumed a low accretion efficiency ($10\%$), relatively close to typical values ($5\%$) of Paper~I. However, our model allows for donor stripping at high initial masses for smaller initial mass ratios at low periods than the models of Paper~I. Our Fig.~\ref{fig-pq30} results in more low-period WR binaries than their Fig.~A.2 (top right). This may explain the lack of WR+O systems with large orbital periods \citep{2024A&A...689A.157S}.

\section{Discussion: Uncertainties}\label{uncert}

In this Section, we discuss the main uncertainties of our results as well as the alternative model with BH kicks.

\subsection{Stability of mass transfer and mass-transfer efficiency}

The key uncertainty of this study is the condition under which a RLO leads to a donor stripping and, in contrast, under which conditions the system mergers. We used the proposition of \citet{2024A&A...691A.174S} to link that to the swelling of the accretor star to the mass-transfer efficiency, which forms the second key uncertainty. Several other conditions have been proposed in the literature. Most prominent is the comparison of the radius evolution of the Roche lobe under mass loss with the adiabatic radius evolution of the donor, most recently investigated by \citet{2010ApJ...717..724G,2015ApJ...812...40G,2020ApJ...899..132G}. These studies find a period dependent minimum mass ratio for donor stripping, which itself has a minimum for a $10\msol$ donor around $q=0.15$ and decreases for higher masses. \citet{2024A&A...691A.174S} discussed that their condition is in general more restrictive than the one of \citet{2010ApJ...717..724G,2015ApJ...812...40G,2020ApJ...899..132G}. While we assume that donors with a deep convective envelope always result in a unstable mass transfer, aforementioned studies find that mass transfer might be stable for systems close to a mass ratio of unity \citep[see also][]{2019A&A...628A..19Q,2021A&A...650A.107M,2024A&A...685A..58E}.

An other physically motivated approach was undertaken by \citet{handle:20.500.11811/7507}, whose method was also used by Paper~I. They stop the deposition of material onto the accretor if it starts to rotate critically. If then the combined luminosity of the two stars is large enough to drive the non-accreted material out of the system, the mass transfer is regarded as stable. The comparison of our Fig.~\ref{fig:Pq} and Sect.~\ref{app:Pq} with their Fig.~2, A.1 and A.2 shows a drastic differences of the parameter space of mass transfer survivors (see Sect.~\ref{sec-dis-xu} for details). Other studies \citep[e.g.][]{1991A&A...241..419P,2002MNRAS.329..897H,2015ApJ...805...20S,2019A&A...624A..66R} use fixed minimal mass ratios for stable mass transfer for each Case~A and~B to decide as the merger criterion, but limit the accretion by the thermal timescale of the accretor, whereby the mass-transfer efficiency becomes a function of period and mass ratio. These mass ratio limits sometimes in tension (e.g. 0.65 for Case~A, see \citet{2007A&A...467.1181D,2013ApJ...764..166D,2014A&A...563A..83C}) and sometimes in agreement (0.25 for Case~B, see \citet{2015ApJ...805...20S,2017NatCo...814906S}) with our models (see Fig.~\ref{fig:Pq} and Sect.~\ref{sec-pq}).

For this study, we assumed a low mass-transfer efficiency for more massive stars. However according to Fig.~\ref{fig:eff} also larger values are in agreement with the observed number of WR+O systems within the uncertainties. In Sect.~\ref{sec-wr} we found that we under-predict the O~star mass compared to that of its WR~companion, which again suggests mass-transfer efficiencies larger than assumed. In addition to this there are uncertainties in IMF, binary fraction, initial mass ratio and initial orbital period distributions, upper stellar mass limit, size of stable mass-transfer window and recent star-formation rate, which affect the synthetic WR+O population.

%mass transfer stability, Pq-diagram, accretion mechanism

%\red{Case~A q-min = 0.65 see de Mink+07,13, Claeys+14, Case~B q-min = 0.25 see Schneider,Izzard+15 incl appendix; https://arxiv.org/abs/2109.10352, stevenson+17 }

\subsection{Case~A RLO}
Case~A mass transfer takes place on the nuclear timescale of the donor and comprises three phases of different structure, which determine the evolution of the stellar masses \citep[][]{2001A&A...369..939W}. The adopted scheme based on the results of \citet{2024A&A...690A.282S} might be too simplistic as their underlying models are restricted to a certain set of assumptions of binary physics. Furthermore we do not resolve the mass transfers phases in detail but jump after a reasonable amount of time directly to the OB+HeS phase. While we are not interested at the models configuration as an Algol system during the mass transfer \citep[e.g.][]{2021arXiv211103329S}, the detailed evolution may have an imprint on the outcome and the subsequent OB+SR phase, especially at initial mass ratios smaller than about 0.5, where the orbit shirks a lot during RLO \citep{2024A&A...690A.282S}.

\subsection{Stellar models} \label{umod}
Our predictions rely on the assumed underlying stellar models. A key uncertainty lies in the HeS models. They influence the evolution of our systems firstly through their lifetime, which determines when a SR forms. This can have an influence on whether a reverse mass transfer occurs. Secondly, their radius evolution determines whether and when a Case~BB/BC RLO takes place. This is important as the assumed SMC HeS models are extrapolated from models with solar metallicity \citep{2018MNRAS.481.1908K} and, most importantly, any thin remaining hydrogen layer was not considered in modelling these stars. As shown by \citet{2020A&A...637A...6L}, the radius evolution of stripped stars has a noticeable effect on the final fate of the system. Furthermore, \citet{2024A&A...685A..58E} demonstrated that also a partial removal of the envelope can happen. The occurrence of Case~BB/BC mass transfer changes not only the orbital period of the system \citep{1997A&A...327..620S}, but also impacts the SN kick (Sect.~\ref{method}) and determines the final mass of the SR.

Also the calculation of the wind-induced spin-down of our mass gainers bears some uncertainty.
Recently, \citep{2025arXiv250212107N} have shown that the numerical treatment of wind angular momentum loss affects the evolution of stellar rotation if stellar winds carry away a significant fraction of the stellar mass. Our approach (Sect.~\ref{sect-rot}) leads to a stronger angular momentum loss compared to the scheme used by Paper~I, and thus to a somewhat smaller fraction of OBe stars. Although under the assumption of rigid rotation, our approach is more physically correct, it appears not to be favoured when compared to the spin rates of Galactic OB~stars \citep{2025arXiv250212107N}. 

%is more similar to the one of \citep{2025arXiv250212107N} with a stronger spin down, we find a weaker impact of wind angular momentum loss than \citet{xu}. On the other hand, our method results in a stronger effect of tidal forces on the stellar rotation than detailed models. As stellar rotation does not strongly impact the evolution of binary stars but rather their observational appearance as OBe~stars, we might overestimate the number of fast rotators at high masses and underestimate them in close systems.

Another uncertainty of the stellar models lies in the occurrence of envelope inflation. \citet{2015A&A...580A..20S} showed that stellar models heavier than about $40\msol$ can exceed the local Eddington limit in their envelope, which leads to a large increase in radius and to a convective envelope. As we assume unstable mass transfer for a donor with a deep convective envelope, including this effect into our models could drastically reduce the production of massive HeS by binary stripping and consequently the formation of WR+O and BH+OB systems.

\subsection{Formation of OBe~stars}
In this study we followed \citet{2004MNRAS.350..189T} and assumed that models with $\vrot/\vcr>0.95$ appear as OBe~stars. It is however neither clear if this is the right value nor whether in general a condition in terms of $\vrot/\vcr$ is appropriate. \citet{2013A&ARv..21...69R} summarises that the average $\vrot/\vcr$ of OBe~stars is around 0.85\footnote{We converted the $W=\vrot/\varv_\mathrm{orb}$ of \citet{2013A&ARv..21...69R} to $\varUpsilon=\vrot/\vcr$ according to their equation 8 and 11.} and the minimum $\vrot/\vcr$ around 0.77. More recent studies confirmed that OBe~stars can rotate substantially sub-critical \citep{2016A&A...595A.132Z,2020MNRAS.493.2528B,2022MNRAS.516.3602E,2022MNRAS.512.3331D}. On the other hand, interferometric observations of \textalpha~Eri \citep{2003A&A...407L..47D,2012A&A...545A.130D} obtained near-critical rotation.

While fast rotation is likely a necessary criterion for a star to become a OBe~star it seems not to be a sufficient condition, as the example of \textalpha~Leo shows \citep{2005ApJ...628..439M}. It rotates with $\vrot/\vcr=0.86$ faster than the aforementioned minimum while it is not a OBe~star but only of spectral type Bn. Several other condition for a rotating star to become a OBe~star such as pulsations and magnetic fields are discussed in \citet{2013A&ARv..21...69R}.

Our simulations yielded that almost all accretors, which rotated faster than $\vrot/\vcr=0.5$ after mass transfer, rotated with at least $\vrot/\vcr=0.95$. Thus the number of OBe~stars will change only weakly if we would change our limit from $0.95$ \citep{2004MNRAS.350..189T} to a lower value as 0.77 \citep{2013A&ARv..21...69R}. A further condition which might need to be fulfilled by a star to become a OBe~star may reduce the number of OBe~stars in our study. This is problematic as our predicted total number is just about in agreement with the observations (see Sect.~\ref{results_var}). 

\subsection{BH kick}\label{sec-BHkick}
In Sect.~\ref{results_fid}, we assumed that the BH receives no birth kick, and thus we find that BH+OB systems do not unbind in our study. In the scenario with flat distribution between $0\kms$ and $200\kms$ (Fig.~\ref{fig:comp}) we find that their number was reduced by a factor of about two. No consensus about the natal kick has been reached in the literature. Many observational works analysing the position and kinematics of low-mass X-ray binaries argue in favour of a high kick scenario \citep{2002ApJ...567..491P,2005ApJ...618..845G,2005ApJ...625..324W,2009ApJ...697.1057F,2012MNRAS.425.2799R,2025A&A...694A.119Y}, while others argue for a low kick \citep{1999A&A...352L..87N,2016MNRAS.456..578M}. \citet{2015MNRAS.453.3341R} even finds clues for both scenarios depending on the systems. \citet{2012ApJ...749...91F} and \citet{2014ApJ...796...37S,2019ApJ...885..151S,2020ApJ...898..143S,2021ApJ...908...67S} use for their population syntheses an approach by which the kick of the initially formed NS is reduced by fallback prior to the BH formation. From the detailed modelling site, \citet{2013MNRAS.434.1355J} is able to explain high BH kicks by an asymmetric SN with fallback. \citet{2022MNRAS.512.4503R} on the other hand proposes a mechanism for a low BH kick based on hydrodynamical simulations while \citet{2020MNRAS.495.3751C} can produce, also with hydrodynamical simulations, both low and high kicks depending on the explosion energy. See \citet{2022ApJ...926....9J} for a recent discussion.

\begin{figure*}
 \centering
 \includegraphics[width=0.64\textwidth]{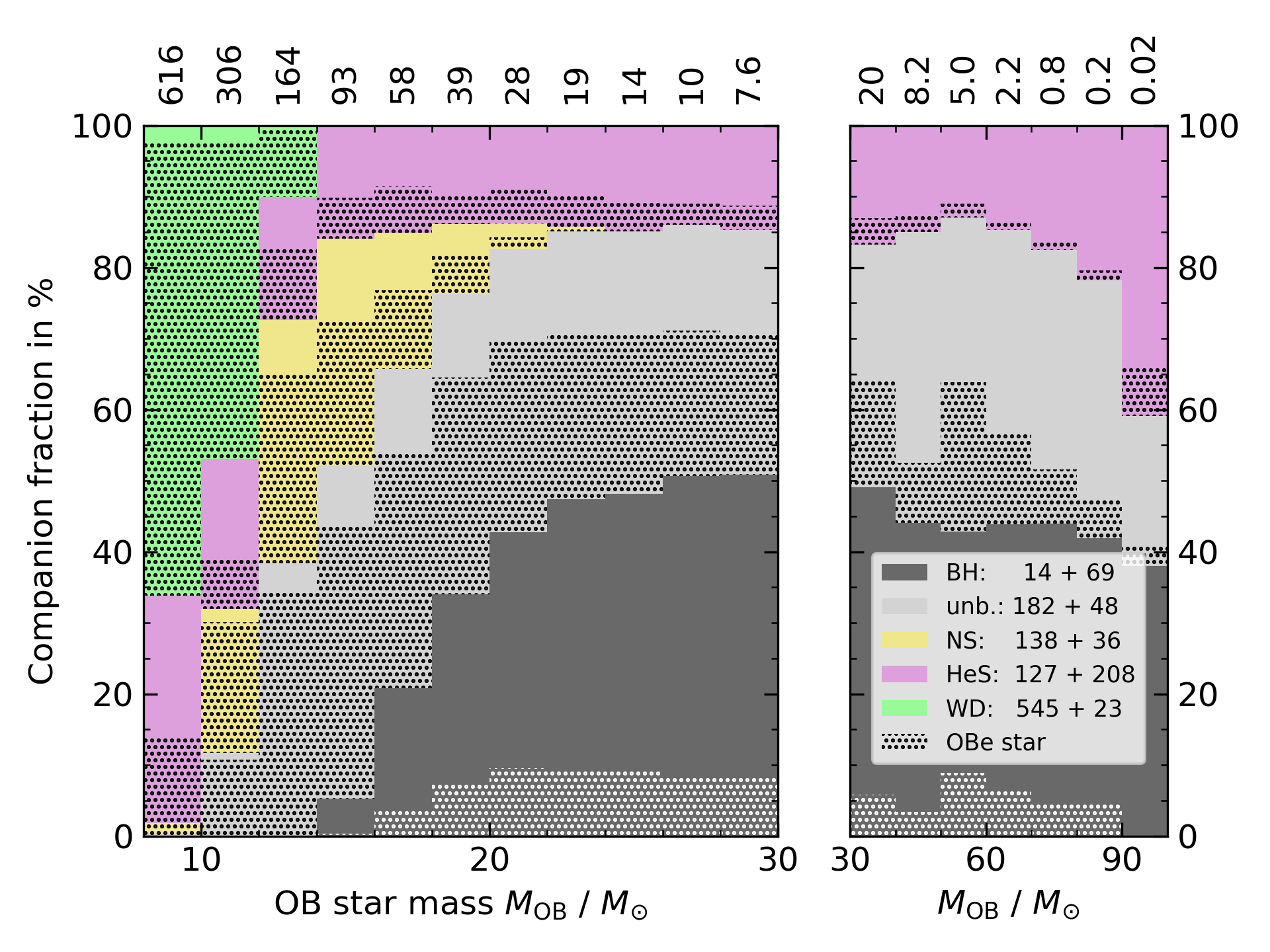}
 \includegraphics[width=0.35\textwidth, trim={35.56mm 0 35.56mm 0}, clip]{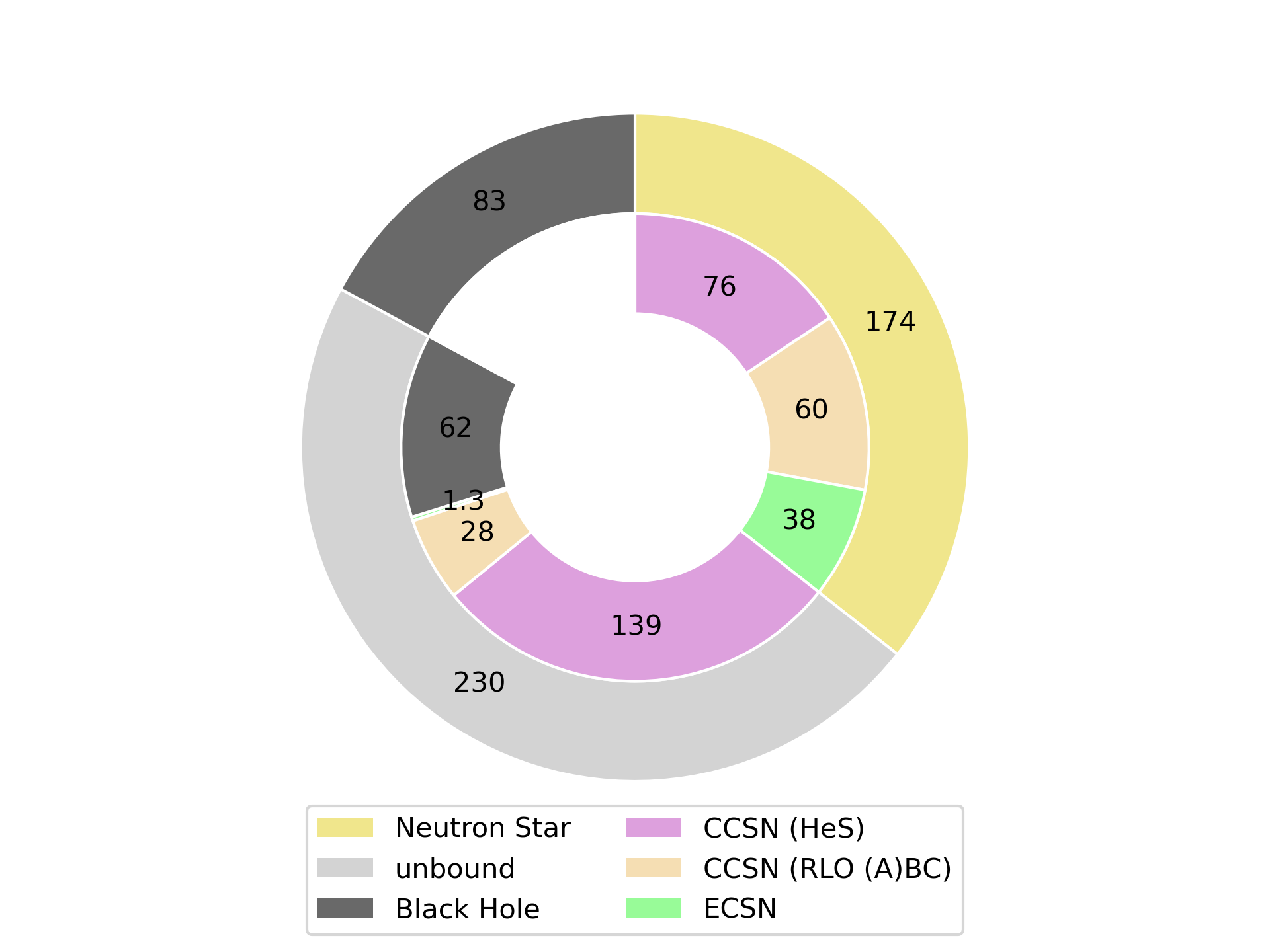}
 \includegraphics[width=\columnwidth]{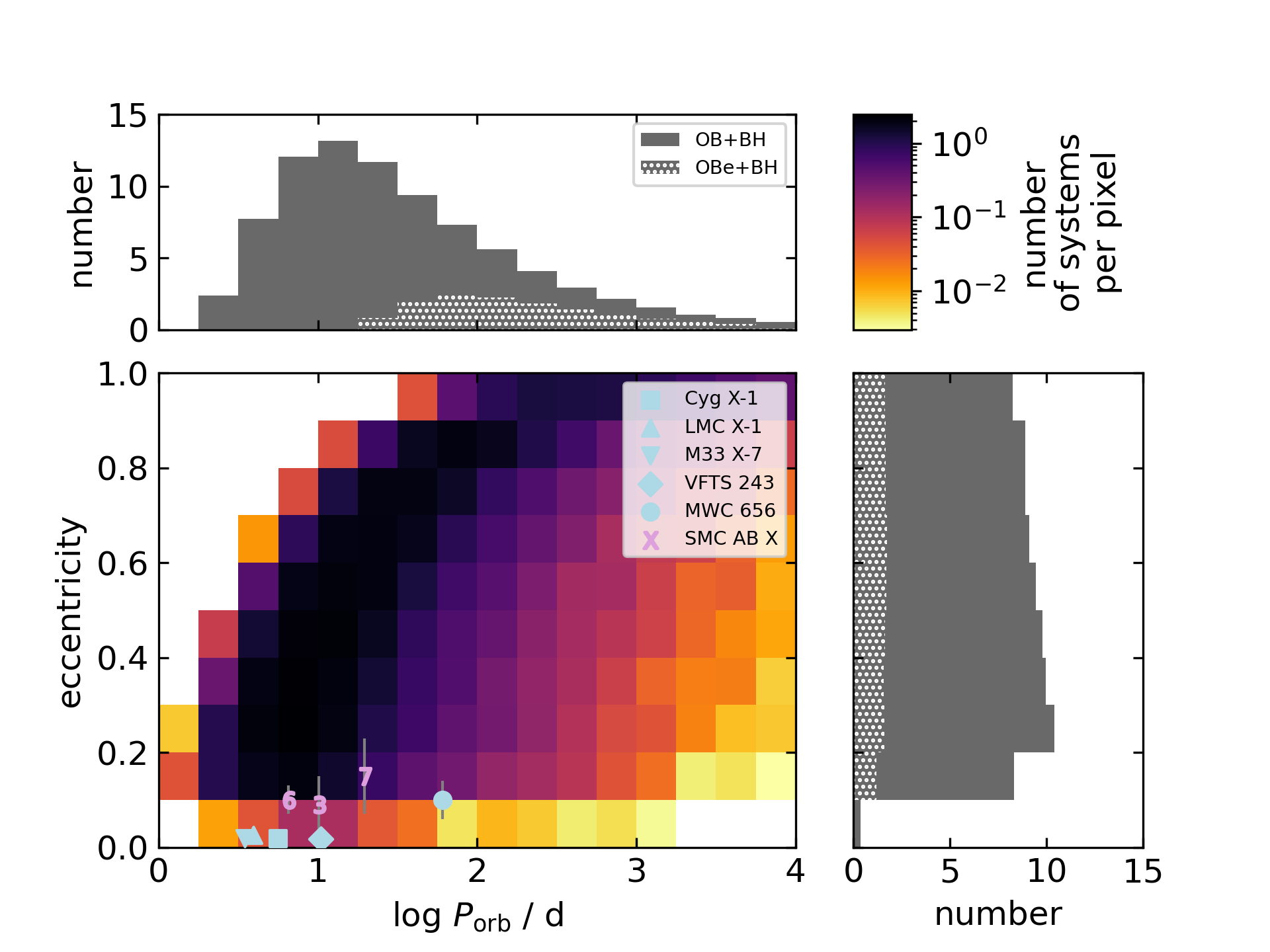}
 \includegraphics[width=\columnwidth]{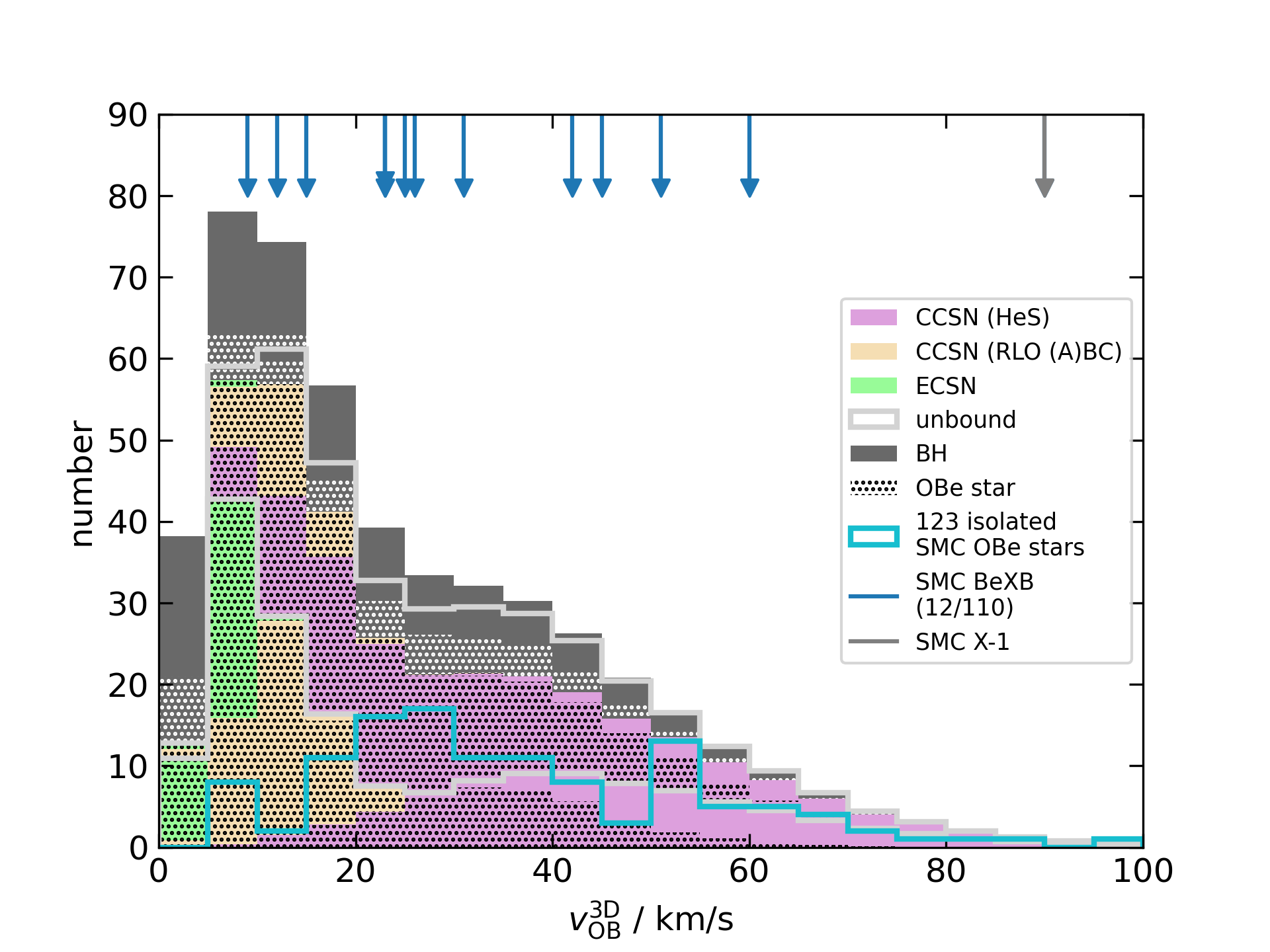}
 \caption{Same as Fig.~\ref{CompFrac}, \ref{fig-3er-orbit} (lower right), and \ref{fig-3er-v3d} (top), but with a BH kick.}
 \label{bhkick}
\end{figure*}

The BH kick does not only affect the total number of OB+BH systems, but also the distributions of orbital parameters such as orbital period and 3D velocity \citep{1998A&A...330.1047T}. We reproduced Fig.~\ref{CompFrac}, \ref{fig-3er-mass} (lower right), and \ref{fig-3er-v3d} with a strong BH kick in Fig.~\ref{bhkick}. About a third of the BH systems unbinds and we only expect 15 OBe+BH systems and 69 regular OB+BH systems. Especially the systems with OBe stars break apart as they are wider then those with regular OB stars. The eccentricity distribution of the unbound systems is flat (like that of the NS systems, Fig.~\ref{fig-3er-NS}), but avoids values below 0.1. The distribution of orbital periods appears similar to the scenario without kicks, but their number decreses not that much towards larger periods. The combined eccentricity-period distribution is similar to that of the NS~systems, but appears to correlate more tightly. The distribution of space velocities with BH kicks allows also for values below $5\kms$. Systems with values above $15\kms$ do not contain an emission-line star. It is quite likely that close OB+BH systems circularise and so the current observations within both eccentricity-period diagrams would allow for both a weak/no kick and a large kick at BH formation.

%SN: Woosly, Janka; kick

\section{More Pq-diagrams}\label{app:Pq}

\begin{figure*}
 \centering
 \includegraphics[width=0.8\textwidth]{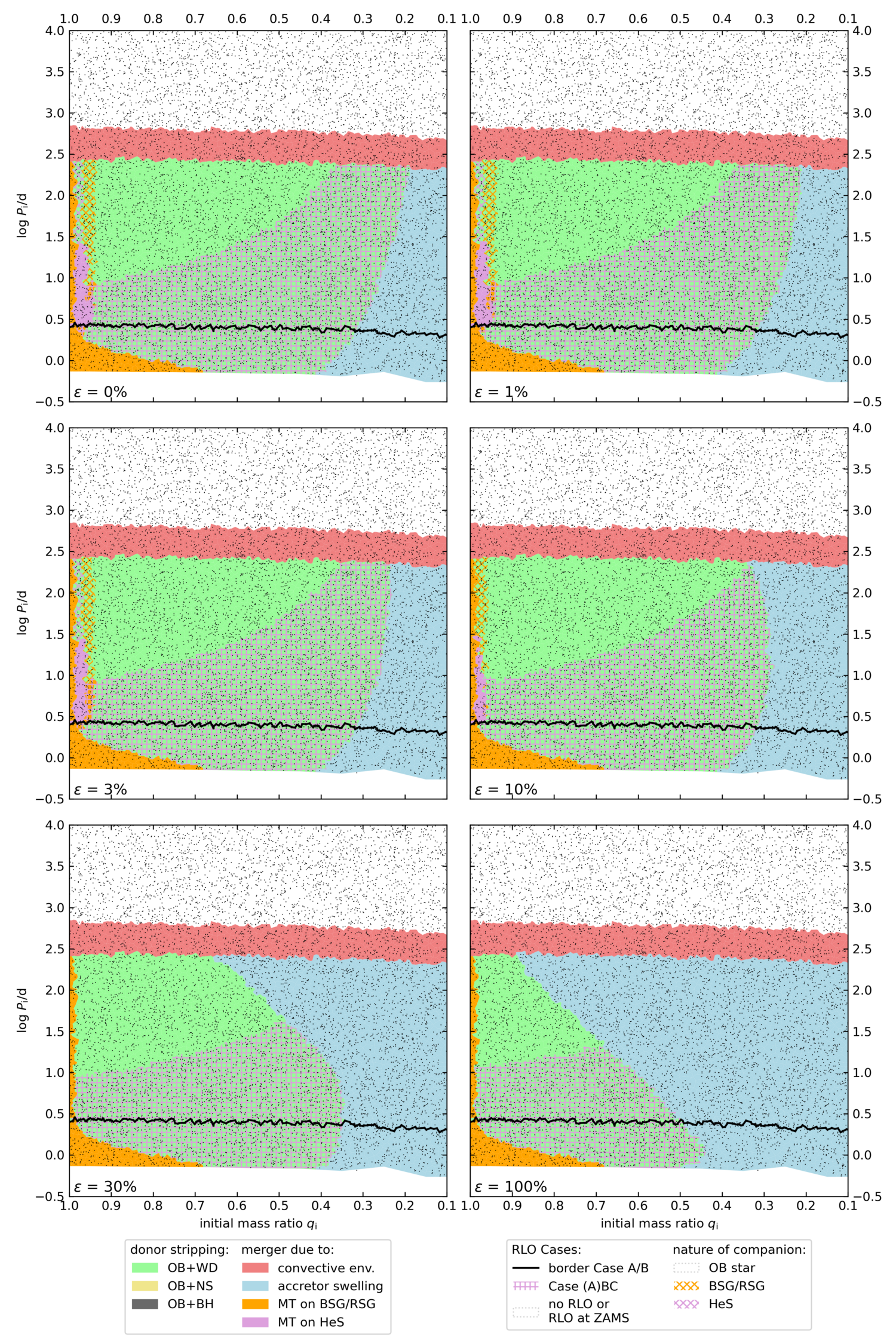}
 \caption{As Fig.\,\ref{fig:Pq}, but for a primary mass of $8\msol$.}
 \label{fig-pq8}
\end{figure*}

Figures~\ref{fig-pq8} to~\ref{fig-pq100} are analogues of Fig.~\ref{fig:Pq} for other initial donor masses.

\begin{figure*}
 \centering
 \includegraphics[width=0.8\textwidth]{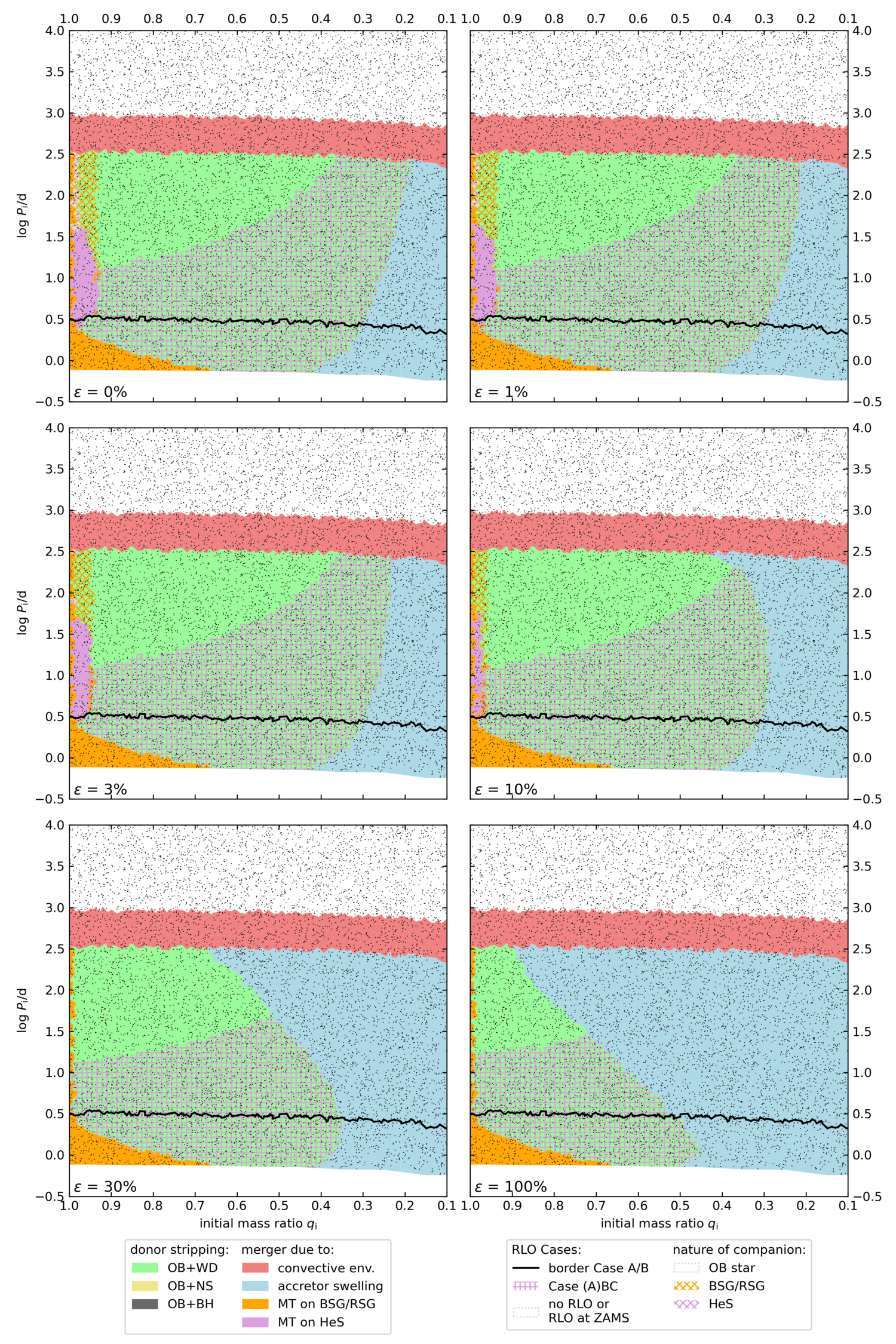}
 \caption{As Fig.\,\ref{fig:Pq}, but for a primary mass of $9\msol$.}
\end{figure*}

\begin{figure*}
 \centering
 \includegraphics[width=0.8\textwidth]{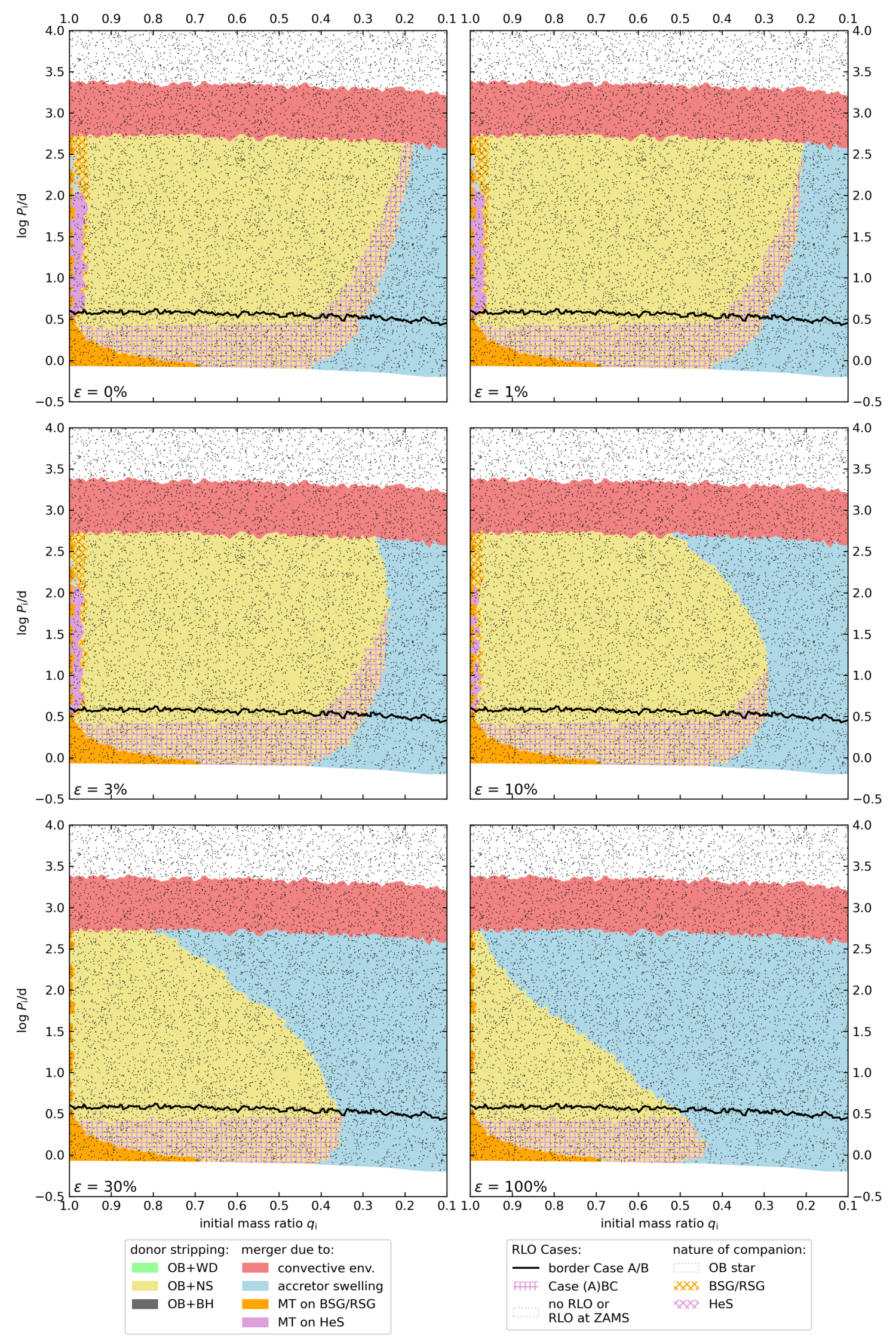}
 \caption{As Fig.\,\ref{fig:Pq}, but for a primary mass of $12\msol$.}
\end{figure*}

\begin{figure*}
 \centering
 \includegraphics[width=0.8\textwidth]{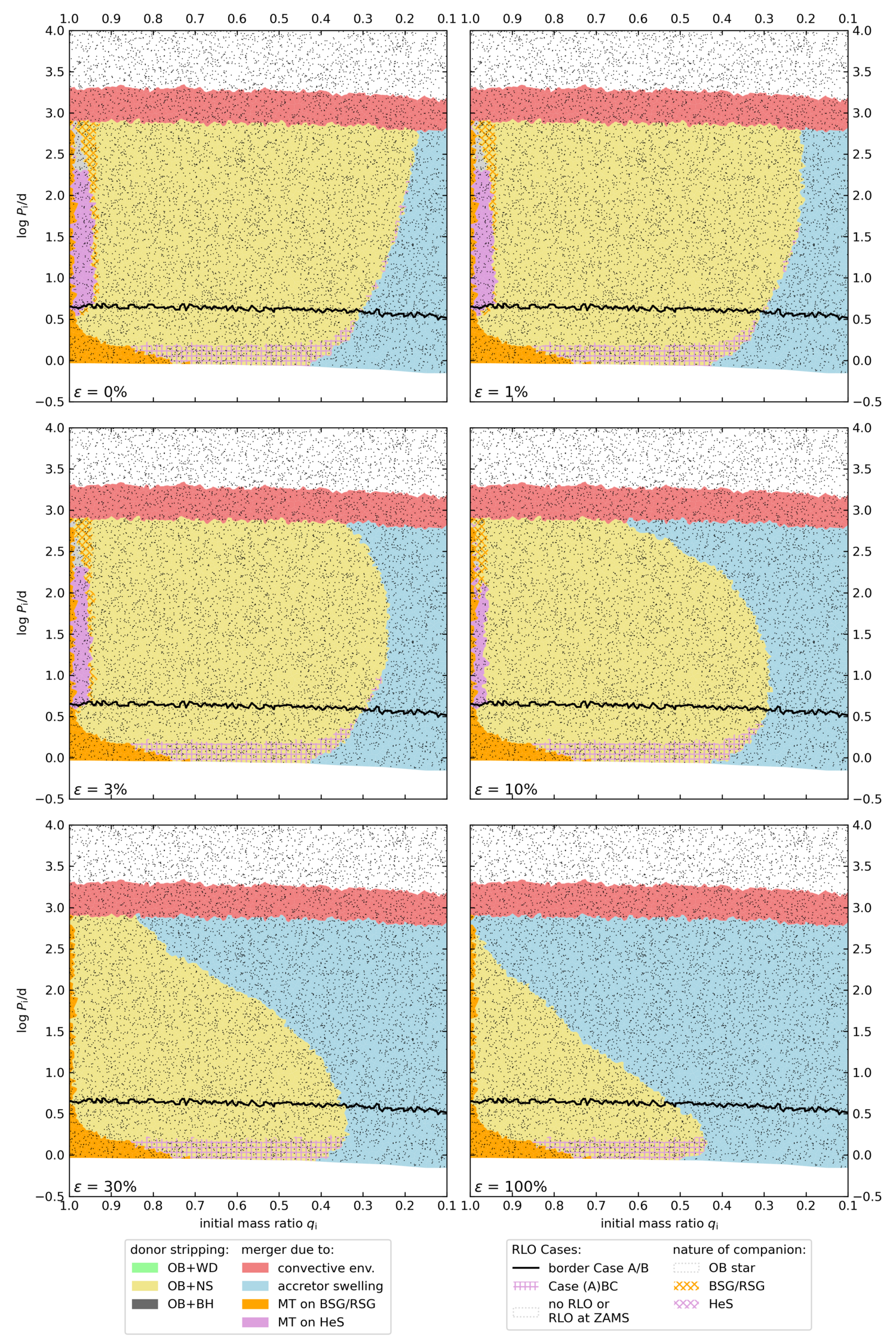}
 \caption{As Fig.\,\ref{fig:Pq}, but for a primary mass of $15\msol$.}
\end{figure*}

\begin{figure*}
 \centering
 \includegraphics[width=0.8\textwidth]{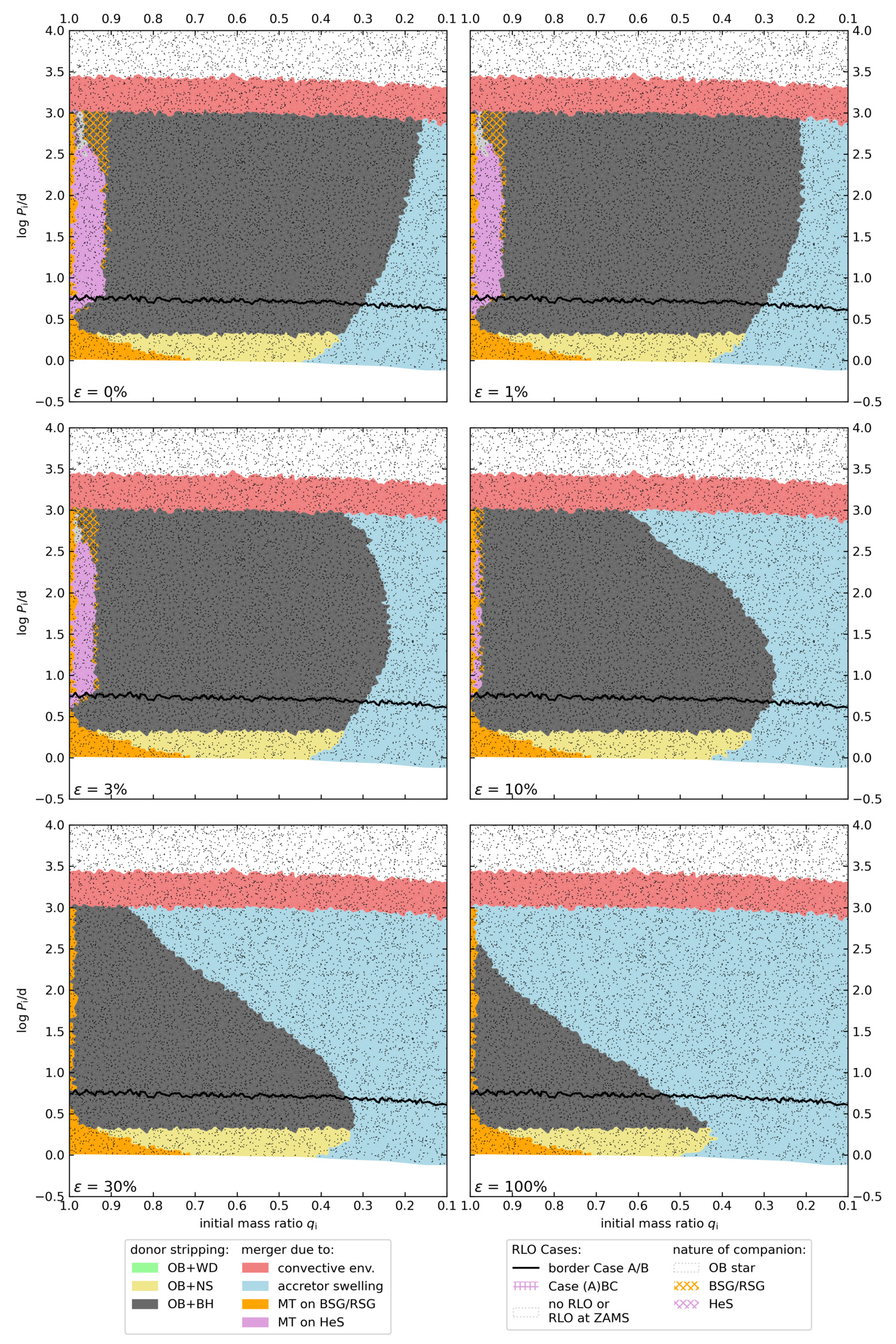}
 \caption{As Fig.\,\ref{fig:Pq}, but for a primary mass of $20\msol$.}
 \label{fig-pq20}
\end{figure*}

\begin{figure*}
 \centering
 \includegraphics[width=0.8\textwidth]{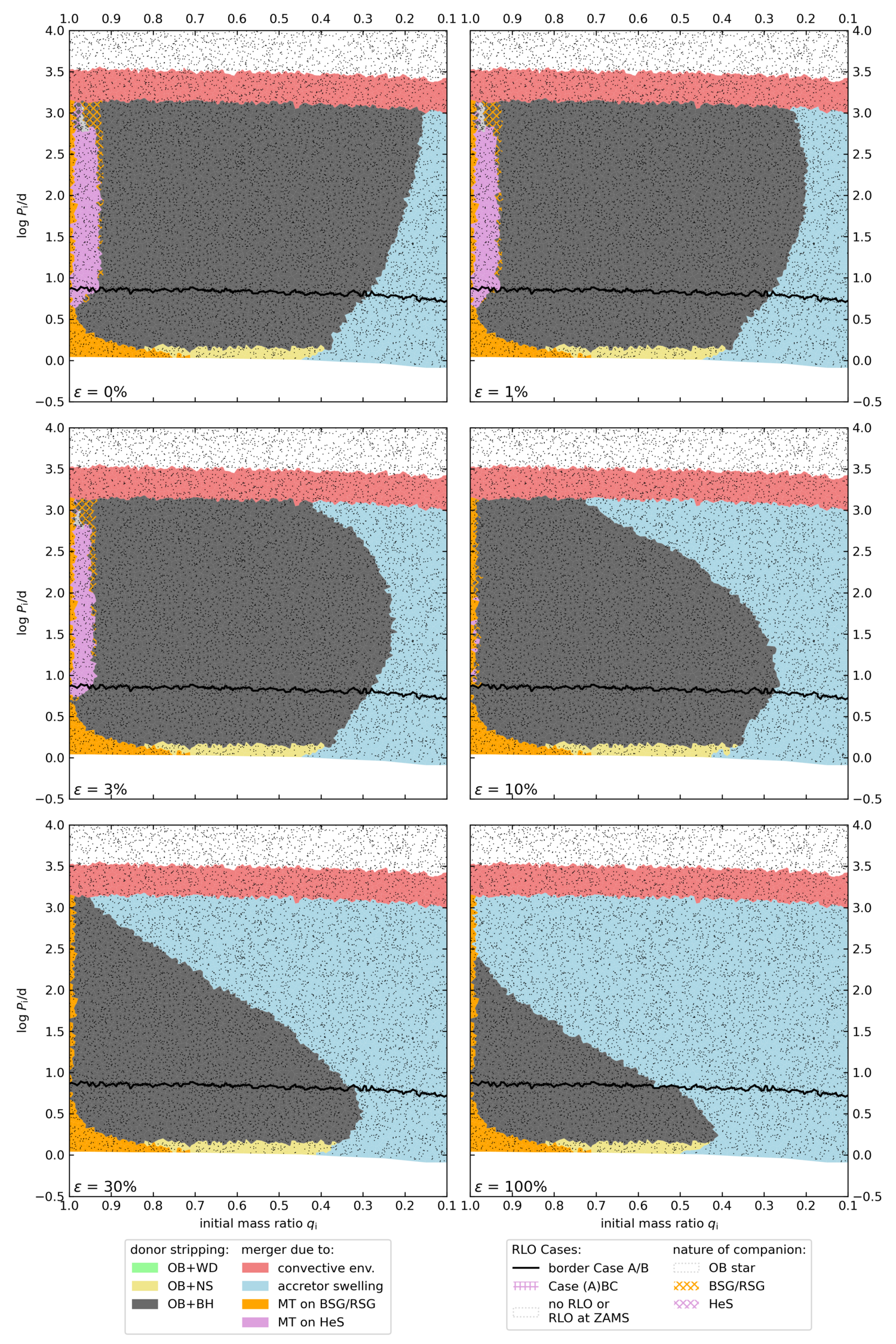}
 \caption{As Fig.\,\ref{fig:Pq}, but for a primary mass of $25\msol$.}
\end{figure*}

\begin{figure*}
 \centering
 \includegraphics[width=0.8\textwidth]{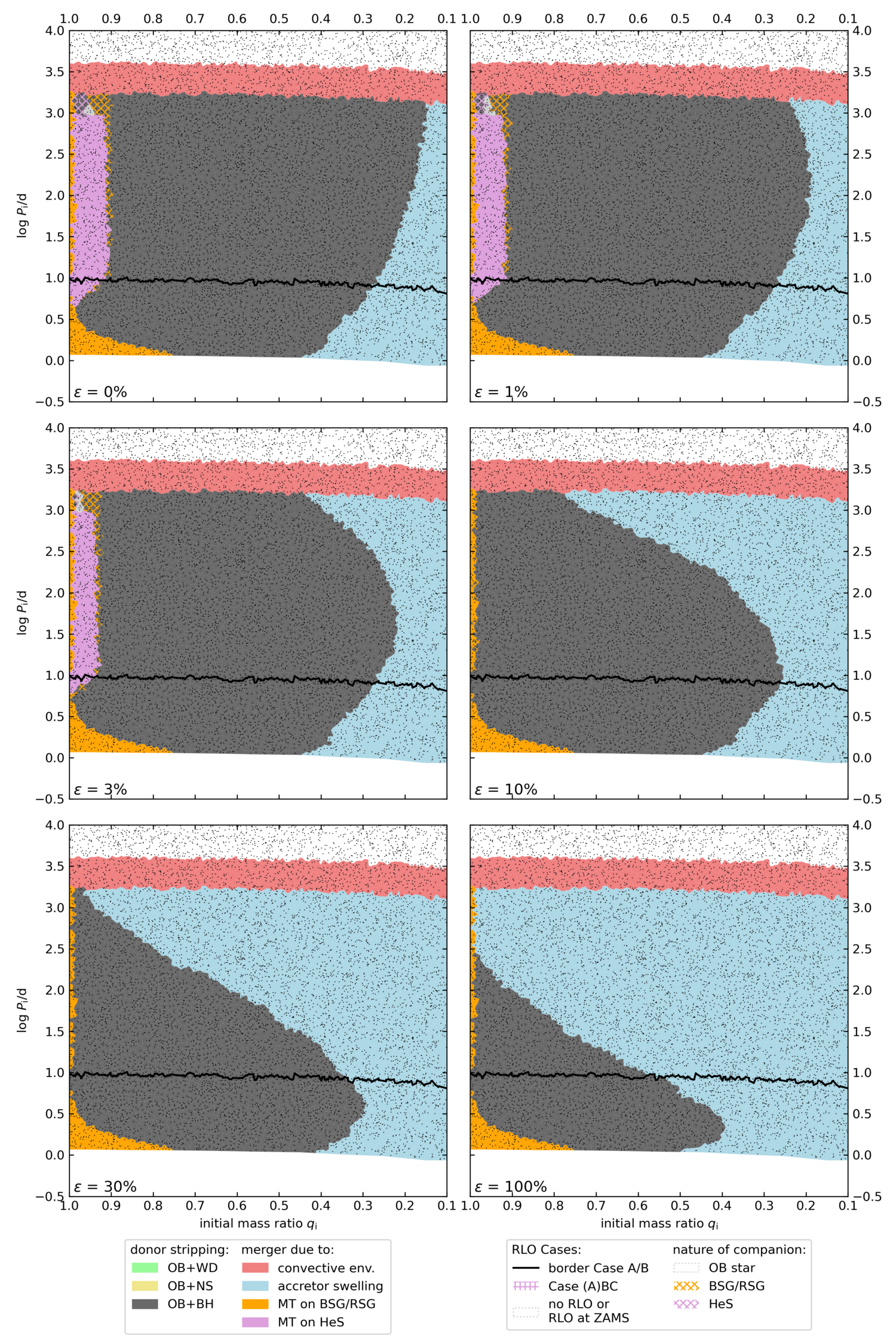}
 \caption{As Fig.\,\ref{fig:Pq}, but for a primary mass of $30\msol$.}
 \label{fig-pq30}
\end{figure*}

\begin{figure*}
 \centering
 \includegraphics[width=0.8\textwidth]{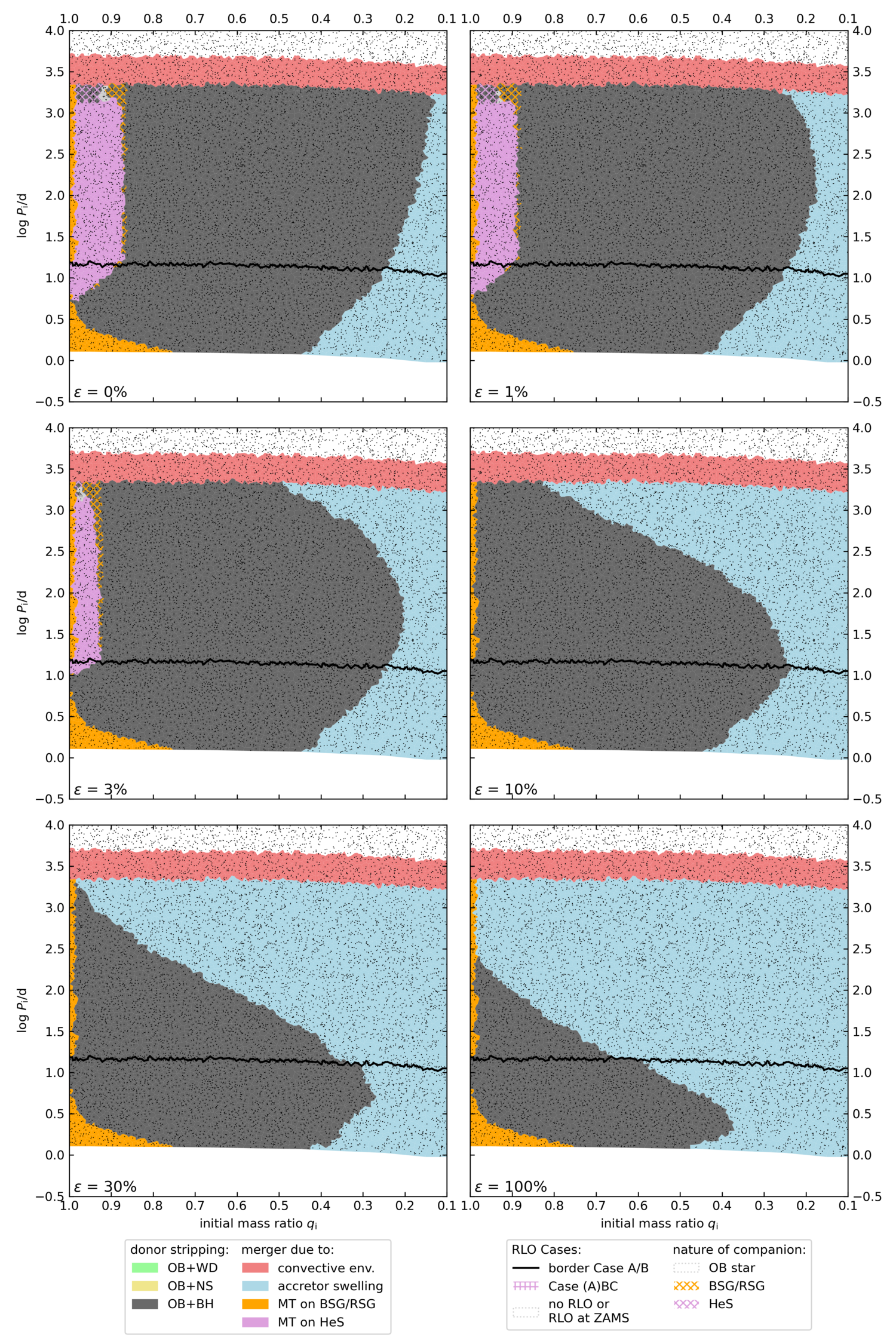}
 \caption{As Fig.\,\ref{fig:Pq}, but for a primary mass of $40\msol$.}
\end{figure*}

\begin{figure*}
 \centering
 \includegraphics[width=0.8\textwidth]{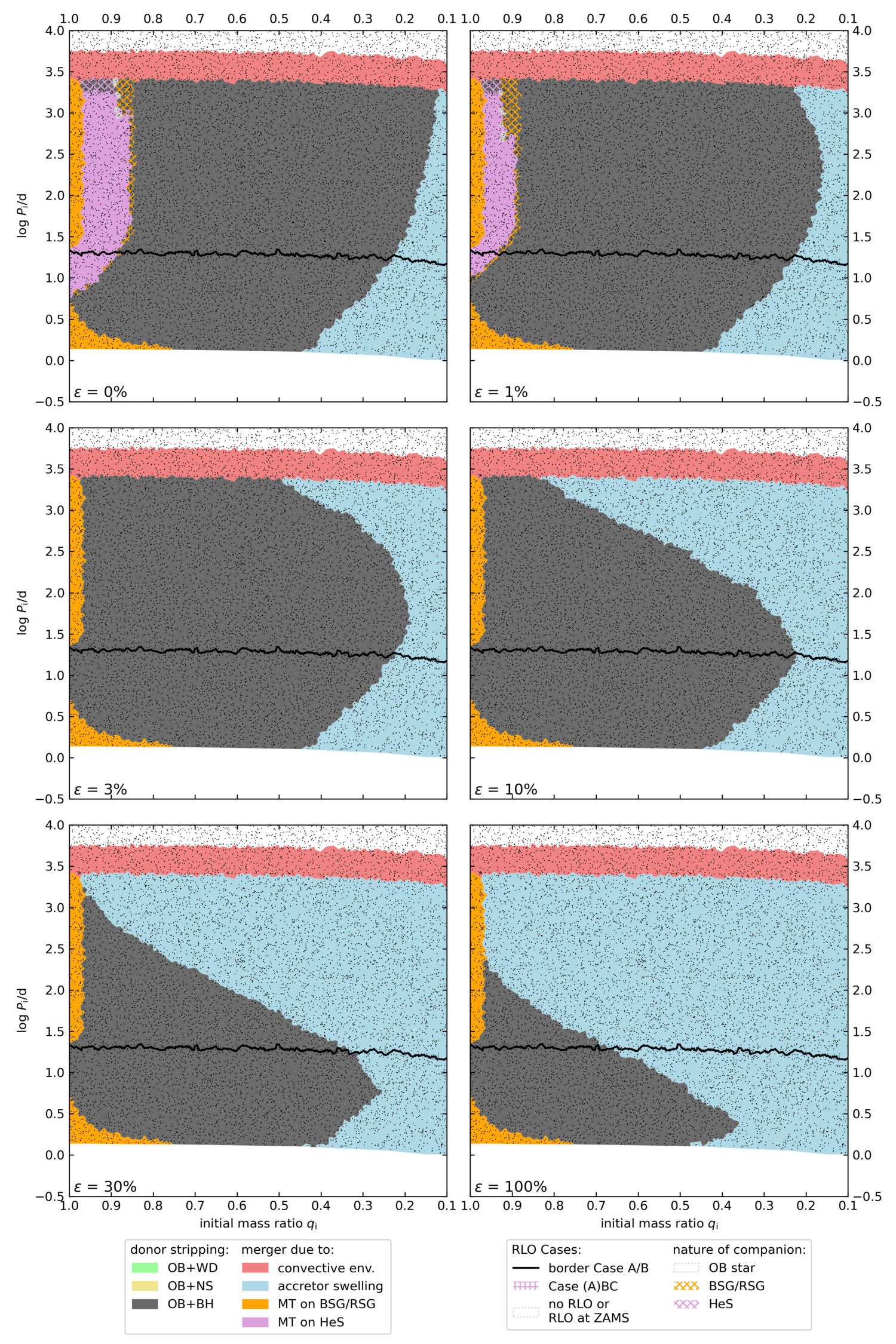}
 \caption{As Fig.\,\ref{fig:Pq}, but for a primary mass of $50\msol$.}
\end{figure*}

\begin{figure*}
 \centering
 \includegraphics[width=0.8\textwidth]{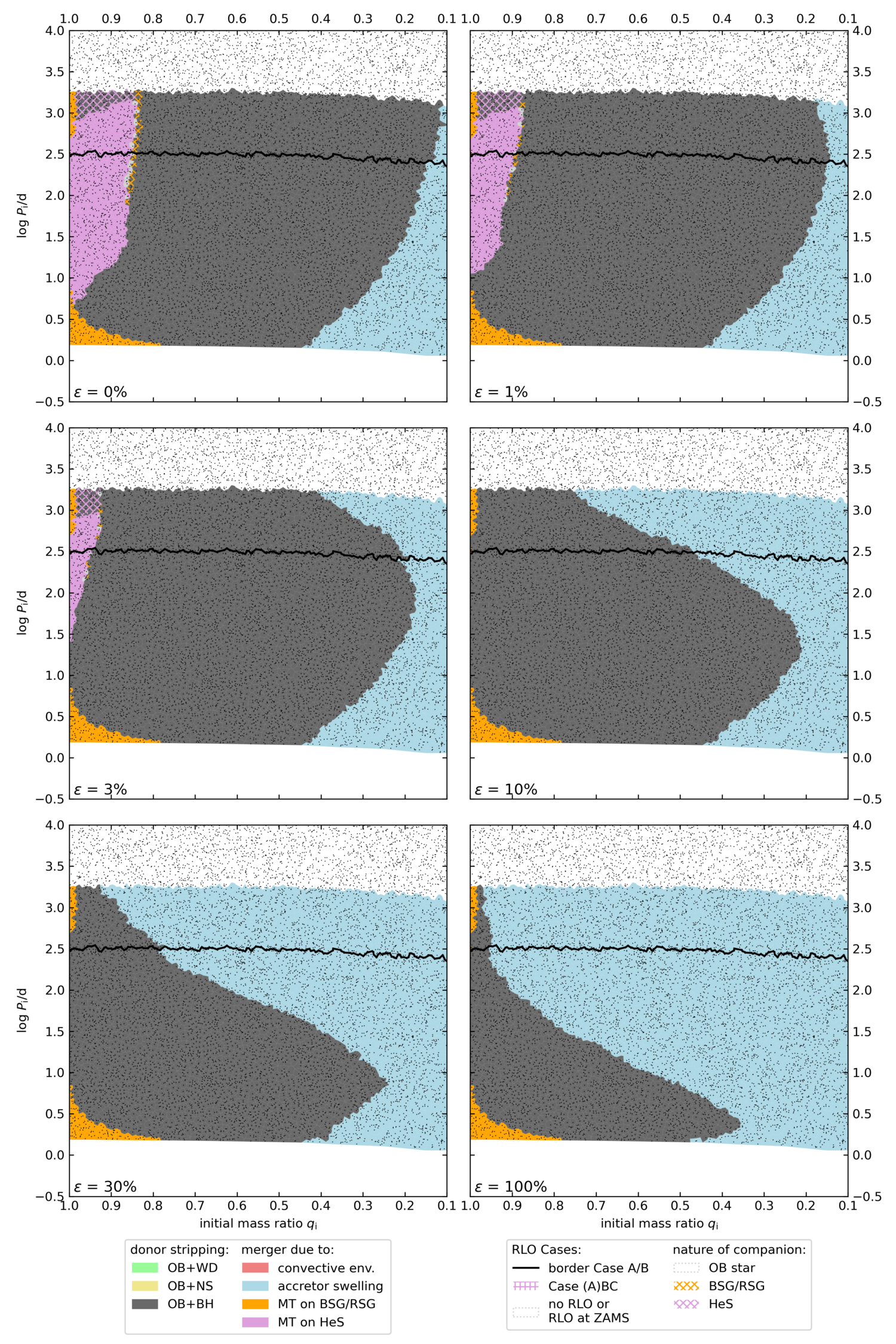}
 \caption{As Fig.\,\ref{fig:Pq}, but for a primary mass of $70\msol$.}
 \label{fig-pq80}
\end{figure*}

\begin{figure*}
 \centering
 \includegraphics[width=0.8\textwidth]{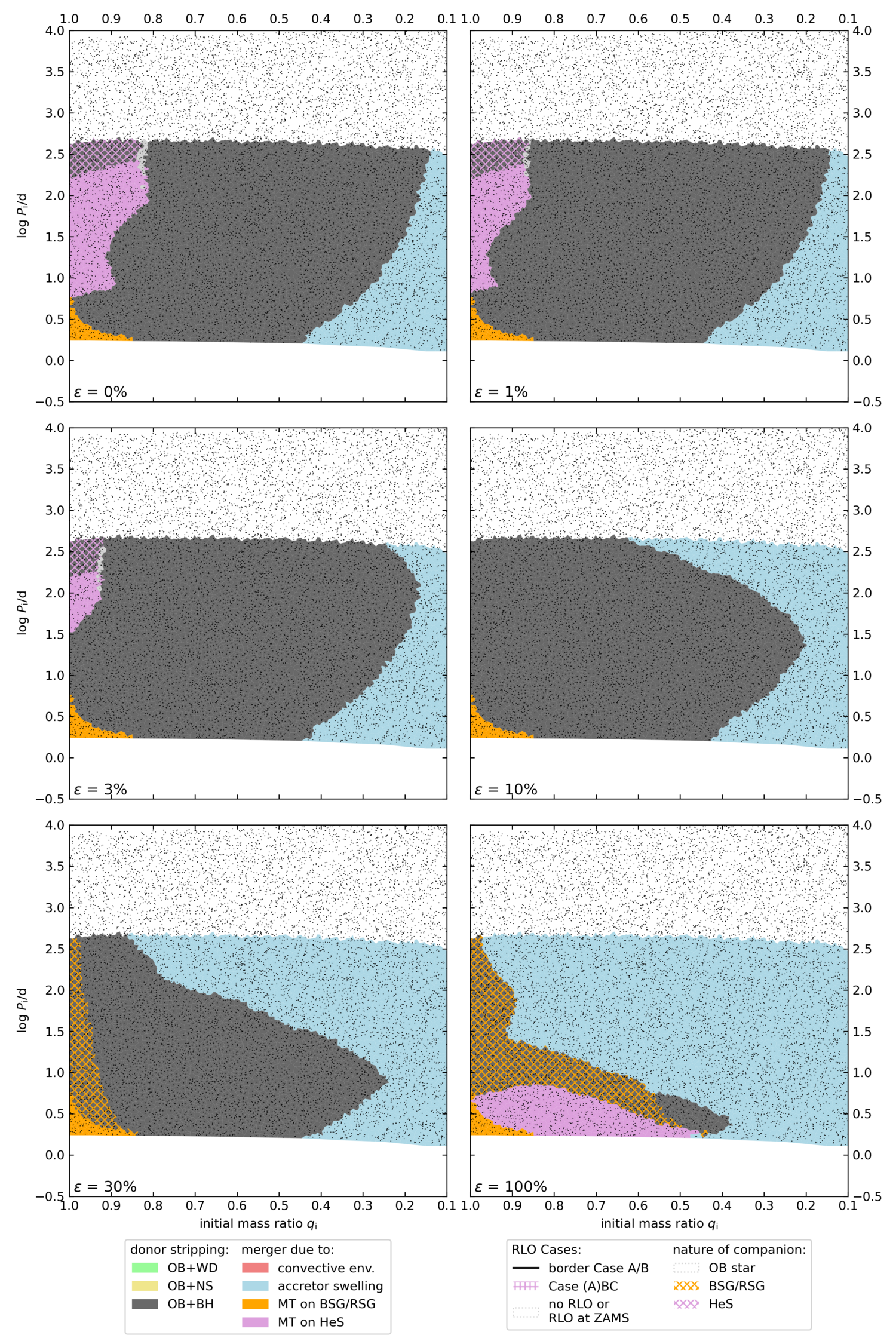}
 \caption{As Fig.\,\ref{fig:Pq}, but for a primary mass of $100\msol$.}
 \label{fig-pq100}
\end{figure*}

\end{document}